\documentclass[pre,reprint,amsmath,amssymb,aps,showpacs,nofootinbib]{revtex4-1}
\usepackage{epsf}
\usepackage{graphicx}
\usepackage{dcolumn}
\usepackage{color}
\usepackage{longtable}
\usepackage{fancyhdr}
\usepackage{amsmath}
\usepackage{boxedminipage}
\usepackage{amssymb}
\usepackage{amsmath}
\usepackage{array}
\usepackage{graphics}
\usepackage{psfrag}
\usepackage{epsfig}
\usepackage{graphicx}
\usepackage{tabularx}
\usepackage{calc}
\newcommand{\be}{\begin{eqnarray}}
\newcommand{\ee}{\end{eqnarray}}
\newcommand{\bal}{\begin{align}}
\newcommand{\eal}{\end{align}}

\newcommand{\Ma}{M \hskip -0.15em a}
\newcommand{\Kn}{K \hskip -0.15em n}

\newcommand{\e}{\text{e}}
\newcommand{\fx}{\boldsymbol{x}}

\newcommand{\fu}{\boldsymbol{\mathbf{u}}}
\newcommand{\fxi}{\boldsymbol{\mathbf{\xi}}}

\newcommand{\fnabla}{\boldsymbol{\mathbf{\nabla}}}
\newcommand{\feta}{\boldsymbol{\mathbf{\eta}}}

\newcommand{\fatj}{\boldsymbol{j}}

\newcommand{\mH}{\mathcal{H}}

\newcommand{\tw}{\mathrm{w}}

\newcommand{\st}[4]{S_{Q#1E#4}^{D#3V#2}}
\newcommand{\minDVMs}{minimal DVMs}
\newcommand{\augDVMs}{augmented DVMs}
\newcommand{\DMBC}{diffuse Maxwell boundary condition}
\newcommand{\LBms}{LB models}
\newcommand{\LBm}{LB model}
\newcommand{\wai}{wall accuracy index}
\newcommand{\wao}{wall accuracy order}
\newcommand{\nn}{\nonumber}
\def\R{{\mathbb{R}}}

\begin{document}
\preprint{}
\title{High-order lattice Boltzmann models for wall-bounded flows at finite Knudsen numbers}
\author{C.\ Feuchter}
\author{W.\ Schleifenbaum}
\affiliation{Department of Mechanical Engineering and Materials Science\\
Aalen University, Beethovenstrasse 1, D-73430 Aalen, Germany}
\date{\today}
\begin{abstract}
We analyze a large number of high-order discrete velocity models for solving the 
Boltzmann--BGK equation for finite Knudsen number flows.
Using the Chapman--Enskog formalism, we prove for isothermal flows a relation  
identifying the resolved flow regimes for low Mach numbers.
Although high-order lattice Boltzmann models recover flow regimes beyond the Navier--Stokes level we 
observe for several models significant deviations from reference results. 
We found this to be caused by their inability to recover the Maxwell boundary condition exactly.
By using supplementary conditions for the gas-surface interaction it is shown how to systematically generate  
discrete velocity models of any order with the inherent ability
to fulfill the diffuse Maxwell boundary condition accurately.
Both high-order quadratures and an exact representation of the boundary condition
turn out to be crucial for achieving reliable results.
For Poiseuille flow, we can reproduce the mass flow and slip velocity up to the 
Knudsen number of $1$. Moreover, for small Knudsen numbers, the Knudsen layer behavior is recovered.
\end{abstract}
\maketitle
\section{Introduction \label{sec1} }
Fluid flow at very small scales has gained an increasing amount of attention recently due to its relevance for engineering applications in micro- and nanotechnologies \cite{HoTai,Karni}, 
e.g.\ microelectromechanical systems (MEMS) and porous media. The characteristic length scale $l_0$ 
of the corresponding geometries is in the range of the mean free path length $\lambda$ of 
the gas molecules. Such flows are often isothermal 
and characterized by extremely small Mach numbers. Nevertheless, these flows can become 
compressible because of considerable pressure gradients caused by viscous effects \cite{Arki}. 
Based on the Chapman--Enskog (CE) expansion \cite{Chapman}, the Knudsen number, defined as $\Kn=\lambda/l_0$, can be 
considered a measure for the deviation of the flow behavior from thermodynamic equilibrium. 
For sufficiently large $\Kn$ non-equilibrium effects become important and the 
validity of the Navier--Stokes equation breaks down. 
In particular, the gas-surface interaction is very complex and cannot be described by the usual no-slip boundary condition. Within the Knudsen layer \cite{Hadji} the notion of the fluid as a continuum is no longer valid.

A fundamental description of hydrodynamics beyond the Navier--Stokes equation is provided by 
the Boltzmann equation valid for all values of $\Kn$ and all flow regimes \cite{ref1.3}. 
Accurate simulations of finite $\Kn$ flows are achieved by low-level methods solving 
the Boltzmann equation numerically, e.g.\ Direct Simulation Monte Carlo (DSMC) which is traditionally used for high Mach number flows \cite{Bird}. However, the application of the 
DSMC method to low Mach number microflows requires a large number of samples to reduce 
statistical errors and is computationally very time consuming.

Therefore, reduced-order models of the Boltzmann equation, such as the Lattice--Boltzmann (LB) approach, have become attractive alternatives \cite{QianLallemand,ChenDoolen,ShanHe,HeLuo}. The LB method is
 based on a reduction  of the molecular velocities to a discrete velocity set in configuration 
space. Standard \LBms, e.g.\ the $D3Q19$ model with 
19 discrete velocities, were developed to describe the Navier--Stokes fluid dynamics. Nowadays, they provide a well-established methodology for the computational modeling of various flow 
phenomena \cite{Succi}. Furthermore, the LB method achieves promising results for microflow 
simulations 
\cite{Lim,NieDoolen,Toschi,Ansumali,Verhaeghe,Reis}. In particular the {\DMBC }  for the 
gas-surface interaction can be implemented at a kinetic level  
\cite{Gatignol,Inamuro,AnsumaliKarlin,SofoneaSekerka,Sofonea,Meng}.

A systematic extension of the LB method to high-order hydrodynamics beyond the Navier--Stokes 
equation has been derived by Shan et al.\ \cite{ShaYuaChe06}. These models are based on an 
expansion of the velocity space using Hermite polynomials in combination with appropriate 
Gauss--Hermite quadratures. 
First analytical solutions of the high-order {\LBm } $D2Q16$ were 
presented by Ansumali et al.\ \cite{AnsumaliKarlinArcidiacono} for Couette flow and by 
Kim et al.\ \cite{KimPitsch} for Poiseuille flow. The collection of LB models in the literature has grown successively, see e.g. Refs. \cite{Shan,ChiKar09}, but makes no claim to be complete in any sense. It was shown \cite{AnsumaliKarlinArcidiacono,KimPitsch} that a higher Gauss--Hermite 
quadrature order significantly improves the simulation accuracy for finite $\Kn$ 
flows compared to standard \LBms. However, several studies \cite{1.10,Izarra,MengZhang1,MengZhang2} 
of high-order LB methods indicate that the accuracy for finite $\Kn$ flows does not 
monotonically increase with a higher Gauss--Hermite order and sensitively depends on the chosen 
discrete velocity set. Generally, it was observed that discrete velocity sets with an even 
number of velocities perform better than sets with the same Gauss--Hermite order but an odd 
number of velocities. Moreover there are some {\LBms }  which show considerable deviations 
from DSMC results for finite $\Kn$ despite a high Gauss--Hermite quadrature order. 
It has been suggested that this is caused by gas-surface interactions 
\cite{MengZhang2,Brookes}. 
The implementation of the {\DMBC } using Gauss--Laguerre off-lattice quadrature models in Ref.\ \cite{Ambrus} shows good results for Couette flow up to $Kn=0.5$.
By using an alternative framework, a high-order LB model with only $27$ discrete velocities has been developed by Yudistiawan et al.\ \cite{Yudi10} and it was shown that the corresponding moment system is quite similar to Grad's $26$-moment system. This off-lattice $D3Q27$ model is able to represent both, Knudsen layer effects and the Knudsen minimum for Poiseuille flow.

\begin{figure}
\centering
\epsfig{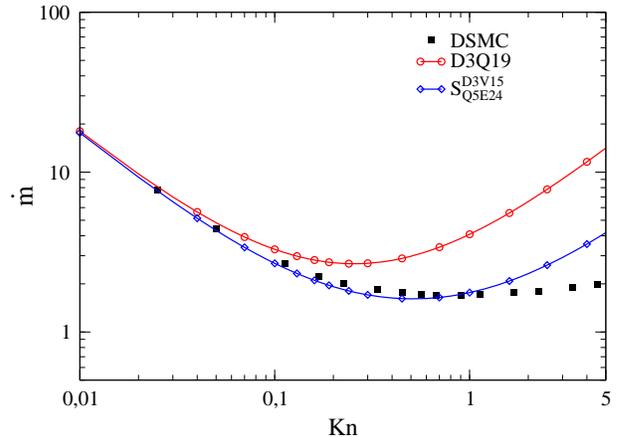}
\caption{\label{fig:intro} Normalized mass flow rate for Poiseuille flow. Both LB models $D3Q19$ and $S_{Q5E24}^{D3V15}$
are of quadrature order $5$ but differ in the accuracy of the wall moment $W_{zxx}$ and, consequently, in their capability
to recover the {\DMBC }.}
\end{figure}

As an introductory example, we consider the most commonly used LB model $D3Q19$ accompanied by the {\DMBC } for unknown distribution functions at solid walls. It is well-known that the $D3Q19$ model shows deviations from reference results (e.g.\ DSMC) for increasing $\Kn$. As shown in Fig.\ \ref{fig:intro}, the normalized mass flow rate for Poiseuille flow becomes inaccurate for $\Kn\gtrsim 0.05$. One reason for this deficiency is the inability of the $D3Q19$ velocity set to represent the {\DMBC } accurately. This drawback can be measured by half-space integrals at the wall \cite{Brookes}. At a solid wall any velocity moment of the distribution function can be decomposed into two parts, a half-space integral over distributions coming from the bulk region and another half-space integral (wall moment) which is given by distributions defined by the Maxwell boundary condition. For instance, the $D3Q19$ model evaluates the third wall moment $W_{zxx}$ (defined in Eq.\ \eqref{defW}), numerically yielding an error of $28\%$. In order to achieve a better representation of the Maxwell boundary condition, however, it is mandatory that this wall moment is captured with a higher accuracy. An alternative LB model $S_{Q5E24}^{D3V15}$, see Appendix \ref{sec:appDVM}, with $15$ velocities and the same quadrature order of $5$ yields an error of only $2\%$ for the wall moment $W_{zxx}$. Obviously, its result for the mass flow rate shown in Fig.\ \ref{fig:intro} is much more accurate compared to $D3Q19$. This indicates that the LB model's representation of the {\DMBC } requires high precision.  Furthermore, due to the strong restriction of the configuration space, the standard LB velocity models capture only the first order of the CE expansion and therefore the applicability for describing finite $\Kn$ flows is limited \cite{Guo}. Both models shown in Fig.\ \ref{fig:intro} are not able to reproduce the Knudsen layer at solid walls where strong non-equilibrium effects are relevant. The description of finite Knudsen number flows must incorporate high-order flow regimes. It is therefore desirable to obtain high-order LB models which additionally are capable of representing the {\DMBC } accurately. 

In this work we systematically develop and investigate a large number of new high-order Gauss--Hermite {\LBms } (on-lattice) which fulfill a constraint guaranteeing an accurate implementation of the {\DMBC }. Consequently these models ensure vanishing errors of the relevant half-space integrals. We prove a theorem using the Chapman--Enskog expansion which specifies for low Mach number flows the recovered flow regimes beyond the Navier--Stokes regime depending on the Gauss--Hermite quadrature order. First simulation results for Poiseuille flow at finite $\Kn$ show that those high-order {\LBms }  which recover the {\DMBC } exactly achieve excellent agreement with DSMC results. In particular these models are able to describe Knudsen layer effects at solid walls. We emphasize that we do not resort to slip-boundary models in order to achieve these results. Finally, we recommend a preferred LB model with $96$ discrete velocities and a $7$th order Gauss-Hermite quadrature order.

The paper is organized as follows. In Sect.\ \ref{sec:dvm}, a brief review of {\LBms } and the generation of high-order discrete velocity models is presented. In Sect.\ \ref{sec:flowRegimes}, we derive the low Mach number theorem mentioned above. In Sect.\ \ref{sec:GasSurf}, we present the systematic generation of high-order {\LBms } 
with the inherent ability to capture the {\DMBC } accurately. In Sect.\ \ref{NM}, we discuss the numerical implementation. In Sect.\ \ref{sec:Poi}, we compare first simulation results for Poiseuille flow at finite $\Kn$ with DSMC data and discuss the ability of the new {\LBms } to describe the Knudsen layer behavior at solid walls. In Sect.\ \ref{C}, a summary of the major conclusions is given.

\section{Definition of Lattice--Boltzmann models}
\label{sec:dvm}
We consider the one-particle velocity distribution function \(f(\fxi)\) governed by the 
Boltzmann equation with the Bhatnagar--Gross--Krook (BGK) collision operator,
\begin{equation}
\label{LBBGK}
\left(\partial_t+\fxi\cdot \fnabla\right)f(\fxi)=-\frac{1}{\tau }\left(f(\fxi)-f^{(0)}(\fxi,\rho,\fu)\right)\; .
\end{equation}
The distribution function \(f(\fxi)\) relaxes to the equilibrium function
\begin{equation}
\label{deff0}
f^{(0)}(\fxi,\rho,\fu)=\frac{\rho }{(2\pi  \theta )^{D/2}}\exp \left(-\frac{(\fu-\fxi)^2}{2 \theta }\right)
\end{equation}
on the time scale \(\tau\). At each point in space and time the macroscopic quantities are 
defined as velocity moments in \(D\)-dimensional space,
\begin{equation}
\label{defM}
M_{i_1\ldots  i_n}=\int d^D\xi\:  f(\fxi)\:\xi _{i_1}\ldots  \xi _{i_n}\; ,
\end{equation}
such as the density \(\rho\), the velocity \(\fu\) and the temperature \(\theta\),
\begin{subequations}
\begin{eqnarray}
\label{defRho}
\rho &=&M\; ,\\
\label{defU}
\rho  u_i&=&M_i\; ,\\
\label{defTheta}
\rho  \left(D \theta +u_iu_i\right)&=&M_{\text{ii}}\; .
\end{eqnarray}
\end{subequations}
Repeated indices are summed over. All quantities are dimensionless and expressed in units of 
characteristic scales, i.e.\ the length scale \(l_0\), the reference density \( \rho_0 \), 
and the isothermal speed of sound \(c_0 = \sqrt{R \theta_0} \) with the specific gas constant \(R\) and
the reference temperature \( \theta_0 \).
Since we focus on flows with a low Mach number $\Ma=|\fu|$, we assume constant
temperature and set \(\theta =1\) henceforth. 
The time $t$ as well as the relaxation time $\tau$ is expressed in units of $t_0=l_0/c_0$.

Following the work of Grad \cite{Grad1,Grad2}, Shan and He \cite{ShanHe} and 
Shan et al.\ \cite{ShaYuaChe06} we discretize the velocity space by expanding 
the distribution 
function in the Hilbert space of tensorial Hermite polynomials up to an 
arbitrary order $N$,
\begin{align}
\label{expandf}
f(\fxi) & \approx f^N(\fxi) = \omega (\fxi)\sum _{n=0}^N \frac{1}{n!}\: a_{i_1\dots i_n}\mH^{(n)}_{i_1\dots i_n}(\fxi) \;, 
\end{align}
where the Hermite polynomials \(\mH^{(n)}_{i_1\dots i_n}(\fxi )\) are introduced by the recurrence relation
\begin{subequations}
\label{defH}
\be
\mH^{(n+1)}_{i_1\dots i_{n+1}}(\fxi )&=&\left(\xi_{i_{n+1}} - \partial_{i_{n+1}}^{(\xi)}\right)\mH^{(n)}_{i_1\dots i_n}(\fxi )\; ,\\ \mH^{(0)}(\fxi)&=&1\; .
\ee
\end{subequations}
The lowest-order term of the expansion \eqref{expandf} is given by the weight function $\omega$,
\begin{equation}
\omega (\fxi )=\frac{1}{(2\pi )^{D/2}\text{}}e^{\left.-\fxi^2\right/2}\; ,
\end{equation}
and the expansion coefficients 
\be
\label{Hermitecoef}
a_{i_1\dots i_n} &=&\int d^D\xi\:f^N(\fxi)\: \mH^{(n)}_{i_1\dots i_n}(\fxi) \; 
\ee
correspond to the moments of the distribution function.
\subsection{Quadratures on Cartesian lattices}
In the configuration space of a discrete velocity model the molecular velocities $\fxi_\alpha$ are 
restricted to a Cartesian lattice \(X\) with uniform spacing $c$, called lattice speed, such 
that the components $\xi_{\alpha i}/c$ are integer-valued. A subset of $X$ containing a 
number $V$ of these velocities, the stencil \(S\subset X\)
\begin{equation}
S=\left\{\left.\fxi_{\alpha }\right|\alpha =1,\ldots ,V\right\}\; ,
\end{equation}
is used with the corresponding weights $\tw_\alpha$ to compute the integral in Eq.\ \eqref{defM} by 
quadrature,
\be
\label{genQua}
M_{i_1\ldots  i_n}&=&\int d^D\xi\:\omega(\fxi)\: P_{i_1\ldots  i_n}(\fxi)\nonumber\\&=&\sum_\alpha \tw_\alpha P_{i_1\ldots  i_n}(\fxi_\alpha)\; ,
\ee
with the function $P_{i_1\ldots  i_n}(\fxi)=f(\fxi)\xi _{i_1}\ldots  \xi _{i_n}/\omega(\fxi)$. 
For a polynomial $P_{i_1\ldots  i_n}(\fxi)$ of degree $Q$
the quadrature \eqref{genQua} is exact as long as it satisfies the orthogonality relation
\begin{align}
\label{orth}
&\int d^D\xi\:\omega(\fxi)\: \mH^{(n)}_{i_1\dots i_n} ({\fxi})
\mH^{(m)}_{j_1\dots j_m} ({\fxi})\nonumber\\
&\quad = \sum_{\alpha=1}^V \tw_{\alpha} \mH^{(n)}_{i_1\dots i_n} ({\fxi}_{\alpha})
\mH^{(m)}_{j_1\dots j_m} ({\fxi}_{\alpha})\nonumber\\ &\quad =   \begin{cases}
     1                                 &  \textrm{if }(i_1,\dots,i_n) = \mathrm{perm}((j_1,\dots, j_m)) \\     0                                 & \mathrm{else} \\
   \end{cases} 
\; ,\nonumber\\&\qquad n+m\leq Q
\end{align}
up to the $Q$th order, where perm($\fatj$) denotes a (odd or even) permutation of the 
vector $\fatj=(j_1, \ldots, j_m)$. This can be readily seen by expanding $P_{i_1\ldots  i_n}(\fxi)$ 
using Hermite polynomials
\begin{align}
\label{scaInt}
& M_{i_1\ldots  i_n}= \sum_{m=0}^Q p_{i_1\dots i_n,j_1\dots j_m} \int d^D\xi\:\omega(\fxi)\mH^{(m)}_{j_1\dots j_m}(\fxi)
\end{align}
with some coefficients $p_{i_1\dots i_n,j_1\dots j_m}$. If Eq.\ \eqref{orth} holds, the $Q$th-order polynomial 
in Eq.\ \eqref{scaInt} is evaluated exactly by the quadrature and hence Eq.\ \eqref{genQua} holds 
as well. We then refer to the stencil $S$ along with its weights $\tw_\alpha$ as a {\it discrete 
velocity model} (DVM) with {\it quadrature order} $Q$. Note that we write the moments \eqref{genQua} 
as sums on discrete velocities, implicitly assuming a sufficiently high quadrature order, unless 
indicated.

A Lattice--Boltzmann (LB) model solves for the variables
\be
\label{deffa}
f_\alpha=\tw_\alpha\frac{f(\fxi_\alpha)}{\omega(\fxi_\alpha)}
\ee
in the LB--BGK equation
\be
\label{LBE}
\left(\partial_t+\xi_{\alpha i}\partial_i\right)f_\alpha=-\frac{1}{\tau}\left(f_\alpha-f_\alpha^{(0)}\right)\: ,
\ee
using a DVM for the evaluation of the macroscopic variables
\begin{subequations}
\label{macroVarDVM}
\begin{eqnarray}
\rho &=&\sum_\alpha f_\alpha\; ,\\
\rho  u_i&=&\sum_\alpha f_\alpha\xi_{\alpha i}\; ,\\
\rho  \left(D \theta +u_iu_i\right)&=&\sum_\alpha f_\alpha\xi_{\alpha i}\xi_{\alpha i}\; .
\end{eqnarray}  
\end{subequations}
The latter serve the {\LBm }  to determine the equilibrium function
\be
\label{equi.hermite}
f_\alpha^{(0)}(\rho,\fu)&=&\tw_\alpha\frac{f^{(0)}(\fxi_\alpha,\rho,\fu)}{\omega(\fxi_\alpha)} \nonumber\\&=& \tw_{\alpha} \sum^{N}_{n=0} 
\frac{1}{n!} a^{(0)}_{i_1 \dots i_n} \mH^{(n)}_{i_1 \dots i_n} ({\fxi}_{\alpha})\; ,
\ee
expanded to a {\it Hermite order} $N$ where the Hermite coefficients for isothermal flows 
are given by
\begin{align}
\label{equi.hermite.coef}
a^{(0)}_{i_1 \dots i_n}  = \rho \: u_{i_1} \dots u_{i_n}\; .
\end{align}

For any stencil \(S\), the weights \(\tw_{\alpha }\) are obtained by
solving the set \eqref{orth} of linear equations. We decompose the stencil $S=\cup _{g=1}^GS_g$  
into a number $G$ of velocity sets ({\it groups}), each group $S_g$ containing $V_g$ velocities 
(s.t.\ $V=\sum_g V_g$) generated by the symmetries of the lattice. These groups $S_g$ can be 
obtained by reflecting a single $\fxi^{(g)}\in S_g$ on those hyperplanes of the lattice which 
reproduce the lattice itself upon reflection. Hence, the velocity weights must be identical in 
each group,\footnote{Note that the metric norm of the velocities, $|\fxi|^2=\xi_i\xi_i$, is not 
useful to identify the weights. E.g.\ the velocities $(3,0,0)$ and $(2,2,1)$ have the same norm, 
however, they are not part of the same group.}
\begin{equation}
\label{wdach}
\tw_{\alpha }=\overline{\tw}_g\;\forall \left\{\alpha \left|\:\fxi _{\alpha }\in S_g\right.\right\}\; .
\end{equation}
It is efficient to rewrite Eq.\ \eqref{orth} into the form
\begin{equation}
\label{defQua}
K_{\alpha \beta }(S)\tw_{\beta }=0\quad  \forall\alpha\; ,\qquad\sum_{\alpha=1}^V \tw_\alpha =1\; ,
\end{equation}
where \(K\) is a symmetric matrix with elements
\begin{equation}
\label{defK}
K_{\alpha \beta }(S)=\sum_{m=1}^QH_{i_1\ldots i_m}^{(m)}(\fxi_\alpha)H_{i_1\ldots i_m}^{(m)}(\fxi_\beta)
\end{equation}
and where the Hermite tensor indices are contracted.
In view of Eq.\ \eqref{scaInt}, 
it is sufficient to solve Eq.\ \eqref{orth} for $n=0$ and $m\leq Q$. Obviously, Eq.\ \eqref{defQua} 
then follows from Eq.\ \eqref{orth} by multiplication of $\mH^{(m)}_{i_1 \ldots i_m}(\fxi_\beta)$ and summing 
on $m$. The reverse is true because Eq.\ \eqref{defQua} can be written, after multiplication by $\tw_\alpha$, as 
\be 0=\tw_\alpha K_{\alpha\beta}(S)\tw_\beta = \sum_{m=1}^Q\left(\sum_\alpha\mH^{(m)}_{i_1\ldots i_m}(\fxi_\alpha)\tw_\alpha\right)^2\nonumber\ee 
which requires that each part of the sum be zero.
To obtain the matrix $K(S)$, the scalar product of two Hermite tensor polynomials is obtained by the recurrence formula (no sum on $m$)
\begin{align}
&\mH^{(m+1)}_{i_1\dots i_{m+1}}(\fxi) \mH^{(m+1)}_{i_1\dots i_{m+1}}(\feta)=
\left(
\xi_{i_{m+1}}\eta_{i_{m+1}}
-\xi_{i_{m+1}}\partial_{i_{m+1}}^{(\eta)}
\right.
\nn\\
&\quad\left.
-\eta_{i_{m+1}} \partial_{i_{m+1}}^{(\xi)}
+\partial_{i_{m+1}}^{(\xi)}\partial_{i_{m+1}}^{(\eta)}
\right)
\mH^{(m)}_{i_1\dots i_{m}}(\fxi)\mH^{(m)}_{i_1\dots i_{m}}(\feta)\; ,
\end{align}
which follows from Eq.\ \eqref{defH}. 

Finding the weights $\tw_\alpha$ for a stencil reduces to the task of finding the null 
space of the symmetric matrix \(K(S)\) and normalizing according to Eq.\ (\ref{defQua}). 
As a special feature, the matrix \(K(S)\) has a parametric dependence on \(c\) (through \(S\)) 
such that for some discrete values $c^*\in\R$ its nullity is increased. This means that by 
populating the stencil with sufficiently many velocity groups, we find for each quadrature order $Q$ a minimal number $G$ of velocity groups such that for $c=c^*$ 
\begin{equation}
\label{dimker}
\dim\ker K(S)=1\: 
\end{equation}
while for $c\neq c^*$ the matrix $K(S)$ is regular and no solution of Eq.\ \eqref{defQua} can be found. The
solutions of Eq.\ \eqref{defQua} which obey Eq.\ \eqref{dimker} with $c=c^*$ are referred to as {\it\minDVMs}.
We only consider DVMs with positive weights.
Fixing the quadrature order $Q$ and the number \(V\) of velocities, there is a countable 
infinity of {\minDVMs}, since \(S\) can be chosen from the (virtually) unbounded lattice \(X\). 
However, by introducing the integer-valued {\it energy} \(E\) of a stencil and limiting it from 
above,
\begin{equation}
E=\frac{1}{2c^2}\sum _{\alpha =1}^V\fxi_{\alpha}\cdot \fxi_{\alpha } \leq E_{\max }
\end{equation}
the number of available $Q$th-order {\minDVMs }  becomes finite. We are thus able to 
determine the complete set of {\minDVMs }  inside the energy sphere defined by $E\leq E_{max}$. 
The notation
\be
\label{Ssymbol}
S_{Q\langle Q\rangle E\langle E\rangle}^{D\langle D\rangle V\langle V\rangle} \; ,\quad\textrm{e.g.\ }D3Q19\rightarrow S_{Q5E15}^{D3V19}
\ee
is introduced for the DVMs. As a reminder, we refer to the spatial dimension as \(D\), the 
quadrature order as \(Q\), the stencil's energy as \(E\) and the number of velocities as \(V\). 

\begin{table}[htp]
\centering
\begin{tabular}{|c|c|c|c|c|c|c|}
\hline
& $D$ & $Q$ & $E_{max}$ & $\#$ & $V_{min}$ &  $\Lambda_{\mathrm{index}}$ \\
\hline
 ${\cal S}_1$ & 3 & 7 & 500 & 4677 & 38 &  0 \\
 ${\cal S}_2$ & 3 & 9 & 625 & 618 & 79 &  0 \\
 ${\cal S}_3$ & 3 & 7 & - & 500 & - & 16 \\
 ${\cal S}_4$ & 3 & 7 & 2500 & 45863 & 80 & 6245\\
 ${\cal S}_5$ & 2 & 7 & 250 & 1188 & 16 & 0 \\
 ${\cal S}_6$ & 2 & 9 & 300 & 592 & 33 & 0 \\
 ${\cal S}_7$ & 2 & 7 & 1000 & 21952 & 20 & 549\\
\hline
\end{tabular}
\caption{\label{tab:dvmSummary} Overview of DVM sets ${\cal S}_n$ which comprise a number $\#$ of minimal DVMs with dimension $D$, quadrature order $Q$, and upper limit $E_{max}$ for the stencil energy. For each set ${\cal S}_n$, the minimal velocity count $V_{min}$ as well as the {\wai } $\Lambda_{\mathrm{index}}$ (for details, see Sect.\ \ref{sec:GasSurf}) is shown. The symbol `-' is used where a quantity was not considered. }
\end{table}

\subsection{Discrete velocity models for bulk flow}
Several sets of DVMs, denoted by ${\cal S}_n$, $n=1,2,\dots$, are introduced in this paper to alleviate the discussion on efficiency and accuracy of the DVMs. An overview of these sets is given in Tab.\ \ref{tab:dvmSummary}. Each set is defined by its dimension $D$, its quadrature order $Q$, the upper limit $E_{max}$ of the stencil's energy and the {\wai } $\Lambda_{\mathrm{index}}$, see Sect.\  \ref{sec:WallModels}. The {\wai } has the value $\Lambda_{\mathrm{index}}=0$ for the DVMs for bulk flow considered here. A set ${\cal S}_n$ contains a number $\#$ of {\minDVMs }  with $V_{min}$ being the minimal velocity count. While in one spatial dimension, it is well-known how to find $V_{min}$ for Gauss-Hermite quadratures, in higher dimensions this is not obvious. For most common quadratures in three dimensions, $V>V_{min}$. Although small values of $V$ are desirable for a high computational performance, it is unclear which choice of the stencil is the most accurate for finite $\Kn$ number flows. It is shown in the next sections what are the essential features of a DVM for the resolution of high-order hydrodynamic regimes. 

\begin{figure}
\centering
\epsfig{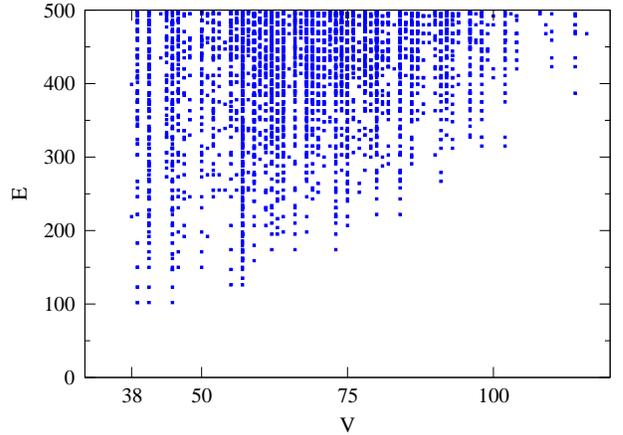}
\caption{\label{fig:scatterQ7} Complete set ${\cal S}_1$ of DVMs for $D=3$, $Q=7$ and $E\leq 500$, shown in the 
diagram of velocity count $V$ and energy $E$. Each dot represents a DVM.}
\end{figure}
\begin{figure}
\centering
\epsfig{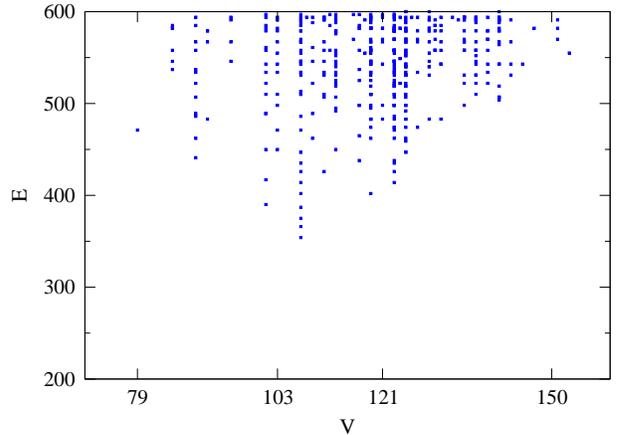}
\caption{\label{fig:scatterQ9} Complete set ${\cal S}_2$ of DVMs for $D=3$, $Q=9$ and $E\leq 625$, 
shown in the diagram of velocity count $V$ and energy $E$. Each dot represents a DVM. Within this energy range one model with a minimal number of $79$ velocities exists.}
\end{figure}

In Fig.\ \ref{fig:scatterQ7}, the $4677$ three-dimensional, $7$th-order DVMs of the set ${\cal S}_1$ are depicted. Each dot represents a DVM. Evidently, the sets of DVMs can only be complete if an upper bound on the stencil energy is provided. Here, we arbitrarily chose $E_{max}=500$ but of course this number may be increased. The lowest velocity count within ${\cal S}_1$ yields $V_{\min}=38$. There are two models of this kind, displayed in Tabs.\ \ref{tab:dvmSet1Amin}-\ref{tab:dvmSet1Aminn}. All tables of DVMs are deferred to the Appendix. In Fig.\ \ref{fig:scatterQ9}, we show the $618$ three-dimensional, $9$th-order DVMs of the set  ${\cal S}_2$. The lowest velocity count yields $V_{\min}=79$. This DVM which is noted in Tab.\ \ref{tab:dvmSet2Amin} complements previously known DVMs with minimal velocities \cite{Shan}. A widely used $9$th-order DVM is the one shown in Tab.\ \ref{tab:dvmSet2standard}, also known as $D3Q121$ \cite{Meng}.

The so-called DVMs for wall-bounded flows of the sets ${\cal S}_3$ and ${\cal S}_4$ ($\Lambda_{\mathrm{index}}>0$) will be discussed in Sect.\ \ref{sec:WallModels}. Candidates of the sets ${\cal S}_5$, ${\cal S}_6$, and ${\cal S}_7$, which concern the popular research area of {\LBms } in $2$ spatial dimensions, are given in Appendix \ref{sec:2D}. In the sequel, we consider three spatial dimensions, $D=3$.

\section{Lattice--Boltzmann hydrodynamics \label{LBH}}
\label{sec:flowRegimes}
In this Section we study the capability of {\LBms} with Gauss--Hermite quadrature order $Q$
and Hermite order $N$ to capture isothermal ($\theta=1$) microflows beyond the 
Navier--Stokes flow regime by using the Chapman--Enskog (CE) expansion.
We restrict our analysis to low $\Ma$ flows. The flow phenomena thus described bear fixed values of 
the Reynolds number $Re$, $\Kn$ and $\Ma$ where $\Ma \ll 1$ is a small number allowing an expansion.

By taking moments of the discrete LB--BGK equation \eqref{LBE} with respect to the discrete 
particle velocities ${\fxi}_{\alpha}$ macroscopic moment equations can be derived. 
For the density $\rho$ and the momentum $\rho {\bf u}$, we obtain the evolution equations
\begin{subequations}
\label{hydro}
\begin{align}
\label{hydro.1}
\partial_t \rho + \partial_i \left( \rho u_i \right) & = 0 \\
\label{hydro.2}
\partial_t \left( \rho u_i \right) + \partial_j \Pi_{ij} & = 0 
\end{align}
\end{subequations}
with the momentum flux tensor
\begin{align}
\label{hydro.3}
\Pi_{ij} & = \sum_{\alpha} f_{\alpha} \xi_{\alpha i} \xi_{\alpha j}\;  .
\end{align}
Equations (\ref{hydro.1}) and (\ref{hydro.2}) represent mass and momentum conservation, respectively, guaranteed by invariants of the BGK collision operator.
Corresponding equations for higher moments, e.g.\ $\Pi_{ij}$, are derived by taking moments 
of Eq.\ \eqref{LBE} with more than one particle velocity $\fxi_{\alpha}$.
\subsection{Chapman--Enskog analysis \label{sec:CE}}
The CE analysis \cite{Chapman} is a perturbative method to solve the Boltzmann equation and is 
based on the assumption that the distribution function $f_{\alpha}$ deviates only slightly 
from the equilibrium $f^{(0)}_{\alpha}$. 
The CE expansion introduces a small parameter $\epsilon$ into the collision time 
$\tau \to \epsilon \tau$ which controls the perturbative analysis and is absorbed into $\tau$ after 
finishing the analysis. 
Physically, the parameter $\epsilon$ may be identified with the Knudsen number, which measures 
the deviation from equilibrium. 
The first, second, and third orders correspond to the Navier--Stokes, the Burnett, and Super--Burnett flow regimes, respectively.
Consequently the LB--BGK equation \eqref{LBE} becomes
\begin{align}
\label{ce.1}
\left( \partial_t + \xi_{\alpha i} \partial_i \right) f_{\alpha} & =  
- \frac{1}{\epsilon \tau } \left( f_{\alpha} - f^{(0)}_{\alpha} \right)
\end{align}
and the distribution function is expanded in powers of $\epsilon$
\begin{align}
\label{ce.2}
f_{\alpha} & = f^{(0)}_{\alpha} + \epsilon f^{(1)}_{\alpha} + \epsilon^2 f^{(2)}_{\alpha}  + \dots \text{  .}
\end{align}
The CE expansion is a multiple-scale expansion of both $f$ and $t$ 
\begin{align}
\label{ce.3}
\partial_t = \partial^{(0)}_t  + \epsilon  \partial^{(1)}_t 
+ \epsilon^2  \partial^{(2)}_t + \dots
\end{align}
with the solvability conditions
\begin{subequations}
\label{ce.4}
\begin{align}
\sum_{\alpha} f^{(n)}_{\alpha} &= 0\;,\quad  n \ge 1\; ,\\
\sum_{\alpha} f^{(n)}_{\alpha} \xi_{\alpha i} &= 0 \;,\quad  n \ge 1\;.
\end{align}
\end{subequations}
Therefore, high-order contributions $f^{(n)}_{\alpha} (n \ge 1)$ of the CE expansion do not 
contribute to the macroscopic density and flow velocity.
By inserting the CE ansatz (\ref{ce.2}) and (\ref{ce.3}) into the Boltzmann 
equation (\ref{ce.1}) we find the general solution for $n \ge 1$
\begin{align}
\label{ce.6}
f^{(n)}_{\alpha} &= - \tau \left[ \left( \partial^{(0)}_t  
+  \xi_{\alpha i} \partial_i  \right) f^{(n-1)}_{\alpha} 
+ \partial^{(1)}_t f^{(n-2)}_{\alpha} 
\right  .
\nn\\
&\qquad\qquad
\left.
+ \dots + \partial^{(n-1)}_t f^{(0)}_{\alpha} \right] \\
& = - \tau \left[ 
\partial_i \xi_{\alpha i} f^{(n-1)}_{\alpha} + \sum_{m=0}^{n-1} \partial^{(m)}_t f^{(n-m-1)}_{\alpha} 
\right] \nonumber .
\end{align}
Furthermore, the solvability conditions \eqref{ce.4} are equivalent to the relations
\begin{subequations}
\label{solva}
\be
\label{ce.7}
  \partial^{(0)}_t
  \begin{bmatrix}
  \rho \\
  u_i
  \end{bmatrix}
  &=&
  \begin{bmatrix}
  - \partial_i \left( \rho u_i \right) \\
  - u_j \partial_j u_i - \frac{1}{\rho} \partial_i \rho
  \end{bmatrix} 
\\
\label{sc2}
   \partial^{(n)}_t
  \begin{bmatrix}
  \rho \\
  u_i
  \end{bmatrix}
  &=&
  \begin{bmatrix}
  0 \\
  - \frac{1}{\rho} \partial_j \Pi^{(n)}_{ij} 
  \end{bmatrix}
\; , 
\quad  n \ge 1 \;,
\ee
\end{subequations}
where
\begin{align}
\label{ce.9}
\Pi^{(n)}_{ij} &= \sum_{\alpha} f^{(n)}_{\alpha} \xi_{\alpha i} \xi_{\alpha j} \; .
\end{align}
The CE analysis is expected to be valid for flow regimes where the system 
is not too far from equilibrium \cite{Harris}.
\subsection{Navier-Stokes and Burnett flow regimes \label{sec:CE:flowReg}}
The Navier--Stokes momentum flux tensor $\Pi^{(1)}_{ij}$ can be calculated by taking 
the second moment of Eq.\ (\ref{ce.6}) for $n=1$
\begin{align}
\label{ce.10}
\Pi^{(1)}_{ij} &=  - \tau \left( \partial^{(0)}_t \Pi^{(0)}_{ij}  
+  \partial_k Q^{(0)}_{ijk} \right) 
\end{align}
with
\begin{align}
\label{ce.10.2}
Q^{(n)}_{ijk} &= \sum_{\alpha} f^{(n)}_{\alpha} \xi_{\alpha i} \xi_{\alpha j} \xi_{\alpha k} \; .
\end{align}
Using Hermite polynomials and the orthogonality relation \eqref{orth} we obtain
\begin{align}
\label{ce.11}
\Pi^{(0)}_{ij} &=  \rho \delta_{ij} + \rho u_i u_j 
\end{align}
and
\begin{align}
\label{ce.12}
Q^{(0)}_{ijk} &=  \sum_{\alpha} f^{(0)}_{\alpha} \xi_{\alpha i} \xi_{\alpha j} \xi_{\alpha k}  \\
             &=  \sum_{\alpha} \tw_{\alpha} \sum^{N}_{n=0} \frac{1}{n!} a^{(0)}_{i_1 i_2 \dots i_n} 
	     \mH^{(n)}_{i_1 i_2 \dots i_n} ({\fxi}_{\alpha}) 
\nn\\
&\qquad\times\left[ \mH^{(3)}_{ijk} ({\fxi}_{\alpha}) 
	     +  \mH^{(1)}_{i} ({\fxi}_{\alpha})  \delta_{jk} 
\right.
\nn\\
&
\left.
\qquad\qquad+  \mH^{(1)}_{j} ({\fxi}_{\alpha})  
	     \delta_{ik} +  \mH^{(1)}_{k} ({\fxi}_{\alpha})  \delta_{ij} \right] \nonumber \\
\label{ce.13}
             &= \rho \left[ u_i \delta_{jk} + u_j \delta_{ik} + u_k \delta_{ij} \right] 
	     + \rho u_i u_j u_k \; .
\end{align}
Inserting these results into Eq.\ (\ref{ce.10}) and evaluating the time derivative with 
respect to Eq.\ (\ref{ce.7}) yields the 
Navier--Stokes momentum flux tensor $\Pi^{(1)}_{ij}$ for isothermal flows,
\begin{align}
\label{ce.14}
\Pi^{(1)}_{ij} &=  - \tau \rho \left( \partial_i u_j + \partial_j u_i \right) .
\end{align}
The standard {\LBms } with accuracy order $Q=5$ (e.g.\ D3Q19) do not capture the last
term in Eq.\ (\ref{ce.13}) and thus cause an error $| \Delta \Pi^{(1)}_{ij} | = \tau \left|  \partial_k \left( \rho u_i u_j u_k \right) \right| = 
\mathcal{O} \left( \Ma^3 \right)$  \cite{Qian}. On the other hand, if a DVM with 
quadrature order $Q\geq 6$ and Hermite order $N=3$ is used, we exactly recover the 
Navier--Stokes momentum flux tensor $\Pi^{(1)}_{ij}$. Evidently, the momentum flux tensor is $\mathcal{O} \left( \Ma \right)$ and higher-order $\Ma$ terms do not contribute when considering low $\Ma$ values.

The Burnett momentum flux tensor $\Pi^{(2)}_{ij}$ can be calculated by taking 
the second moment of Eq.\ (\ref{ce.6}) for $n=2$
\begin{align}
\label{ce.15}
\Pi^{(2)}_{ij} &=  - \tau \left( \partial^{(0)}_t \Pi^{(1)}_{ij} + \partial^{(1)}_t \Pi^{(0)}_{ij}  
+  \partial_k Q^{(1)}_{ijk} \right) .
\end{align}
The third moment of Eq.\ (\ref{ce.6}) for $n=1$ yields
\begin{align}
\label{ce.16}
Q^{(1)}_{ijk} &=  - \tau \left( \partial^{(0)}_t Q^{(0)}_{ijk} 
+  \partial_n R^{(0)}_{ijkn} \right) 
\end{align}
with
\begin{align}
\label{ce.17}
R^{(0)}_{ijkn} &=  \sum_{\alpha} f^{(0)}_{\alpha} \xi_{\alpha i} \xi_{\alpha j} \xi_{\alpha k}  \xi_{\alpha n} \; .
\end{align}
By applying the Hermite polynomial the tensor $R^{(0)}_{ijkn}$ can be expressed as
\begin{align}
\label{ce.18}
R^{(0)}_{ijkn} &=   \rho \left( \delta_{ij} \delta_{kn} + \delta_{ik} \delta_{jn} 
+ \delta_{in} \delta_{jk} \right) \nonumber\\
&\quad + \rho (  u_i u_j \delta_{kn} + u_i u_k \delta_{jn} + u_i u_n \delta_{jk}
+ u_j u_k \delta_{in} 
\nn\\
&\quad + u_j u_n \delta_{ik} + u_k u_n \delta_{ij} ) + \rho u_i u_j u_k u_n \; .
\end{align}
Inserting this result together with relation (\ref{ce.13}) into Eq.\ (\ref{ce.16}) we can 
compute the tensor $Q^{(1)}_{ijk}$, which is required for the evaluation of $\Pi^{(2)}_{ij}$.
After an extensive calculation of all terms in Eq.\ \eqref{ce.15} by considering the
multiple-scale time derivatives $\partial^{(0)}_t$ and $\partial^{(1)}_t$ we obtain the
Burnett momentum flux tensor for isothermal flows,
\begin{align}
\label{ce.19}
\Pi^{(2)}_{ij} &= 2 \rho \tau^2 \bigg[ 
\left( \partial_k u_i \right) \left( \partial_k u_j \right) 
\nn\\
&\quad\quad+ \left.\frac {1}{\rho^2} \left( \partial_i \rho \right) \left( \partial_j \rho \right) 
- \frac {1}{\rho}  \partial_i \partial_j \rho 
\right] .
\end{align}
The Burnett tensor for an incompressible flow field is $\Pi^{(2)}_{ij}={\cal O}( \Ma^2)$
and thus does not contribute to the momentum dynamics in the low $\Ma$ regime.
However, for finite $\Kn$ the flow behavior can become compressible even 
for low $\Ma$ \cite{Arki} and in that case we obtain contributions from the
last term of Eq.\ \eqref{ce.19}.
We will show in the following Section that these low $\Ma$ terms are recovered 
even for {\LBms }  with accuracy order $Q=5$.
\subsection{Low--Mach truncation error \label{LBH.3}}
Depending on the quadrature accuracy order $Q$ a {\LBm } is able to recover different 
flow regimes. In this Section we analyze the quadrature error with 
respect to the power of $\Ma$ and discuss the ability to capture high-order flow regimes
especially for low $\Ma$ flows. For this purpose we consider the $k$th moment of the 
$n$th CE level, which is defined by
\begin{align}
\label{ce.20}
M^{(n)}_{i_1 \dots i_k} &= \sum_{\alpha} f^{(n)}_{\alpha} \xi_{\alpha i_1} \dots \xi_{\alpha i_k}\; ,
\end{align}
where $k \ge 2$ because of the solvability conditions (\ref{ce.4}).
An equation for this moment can be derived by taking the $k$th moment of Eq.\ (\ref{ce.6})
with respect to the discrete particle velocities ${\fxi}_{\alpha}$
\begin{align}
\label{ce.21}
M^{(n)}_{i_1 \dots i_k} &= - \tau \left[  \partial^{(0)}_t  M^{(n-1)}_{i_1 \dots i_k}
+ \partial^{(1)}_t  M^{(n-2)}_{i_1 \dots i_k} + \dots 
\right.\nn\\
&\left.\quad\quad + \partial^{(n-1)}_t  M^{(0)}_{i_1 \dots i_k} 
+ \partial_j M^{(n-1)}_{i_1 \dots i_k j} \right] .
\end{align}
Based on this relation it can be easily shown by induction that the moment $M^{(n)}_{i_1 \dots i_k}$ is completely
determined by moments of the equilibrium distribution.\footnote{
For example  $M^{(1)}_{i_1 \dots i_k}$ is affected by two equilibrium moments
\begin{align}
\label{ce.22}
M^{(1)}_{i_1 \dots i_k} &= - \tau \left[  \partial^{(0)}_t  M^{(0)}_{i_1 \dots i_k}
+ \partial_{j_1} M^{(0)}_{i_1 \dots i_k j_1} \right] 
\end{align}
and accordingly the next CE moment ($n=2$)
\begin{align}
\label{ce.23}
M^{(2)}_{i_1 \dots i_k} &= - \tau \left[  \partial^{(0)}_t  M^{(1)}_{i_1 \dots i_k}
+ \partial^{(1)}_t  M^{(0)}_{i_1 \dots i_k}
+ \partial_{j_2} M^{(1)}_{i_1 \dots i_k j_2} \right] 
\end{align}
is also determined by equilibrium moments, where the highest moment 
($M^{(0)}_{i_1 \dots i_k j_1 j_2}$) occurs in the last term on the right 
hand side of Eq.\ (\ref{ce.23}) and contains $(k+2)$ velocities due to Eq.\ (\ref{ce.22}).
Continuing this argumentation we find that the highest equilibrium moment of the $n$-th CE moment
$M^{(n)}_{i_1 \dots i_k}$ is the $(k+n)$-th equilibrium moment $M^{(0)}_{i_1 \dots i_k j_1 \dots j_n}$ 
which is included in the last term on the right-hand side of Eq.\ (\ref{ce.21}).
This can be easily shown by induction.} 
We assume that the moment 
$M^{(n)}_{i_1 \dots i_k} (\partial^{(0)} \rho,\partial_{j_1}^{(1)} \rho, \partial_{j_1 j_2}^{(2)}\rho, ... , \partial^{(0)} u_i,\partial_{j_1}^{(1)} u_i, \partial_{j_1 j_2}^{(2)}u_i, ... )$ 
is an algebraic expression of $\rho$, $u_i$ and
their spatial derivatives, where $\partial^{(0)} = 1$ and $\partial_{j_1 \dots j_q}^{(q)} = \partial_{j_1} \dots \partial_{j_q}$.
By using the relations \eqref{solva} for the multiple-scale time derivatives the 
equation for $M^{(n)}_{i_1 \dots i_k}$ can be written as
\begin{align}
\label{ce.21b}
M^{(n)}_{i_1 \dots i_k} = &- \tau \bigg[  \partial^{(0)}_t  M^{(n-1)}_{i_1 \dots i_k} \nn\\
&- \frac{1}{\rho} \sum_{m=1}^{n-1} \sum_{q} 
\frac{\partial M^{(n-m-1)}_{i_1 \dots i_k}}{\partial ( \partial_{j_1 \dots j_q}^{(q)} u_s)}
\partial_{j_1 \dots j_q}^{(q)} \left( \partial_r \Pi^{(m)}_{rs} \right)  \nn\\
&+ \partial_j M^{(n-1)}_{i_1 \dots i_k j} \bigg] .
\end{align}
The sum on $q \ge 0$ in Eq.\ \eqref{ce.21b} accounts for all derivatives of $\rho$ and $u_i$ the 
moment $M^{(n)}_{i_1...i_k}$ depends on.
Because of a finite accuracy order of the Gauss--Hermite quadrature higher moments of the CE expansion
are only approximately captured and therefore high-order flow regimes are not accurately
recovered. In order to discuss such a truncation error for a moment $M^{(n)}_{i_1 \dots i_k}$  we
define the quadrature error
\begin{align}
\label{ce.24}
\Delta M^{(n)}_{i_1 \dots i_k} &= M^{(n)}_{i_1 \dots i_k} - M^{(n) \mathrm{DVM}}_{i_1 \dots i_k} ,
\end{align}
where $M^{(n) \mathrm{DVM}}_{i_1 \dots i_k}$ is a possibly inaccurate moment, calculated by a $\mathrm{DVM}$.

Due to the fact that a general CE moment $M^{(n)}_{i_1 \dots i_k}$ can be expressed by 
equilibrium moments it is important to consider, in a first step, the quadrature error 
for equilibrium moments
\begin{align}
\label{ce.25}
M^{(0)}_{i_1 \dots i_k} &= \sum_{\alpha} f^{(0)}_{\alpha} \xi_{\alpha i_1} \dots \xi_{\alpha i_k} ,
\end{align}
with the equilibrium function given by Eq.\ \eqref{equi.hermite}.
Note that the Hermite coefficients, see Eq.\ \eqref{equi.hermite.coef}, yield
\begin{align}
\label{ce.27}
a^{(0)}_{i_1 \dots i_n} = \mathcal{O} \left( \Ma^n \right)\; .
\end{align}
The product $\xi_{\alpha i_1} \dots \xi_{\alpha i_k}$ in (\ref{ce.25}) can be expressed by the 
Hermite polynomial $\mH^{(k)}_{i_1 \dots i_k} (\fxi_{\alpha})$, by $\mH^{(k-2)}_{i_1 \dots i_{k-2}} (\fxi_{\alpha})$, and terms with lower-order Hermite
polynomials
\begin{align}
\label{ce.28}
&\xi_{\alpha i_1} \dots \xi_{\alpha i_k} = \mH^{(k)}_{i_1 \dots i_k} (\fxi_{\alpha})
\nn\\&\quad+ \sum_{r<s} \mH^{(k-2)}_{i_1\dots i_{r-1}i_{r+1}\dots i_{s-1}i_{s+1}\dots i_{k}} (\fxi_{\alpha})
\:\delta_{i_{r} i_{s}}+ \dots \; .
\end{align}
Therefore we obtain 
\begin{widetext}
\begin{align}
\label{ce.29}
M^{(0)}_{i_1 \dots i_k} = \sum^{N}_{n=0} \frac{1}{n!} a^{(0)}_{j_1 \dots j_n} \sum_{\alpha}
\tw_{\alpha}\left[ \mH^{(n)}_{j_1 \dots j_n} (\fxi_{\alpha})
\mH^{(k)}_{i_1 \dots i_k} (\fxi_{\alpha})
+\sum_{r<s}\delta_{i_{r} i_{s}}\mH^{(n)}_{j_1 \dots j_n} (\fxi_{\alpha}) 
\mH^{(k-2)}_{i_1\dots i_{r-1}i_{r+1}\dots i_{s-1}i_{s+1}\dots i_{k}} (\fxi_{\alpha})
+ \dots \right] .
\end{align}
\end{widetext}
An exact evaluation of the moment $M^{(0)}_{i_1 \dots i_k} $ requires a sufficiently high Hermite order
$N \ge k$ of the equilibrium \eqref{equi.hermite} to ensure that the contributions of all Hermite 
polynomials in the brackets are included \cite{Shan}.
Due to the orthogonality relation \eqref{orth} higher Hermite polynomials in the equilibrium, 
$\mH^{(n)}_{i_1 \dots i_n}$ with $n>k$, do not contribute to the moment $M^{(0)}_{i_1 \dots i_k}$.
In addition, we need an adequate Gauss--Hermite quadrature with an accuracy order $Q \ge k+N$ 
to guarantee the correct evaluation of all Hermite contractions in Eq.\ \eqref{ce.29}.
Otherwise we obtain quadrature errors, which are analyzed in detail in Appendix \ref{ap1}.
The result of this error estimate for equilibrium moments $M^{(0)}_{i_1 \dots i_k}$ is given by
\begin{widetext}
\begin{align}
\label{ce.34}
   \Delta M^{(0)}_{i_1 \dots i_k} =
   \begin{cases}
     0                                 & \text{for   }  k+N \le Q \text{   and   } k \le N \\
     \mathcal{O} \left( \Ma^{N+1} \right) & \text{for   }  k+N \le Q \text{   and   } k > N \\
     \mathcal{O} \left( \Ma^{Q-k+1} \right) & \text{for   } k-1 \le Q < k+N \\
     \mathcal{O} \left( \Ma^0 \right) & \text{for   } Q < k-1 \; .
   \end{cases} 
\end{align}
\end{widetext}
At this point, it is important to discuss the optimal choice of the Hermite order $N$ 
of the equilibrium function \eqref{equi.hermite} in order to 
ensure an exact evaluation of all $M^{(0)}_{i_1 \dots i_k}$ with maximal $k$.
Using Eq.\ \eqref{ce.34} this is achieved for $N = \frac{1}{2} Q$.
Usually $Q$ is an odd number and $N$ is an integer, which allows either 
$N = \frac{1}{2} \left(Q+1 \right)$ or $N = \frac{1}{2} \left(Q-1 \right)$.

For an odd Gauss--Hermite quadrature order $Q$ we fix in the following the Hermite order
of the equilibrium \eqref{equi.hermite} to $N = \frac{1}{2} \left(Q-1 \right)$. 
For this particular choice we obtain for the quadrature error (\ref{ce.34}) the relation
\begin{align}
\label{ce.35}
   \Delta M^{(0)}_{i_1 \dots i_k} &=
   \begin{cases}
      0                                   & \text{for   }  2k + 1 \le Q  \\
      \mathcal{O} \left( \Ma^{Q-k+1} \right) & \text{for   }  k-1 \le Q < 2k+1 \\
      \mathcal{O} \left( \Ma^0 \right)       & \text{for   }  Q < k-1
   \end{cases} 
\nn\\
   &= \Theta \left( 2k-Q + \frac{1}{2} \right) \; \mathcal{O} \left( \Ma^{\max(Q-k+1,0)} \right) \; ,
\end{align}
where $\Theta$ is the Heaviside step function.
If we choose the other possibility, $N = \frac{1}{2} \left(Q+1 \right)$, we get for the 
estimate of the quadrature
error the same result (\ref{ce.35}). But we want to point out that the quadrature error for 
$N = \frac{1}{2} \left(Q- 1 \right)$ is different from the choice 
$N = \frac{1}{2} \left(Q+1 \right)$, however, the leading power of  $\Ma$ in the error term is the same.

The dominant quadrature error (with respect to $\Ma$) of 
a quantity consisting of several equilibrium moments is determined by the 
highest equilibrium moment $M^{(0)}_{i_1 \dots i_k}$. 
In order to discuss the 
relevant truncation error for a CE moment $M^{(n)}_{i_1 \dots i_k}$
we show in the following that the multiple-scale time derivative $\partial_t^{(0)}$ in 
Eq.\ (\ref{ce.21b}) does not alter the $\Ma$ power of the truncation error 
of an inaccurate CE moment. For this purpose, we consider a CE moment 
$M^{(n)}_{i_1 \dots i_k}$ which is recovered by a 
Gauss--Hermite quadrature up to an error
\begin{align}
\label{ce.45}
\Delta M^{(n)}_{i_1 \dots i_k} & = \mathcal{O} \left( \Ma^m \right)
\end{align}
with $\Ma$ power $m$. The time derivative $\partial^{(0)}_t$ of 
$M^{(n)}_{i_1 \dots i_k}$ is given by
\begin{align}
\partial^{(0)}_t & M^{(n)}_{i_1 \dots i_k} = 
\partial^{(0)}_t \left( M^{(n)\mathrm{DVM}}_{i_1 \dots i_k} + \Delta M^{(n)}_{i_1 \dots i_k} \right)
\nonumber \\
\label{ce.46}
& = \partial^{(0)}_t M^{(n) \mathrm{DVM}}_{i_1 \dots i_k}
- \sum_{q} \frac {\partial \Delta M^{(n)}_{i_1 \dots i_k}}{\partial ( \partial_{j_1 \dots j_q}^{(q)} \rho)} 
\partial_{j_1 \dots j_q}^{(q)} \left( \partial_j \rho u_j \right)
\nn\\& 
- \sum_{q} \frac {\partial \Delta M^{(n)}_{i_1 \dots i_k}}{\partial (\partial_{j_1 \dots j_q}^{(q)} u_j)}
\partial_{j_1 \dots j_q}^{(q)} \left( u_s \partial_s u_j + \frac{1}{\rho} \partial_j \rho \right)
\end{align}
which implies
\begin{align}
\label{ce.46b}
\Delta \partial^{(0)}_t & M^{(n)}_{i_1 \dots i_k} = 
- \sum_{q} \underbrace{\frac {\partial \Delta M^{(n)}_{i_1 \dots i_k}}{\partial (\partial_{j_1 \dots j_q}^{(q)} \rho)}}_{\mathcal{O} \left( \Ma^m \right)}  
\partial_{j_1 \dots j_q}^{(q)} \partial_j \left( \rho u_j \right)
\nn\\&\quad - \sum_{q} \underbrace{\frac {\partial \Delta M^{(n)}_{i_1 \dots i_k}}{\partial (\partial_{j_1 \dots j_q}^{(q)} u_j)}}_{\mathcal{O} \left( \Ma^{m-1} \right)} \partial_{j_1 \dots j_q}^{(q)}
\left( \underbrace{u_s \partial_s u_j}_{\mathcal{O} \left( \Ma^2 \right)} 
+ \frac{1}{\rho}  \underbrace{\partial_j \rho}_{\mathcal{O} \left( \Ma \right)} \right) .
\end{align}
For finite $\Kn$ flows density gradients $\partial_j \rho$ can be of order $\mathcal{O} \left( \Ma \right)$,
even in the low $\Ma$ flow regime.
Consequently, we find that the time derivative $\partial^{(0)}_t$  does not lower the order in $\Ma$ number of the error term \eqref{ce.45}
\begin{align}
\label{ce.47}
\Delta \partial^{(0)}_t M^{(n)}_{i_1 \dots i_k} = 
\partial^{(0)}_t \Delta M^{(n)}_{i_1 \dots i_k} = \mathcal{O} \left( \Ma^m \right) .
\end{align}
The error of the time derivatve of a moment equals the time derivative of the moment error.
Based on the recurrence relation \eqref{ce.21b} for a CE moment $M^{(n)}_{i_1 \dots i_k}$,
the error estimate \eqref{ce.35} for equilibrium moments and relation \eqref{ce.47} for
the multiple-scale derivative $\partial^{(0)}_t$, we are able to prove the following theorem:
\begin{widetext}
\vspace{2cm}
\textbf {Theorem} :
\textit{For any {\LBm } with an odd Gauss--Hermite quadrature order $Q$ and a Hermite order 
$N =  \left(Q-1 \right)/2$ of the equilibrium \eqref{equi.hermite}
the truncation error of the $k$th velocity moment $M^{(n)}_{i_1 \dots i_k}$ in the $n$th CE level
is given, for low $\Ma$ values, by}
\begin{align}
\label{ce.50}
\Delta M^{(n)}_{i_1 \dots i_k} &= \left( - \tau \right)^n  
\partial_{j_1} \dots \partial_{j_n} \Delta M^{(0)}_{i_1 \dots i_k j_1 \dots j_n} 
+ \textit{subleading terms} 
\end{align}
\textit{and can be estimated by}
\begin{align}
\label{ce.66}
\Delta  M^{(n)}_{i_1 \dots i_k} &= \Theta \left( 2(k+n)-Q + \frac{1}{2} \right)
\; \mathcal{O} \left( \Ma^{\max(Q-k-n+1,0)} \right) \; ,
\end{align}
\textit{where $k \ge 2$.}
\end{widetext}
The theorem is proved by induction in Appendix \ref{ap2}. 
It is in agreement with the accuracy determinations of LB models given by 
Shan et al.\ \cite{ShaYuaChe06}. 
Moreover, the theorem presented here enables to identify the recovered flow regimes 
for low $\Ma$ values by analyzing the truncation error.

In the following we discuss the macroscopic momentum dynamics
\begin{align}
\label{ce.67}
\partial_t \left( \rho u_i \right) + \partial_j \Pi_{ij} & = 0 
\end{align}
with the momentum flux tensor
\begin{align}
\label{ce.68}
\Pi_{ij} & = \sum_n \Pi^{(n)}_{ij} 
\end{align}
and analyze the recovered flow regime of a {\LBm } with Gauss--Hermite accuracy order $Q$
by using the theorem. Based on the error estimate (\ref{ce.66}) 
the relevant error of the momentum flux tensor of the $n$th flow regime $\Pi^{(n)}_{ij}$ 
with respect to the $\Ma$ power is given by
\begin{align}
\label{ce.69}
\Delta  \Pi^{(n)}_{ij} &= \Theta \left( 2n+4-Q + \frac{1}{2} \right)
\; \mathcal{O} \left( \Ma^{\max(Q-n-1,0)} \right) .
\end{align}
For {\LBms }  with quadrature order $Q=5$ the Navier-Stokes momentum flux tensor $\Pi^{(1)}_{ij}$ 
is not evaluated exactly for all $\Ma$ numbers, because the highest 
equilibrium moment $Q^{(0)}_{ijk}$ is only recovered up to an error $\sim \Ma^3$ in
accordance with relation (\ref{ce.66}). The leading momentum flux error is given by 
$\Delta  \Pi^{(1)}_{ij} = -\tau \partial_k \Delta Q^{(0)}_{ijk} 
= \mathcal{O} ( \Ma^{3} )$.
It is interesting to notice that even {\LBms }  with accuracy $Q=5$ recover the
Burnett momentum flux tensor for low $\Ma$ values. Because of relation (\ref{ce.69})
we obtain $\Delta  \Pi^{(2)}_{ij}  = \mathcal{O} ( \Ma^{2} )$, which indicates that
low $\Ma$ contributions are recovered. Considering the isothermal Burnett tensor
we observe that only the last term on the right-hand side of Eq.\ \eqref{ce.19}
includes low $\Ma$ contributions for compressible finite $\Kn$ flows.

{\LBms }  with $Q=7$ recover the Navier-Stokes momentum flux tensor \eqref{ce.14} exactly whereas
the Burnett tensor includes an error of 
$\Delta  \Pi^{(2)}_{ij} = \mathcal{O} ( \Ma^{4} )$ according to relation \eqref{ce.69}.
This is caused by an inaccurate evaluation of the equilibrium moment $R^{(0)}_{ijkn}$. 
Furthermore, the momentum flux tensor $\Pi^{(3)}_{ij}$ is captured up to an $\mathcal{O} ( \Ma^{3} )$ 
error and  $\Pi^{(4)}_{ij}$ up to an $\mathcal{O} ( \Ma^{2} )$ error.
Therefore, {\LBms }  with quadrature accuracy order $Q=7$ recover the momentum dynamics
up to the $4$th flow regime ($\Pi^{(4)}_{ij}$) for low $\Ma$ flows.

In general the error estimate \eqref{ce.69} states that for a low $\Ma$ flow {\LBms } 
with Gauss--Hermite quadrature order $Q$ capture the correct momentum dynamics up to the 
$(Q-3)$th flow regime, because of $\Delta  \Pi^{(Q-3)}_{ij} = \mathcal{O} ( \Ma^{2} )$.
On the other hand, flow regimes higher than $Q-2$ are not described correctly. 
For example, the error of $\Delta \Pi^{(Q-2)} = \mathcal{O}(\Ma)$
corrupts the leading order terms of the momentum dynamics. 
Fortunately, such effects are suppressed by a term $~ \Kn^{Q-2}$ which 
is sufficiently small for $\Kn<1$.
\section{Gas--surface interaction \label{sec:GasSurf}}
For wall-bounded flows at finite $\Kn$ the gas-surface interaction
is of crucial importance because of the influence of the Knudsen
layer.
It is known that the diffuse Maxwell reflection model \cite{ref1.3} 
is sufficiently accurate to describe flows for a wide range of $\Kn$.
Its numerical implementation in the LB framework has been reported in 
Refs.\ \cite{AnsumaliKarlin,SofoneaSekerka,Sofonea,Meng,Sbragaglia}.
Although high-order LB models recover flow regimes beyond the Navier--Stokes level, 
there are some models, e.g.\ $D3Q121$, which show significant deviations to reference results
for wall-bounded flows at finite $\Kn$ (see Sect.\  \ref{PF}).
The reason for this failure will be shown to be caused by the inability of these models to recover
the diffuse Maxwell boundary condition accurately.
We define wall moments and introduce a wall  
accuracy order and thus assess the capability of a LB model to capture the diffuse 
Maxwell boundary condition.

In this Section we discuss a systematic way to generate DVMs which inherently exhibit 
the diffuse Maxwell boundary condition. 
\subsection{Diffuse Maxwell boundary condition \label{sec:Maxwell}}
The diffuse Maxwell reflection model suggests that particles emitted from the solid 
surface do not depend on anything prior to their surface impact and their velocities
are normalized by the equilibrium distribution.
This notion infers that the scattering kernel only depends on the emitted velocities. 
We can then write \cite{AnsumaliKarlin} the distribution function particles emitted by the wall as 
\begin{align}
\label{defPsi}
& f(\fxi) = \Psi f^{(0)}(\fxi,\rho_w,0) \hspace{0,5cm} \text{for} \hspace{0,5cm} {\bf n}\cdot\fxi > 0
\end{align}
where ${\bf n}={\bf e}_z$ is the inward wall normal vector and the scalar $\Psi$ ensures 
mass conservation across the (impermeable and stationary) surface,
\begin{align}
\label{Psi}
& \Psi = \frac {\int_{{\bf n}\cdot\fxi' < 0} d^3 \xi' f(\fxi') \left|{\bf n}\cdot\fxi'\right| }
{\int_{{\bf n}\cdot\fxi > 0} d^3 \xi f^{(0)}(\fxi,\rho_w,0)\left|{\bf n}\cdot\fxi\right| } \quad .
\end{align}
The equilibrium function $f^{(0)}(\fxi,\rho_w,0)$ is evaluated for the density $\rho_w$ 
and the macroscopic velocity $\fu=0$ at the wall. It was shown in Ref.\ \cite{1.10} that 
for steady unidirectional flows $\Psi = 1$. The discussion of this Section allows $\Psi$ to be arbitrary.
\subsection{Equilibrium wall moments \label{sec:WallMoms}}
At the wall, the velocity moments \eqref{defM} are decomposed into a part connected to 
the fluid domain and a part which is influenced by the wall
\begin{align}
\label{4.2}
M_{i_1 \dots i_n} &= \int_{{\bf n}\cdot\fxi < 0} d^3 \xi\: f(\fxi)\: \xi_{i_1} \dots \xi_{i_n} 
\nn\\&\quad+ \int_{{\bf n}\cdot\fxi > 0} d^3\xi\: f(\fxi)\: \xi_{i_1} \dots \xi_{i_n}\; .
\end{align}
The second integral is completely determined by 
distributions coming from the wall representing the gas--surface interaction.
Using Eqs.\ \eqref{defPsi} and \eqref{Psi}, we can write
\begin{align}
\label{Mbrief}
M_{i_1 \dots i_n} = & \int_{{\bf n}\cdot\fxi < 0} d^3 \xi\: f(\fxi)\: \xi_{i_1} \dots \xi_{i_n} 
+ \Psi \rho_w W_{i_1 \dots i_n}
\end{align}
where the equilibrium wall moments $W_{i_1 \dots i_n}$ were introduced as
\begin{align}
\label{defW}
W_{i_1 \dots i_n} = \int d^3\xi\: \Theta({\bf n}\cdot\fxi)\: \omega(\fxi)\: \xi_{i_1} \dots \xi_{i_n}\; .
\end{align}
The capability of a DVM to capture the Maxwell boundary condition can now be investigated by analyzing the equilibrium wall moments. It is straightforward to obtain exact solutions for  $W_{i_1 \dots i_n}$, expressing them by 
the scalar integrals
\begin{align}
\label{Wscalar}
&\int_{-\infty}^\infty d\xi_x\int_{-\infty}^\infty d\xi_y\int_0^\infty d\xi_z\:\omega(\fxi)\xi_z^{m_z}\xi_y^{m_y}\xi_x^{m_x}\nonumber\\
&\quad=\frac{\sqrt{2}^{\: m_x+m_y+m_z}}{8\sqrt{\pi}^3}\left(1+(-1)^{m_x}\right)\left(1+(-1)^{m_y}\right)\nn\\&\quad\quad\times\Gamma\left(\frac{m_x+1}{2}\right)\Gamma\left(\frac{m_y+1}{2}\right)\Gamma\left(\frac{m_z+1}{2}\right)\: ,
\end{align}
with the Euler Gamma function $\Gamma$. We will now show how to evaluate the equilibrium wall 
moments \eqref{defW} numerically with a discrete velocity model.
\subsection{Discrete velocity models for wall-bounded flows \label{sec:WallModels}}
In the discrete velocity space the wall moments \eqref{defW} are computed by quadrature. Note, however, that the evaluation of $n$th wall moments $W_{i_1 \dots i_n}$ cannot be performed in an exact manner by Gauss-Hermite quadratures with order $Q=n$. This is due to the non-analyticity of the Heaviside function at the wall where an expansion of the integrand does not exist. Although alternative quadratures can be introduced using functions orthogonal in the half-space  \cite{FreGibFra09,Ambrus}, we would like to cope with this difficulty within the conventional framework of Gauss-Hermite quadratures. We employ the quadrature prescription
\begin{align}
\label{wallQuad}
W^{\mathrm{DVM}}_{i_1 \dots i_n} = \sum_{{\bf n}\cdot\fxi_\alpha>0}  \tw_{\alpha} \xi_{\alpha i_1}\dots \xi_{\alpha i_n} \quad  ,
\end{align}
and demand that---while defining the stencil---the supplementary conditions
\begin{align}
\label{wallCond}
W^{\mathrm{DVM}}_{i_1 \dots i_n} = W_{i_1 \dots i_n} .
\end{align}
be fulfilled, aside from the orthogonality condition \eqref{orth}. We thus ensure that the wall moments \eqref{defW} are computed exactly. To this end the kernel of the matrix $K$, see Eq.\ \eqref{defK}, is augmented 
by adding groups of velocities, yielding 
$ \{ \tw_{\alpha}^{(1)}, \tw_{\alpha}^{(2)}, \dots , \tw_{\alpha}^{(J)} \}$ as an 
orthogonal basis of the $J$-dimensional null space. Any weight vector
\begin{align}
\label{9.13}
\tw_{\alpha} = x_j \tw_{\alpha}^{(j)} 
\end{align}
in this null space then defines a quadrature where the freedom in the 
coefficients $x_j$ is used to implement the wall conditions \eqref{wallCond}. 
We achieve this by solving the set of linear equations
\begin{align}
\label{wallEq}
A_{kj} x_{j} = b_k
\end{align}
with the matrix $A\in\mathbb{R}^{L \times J}$ given by
\begin{align}
\label{defA}
A_{kj} = \sum_{{\bf n}\cdot\fxi_\alpha>0} T^\alpha_k \tw_\alpha^{(j)}\: ,
\end{align}
where the symbols $T^\alpha_k$ are explained below.
The index $k$ in Eq.\ \eqref{wallEq} counts the $L$ non-trivial equilibrium wall moments 
of Eq.\ \eqref{wallCond}. These moments are determined for $n\leq Q$, 
i.e.\ the quadrature's accuracy at the wall is not required to exceed the 
accuracy in the bulk. Note that the integrals in Eq.\ \eqref{Wscalar} vanish 
by symmetry if $m_x$ or $m_y$ are odd. Since this symmetry is preserved by 
the stencils, the quadrature then automatically yields correct results. For even $m_x$ and $m_y$, the equilibrium wall moments \eqref{defW} are automatically exact for even $m_z=2,4,\dots$ since the contribution of wall-parallel velocities in Eq.\ \eqref{halfQuad} is suppressed and the invariance of $P(\fxi)$ for the transformation \eqref{wallmirror} applies.
We thus find non-trivial equilibrium wall moments to be imposed as constraints for 
the following integer values of $m_x$, $m_y$, and $m_z$ in  Eq.\ \eqref{Wscalar},
\begin{align}
\label{nontriv}
&m_x\in\{0,2,4,\dots\}\nn\\
&m_y\in\{0,2,4,\dots |\: m_y\leq m_x\}\nn\\
&m_z\in\{0,1,3,5,\dots\}\nn\\
&m_x+m_y+m_z\leq Q\; .
\end{align}
For $Q=7$, there are $L=19$ components 
$T^\alpha_k$ of the vector ${\bf T}^\alpha$ (no sum on $\alpha$),
\be
\label{defT}
{\bf T}^\alpha&=&(
1,\:
\xi_{\alpha z},\:
\xi_{\alpha x}^2,\:
\xi_{\alpha z}^3,\:
\xi_{\alpha z}\xi_{\alpha x}^2,\:
\xi_{\alpha x}^4,\:
\xi_{\alpha x}^2\xi_{\alpha y}^2,\:
\nn\\&&\:
\xi_{\alpha z}^5,\:
\xi_{\alpha z}^3\xi_{\alpha x}^2,\:
\xi_{\alpha z}\xi_{\alpha x}^4,\:
\xi_{\alpha z}\xi_{\alpha x}^2\xi_{\alpha y}^2,\:
\xi_{\alpha x}^6,\:
\nn\\&&\:
\xi_{\alpha x}^4\xi_{\alpha y}^2,\:
\xi_{\alpha z}^7,\:
\xi_{\alpha z}^5\xi_{\alpha x}^2,\:
\xi_{\alpha z}^3\xi_{\alpha x}^4,
\nn\\&&
\xi_{\alpha z}^3\xi_{\alpha x}^2\xi_{\alpha y}^2,\:
\xi_{\alpha z}\xi_{\alpha x}^6,\:
\xi_{\alpha z}\xi_{\alpha x}^4\xi_{\alpha y}^2
)\; ,
\ee
whereas $b_k$ are the components of the vector
\be
\label{defb}
{\bf b}&=&
(
W,\:
W_{z},\: 
W_{xx},\: 
W_{zzz},\: 
W_{zxx},\: 
W_{xxxx},\: 
W_{xxyy},\: 
\nn\\&&\:
W_{zzzzz},\: 
W_{zzzxx},\: 
W_{zxxxx},\: 
W_{zxxyy},\:
W_{xxxxxx},\: 
\nn\\&&\:
W_{xxxxyy},\: 
W_{zzzzzzz},\: 
W_{zzzzzxx},\: 
W_{zzzxxxx},\: 
\nn\\&&\:
W_{zzzxxyy},\: 
W_{zxxxxxx},\: 
W_{zxxxxyy}
)
\; .
\ee
For $Q>7$, additional components are appended to the latter vectors, 
in the nested order of increasing $n$, decreasing $m_z$ and decreasing 
$m_x$, see Eq.\ \eqref{nontriv}.

By solving the supplementary Eq.\ \eqref{wallEq} we obtain stencils 
suitable as $Q$th-order quadratures for the gas-surface interaction 
as well as the bulk flow. 
As shown in Sect.\ \ref{sec:Poi}, {\LBms }  of a given quadrature order $Q$ 
show strongly different behavior depending on the quantities
\begin{align}
\label{wallError}
\sigma_{i_1 \dots i_n}=\left.\left( W_{i_1 \dots i_n} - W^{\mathrm{DVM}}_{i_1 \dots i_n} \right)\right/ W_{i_1 \dots i_n}
\end{align}
which measure the numerical error of the gas-surface interaction. 
It is convenient to use the expression
\be
\label{sigmaSigma}
\sigma_\Sigma=\frac{
\sum_{n=0}^Q\sum_{i_1\dots i_n}
|\sigma_{i_1 \dots i_n}|\e^{-n}
}
{
\sum_{n=0}^Q\sum_{i_1\dots i_n}
\e^{-n}
}
\ee
assessing the net effect of non-trivial wall errors. 

The two DVM sets ${\cal S}_3$ and ${\cal S}_4$ shown in Tab.\ \ref{tab:dvmSummary} 
are designed for wall-bounded flows, taking into account some of the 
equilibrium wall moments \eqref{wallCond}. In particular, the DVM set ${\cal S}_3$ 
comprises $500$ stencils for $D=3$ and $Q=7$ with the supplementary condition
\be
\label{wallConstraintSigma3}
\sigma_{zxx}=0\: .
\ee
As shown in Sec.\ \ref{sec:Poi}, a minimal equilibrium wall moment error $|\sigma_{zxx}|$ is crucial 
to the resolution of correct mass flow and slip velocity in finite $\Kn$ flows. 
Further optimization (see below) yields the stencils $\st{7}{77}{3}{672}$ 
and $\st{7}{107}{3}{1023}$ shown in Tables \ref{tab:dvmD3Q7Set3} and \ref{tab:dvmD3Q7Set3b}. Note that 
these are {\it\augDVMs}: they are based on DVMs for {\minDVMs }  and the last 
two velocity groups are added to account for the wall constraint \eqref{wallConstraintSigma3}.

We put forward the so-called {\it {\wai }} $\Lambda_{\mathrm{index}}$ for 
indicating a DVM's ability to evaluate the wall integrals \eqref{defW}. 
The {\wai } is defined as the binary number
\be
\label{defbin}
\Lambda_{\mathrm{index}}=(\lambda_L\dots\lambda_2\lambda_1)_2
\ee
where the $\lambda_k$ are numerical booleans which determine whether 
the $k$th wall constraints in Eq.\ \eqref{wallEq} is satisfied,
\be
\lambda_k=  \begin{cases}
0 & \text{for   }  A_{kj} x_{j} \neq b_k \\
1 & \text{for   }  A_{kj} x_{j} = b_k
\end{cases}
\ee
For instance, the DVM set ${\cal S}_3$ with the wall constraint \eqref{wallConstraintSigma3} 
can be characterized by the {\wai }
\be
\label{lambdaS3}
\Lambda_{\mathrm{index}}=0000000000000010000_2 = 16\; .
\ee
In Eq.\ \eqref{lambdaS3}, it is shown that the {\wai } can be given by a decimal number as well. We thus use $\Lambda_{\mathrm{index}}$ as an abbreviation bearing lots of information on the validity of the wall 
constraints \eqref{wallCond}.

Furthermore, we define the {\it {\wao }} $\Lambda$ by the highest rank 
of equilibrium wall moments \eqref{defW} up which the quadrature is exact,
\be
\label{defLambda}
\sigma_{i_1\dots i_n}=0\quad\forall\:\left\{ n ,i_1 \dots i_{n}\: |\: n=0,1,\dots, \Lambda\right\}\: .
\ee
If Eq.\ \eqref{defLambda} is false for any $\Lambda\geq 0$, we set 
$\Lambda=-1$ which applies for all DVMs for bulk flow. Note that 
for ${\cal S}_3$ we get $\Lambda=-1$ as well.

For the DVM set ${\cal S}_4$, where $D=3$ and $Q=7$, we split the 
domain of numerical integration across the wall, i.e.\ the stencils 
do not contain velocities parallel to the wall, 
${\bf n}\cdot\fxi_\alpha\neq 0\:\forall\alpha$. A noteworthy property of these models is that they enforce wall interaction 
of the stencil nearest to the wall. The occurrence of ballistic 
particles as described in Ref.\ \cite{Toschi} is thus avoided. We refer 
to the stencils without wall-parallel velocities as {\it scattering stencils}.

Considering $Q$th order polynomials $P(\fxi)$ invariant for 
\be
\label{wallmirror}
{\bf n}\cdot\fxi\longrightarrow -{\bf n}\cdot\fxi
\ee
(even functions included) the relation
\begin{align}
&\int d^3\xi\: \Theta({\bf n}\cdot\fxi) \omega(\fxi)P(\fxi)=\frac{1}{2}\int d^3\xi\: \omega(\fxi)P(\fxi)\nonumber\\
&\qquad=\frac{1}{2}\sum_\alpha \tw_\alpha P(\fxi_\alpha)\nonumber\\
\label{halfQuad}
&\qquad=\frac{1}{2}\sum_{{\bf n}\cdot\fxi_\alpha=0}\tw_\alpha P(\fxi_\alpha)+\sum_{{\bf n}\cdot\fxi_\alpha>0}\tw_\alpha P(\fxi_\alpha)
\end{align}
is exact for quadratures with order $Q$. For scattering stencils, the first sum in Eq.\ \eqref{halfQuad} is absent by definition (${\bf n}\cdot\fxi_\alpha\neq 0$) and therefore the customary wall quadrature prescription \eqref{wallQuad} yields exact results for even wall moments, 
\be
\label{S4prop}
\sigma_{i_1 \dots i_{2n}}=0\quad\forall\:\left\{ n ,i_1 \dots i_{2n}\: |\: 0\leq2n\leq Q\right\}\: .
\ee
On the other hand, wall-parallel and zero velocities infer a quadrature error using Eq.\ \eqref{wallQuad} because their contribution to Eq.\ \eqref{halfQuad} is neglected. This circumstance is an advantage of scattering stencils such as $D2Q16$ and it also explains why DVMs with an even number of velocities are superior to others in the context of wall-bounded flows. Such stencils necessarily lack the rest velocity $\fxi=0$ which typically comes with a relatively large weight $\mathrm{w}_0$ in the quadrature. The rest velocity causes a considerable error neglecting the term $\frac{1}{2}\mathrm{w}_0P(0)$ in Eq.\ \eqref{halfQuad} whilst using the wall quadrature prescription \eqref{wallQuad}. This error is suppressed by stencils with an even number of velocities. 

The {\wao } of ${\cal S}_4$ is found, according to Eq.\ \eqref{defLambda} 
as $\Lambda=0$. On the other hand, the {\wai } yields
\be
\label{lambdaS4}
\Lambda_{\mathrm{index}}=0000001100001100101_2 = 6245 \; .
\ee

The energies of the stencils in ${\cal S}_4$ are limited by $E_{max}=2500$  
which yields 45863 DVMs with the property \eqref{S4prop}. Among those, 
we picked out the DVMs $S_{Q7E1932}^{D3A96}$ and $S_{Q7E1764}^{D3A112}$, 
for reasons explained in Sec.\ \ref{sec:Poi}, and we present them in Tables \ref{tab:dvmD3Q7W0} and \ref{tab:dvmD3Q7W0b}.

\subsection{Maxwell boundary factor $\Psi$}
\label{sec:wallCorrolary}
We now turn to the evaluation of the factor $\Psi$, see Eq.\ \eqref{Psi}, of the \DMBC. Assuming that the CE expansion, cf.\ Eq.\ (\ref{ce.2}), is applicable we write
\begin{align}
\label{psi.1}
\Psi = \sum_{n=0} \Psi^{(n)}
\end{align}
with
\begin{align}
\label{psi.2}
\Psi^{(n)} = \frac{\sum_{{\bf n}\cdot{\fxi}_{\alpha} < 0 }  f^{(n)}_{\alpha} \left| {\bf n}\cdot{\fxi}_{\alpha} \right|}
{\sum_{ {\bf n}\cdot{\fxi}_{\beta} > 0 }  f^{(0)}_{\beta} (\rho_w,0) \left| {\bf n}\cdot{\fxi}_{\beta}  \right|} 
\end{align}
where we have set the expansion parameter $\epsilon = 1$. The denominator in Eq.\ (\ref{psi.2}),
\begin{align}
\label{psi.3}
Z = \sum_{{\bf n}\cdot{\fxi}_{\beta}  > 0 }  f^{(0)}_{\beta} (\rho_w,0) \left| {\bf n}\cdot {\fxi}_{\beta} \right| 
=\rho_w n_i W^{\mathrm{DVM}}_{i}\; ,
\end{align}
is captured exactly by the DVM if the wall accuracy order yields $\Lambda \ge 1$. The numerator in Eq.\ (\ref{psi.2}) can be
analyzed by using Eq.\ (\ref{ce.6}),
\begin{subequations}
\label{psi.4}
\begin{align}
\Psi^{(0)}=  &- \frac{n_i}{Z} Y^{(0)}_i \\
\Psi^{(1)}=  &- \frac{n_i}{Z} (- \tau) \left[ \partial^{(0)}_t Y^{(0)}_i + \partial_j Y^{(0)}_{ij} \right] \\
\Psi^{(2)}=  &- \frac{n_i}{Z}  \Big[
(- \tau)^2 \Big( \partial^{(0)}_t \partial^{(0)}_t Y^{(0)}_i 
+ 2 \partial^{(0)}_t \partial_j Y^{(0)}_{ij}
\nn\\
&+ \partial_j \partial_k Y^{(0)}_{ijk} \Big) + (- \tau) \partial^{(1)}_t Y^{(0)}_i \Big] \\
\cdots \nn\\
\Psi^{(n)}=  &\:\Psi^{(n)} \left( Y^{(0)}_{i_1} \dots Y^{(0)}_{i_1 \dots i_{n+1}} \right)
\end{align}
\end{subequations}
where we have introduced the moments
\begin{align}
\label{psi.5}
Y^{(0)}_{i_1 \dots i_k} = \sum_{{\bf n}\cdot{\fxi}_{\alpha}  < 0 }  
f^{(0)}_{\alpha}(\rho_w,\fu)\:  {\xi}_{\alpha i_1} \dots {\xi}_{\alpha i_k} \; .
\end{align}
The equilibrium (\ref{equi.hermite}) expanded up to the Hermite order $N$ 
can be expressed by
\begin{align}
\label{psi.6}
f_\alpha^{(0)}(\rho_w,\fu)&=\tw_{\alpha} \sum^{N}_{n=0} \frac{1}{n!} a^{(0)}_{i_1 \dots i_n} \mH^{(n)}_{i_1 \dots i_n} 
({\fxi}_{\alpha}) 
\nn\\
&= \tw_{\alpha} \sum^{N}_{n=0} b_{i_1 \dots i_n} {\xi}_{\alpha i_1} \dots {\xi}_{\alpha i_n}
\end{align}
with some coefficients $b_{i_1 \dots i_n}$. Thus, the moments $Y^{(0)}_{i_1 \dots i_k}$ are determined
by the equilibrium wall moments
\begin{align}
\label{psi.YbyW}
Y^{(0)}_{i_1 \dots i_k} = \sum^{N}_{n=0} (-1)^{n+k} b_{j_1 \dots j_n} W^{\mathrm{DVM}}_{j_1 \dots j_n i_1 \dots i_k} \; .
\end{align}
In general, the $n$th CE contribution of $\Psi$ is a function of the moments $Y^{(0)}_{i_1 \dots i_k}$ with $k\leq n+1$, see Eq.~\eqref{psi.4}, which can be directly expressed by the equilibrium wall moments $W^{\mathrm{DVM}}_{i_1 \dots i_m}$ with $m\leq N+n+1$. As a consequence, a sufficiently high accuracy of the
equilibrium wall moments, characterized by the wall accuracy order $\Lambda$, guarantees an exact evaluation of the Maxwell boundary factor $\Psi$ up to the $n$th CE level with
\be
\label{wallnmax}
n = \Lambda-N-1\; .
\ee
\section{Numerical method \label{NM}}
For the numerical solution of our test case shown below, we include an external force $F_{\alpha}$ 
in the LB--BGK equation \eqref{LBE} by writing
\begin{align}
\label{nm.1}
\left( \partial_t + \xi_{\alpha i} \partial_i \right) f_{\alpha} =  - \frac{1}{\tau} \left( f_{\alpha} - f^{eq}_{\alpha}  \right)
\end{align}
where we introduced the generalized equilibrium function
\begin{align}
\label{nm.2}
f^{eq}_{\alpha} = f^{(0)}_{\alpha} + \tau F_{\alpha}
\end{align}
which must be expanded using Hermite polynomials \cite{ShaYuaChe06}. Using the Hermite 
order $N=3$, this yields for the equilibrium
\begin{align}
\label{nm.3}
f^{(0)}_{\alpha} (\rho,{\bf u}) &= \rho \tw_{\alpha} \left[ 1 + \left( u_i \xi_{\alpha i} \right)
+ \frac{1}{2} \left( \left( u_i \xi_{\alpha i} \right)^2 - u_i u_i \right)
\right.
\nn\\
&\left.
\quad+ \frac{1}{6} \left( u_i \xi_{\alpha i} \right) 
\left( \left( u_j \xi_{\alpha j} \right)^2 - 3 u_j u_j \right)
\right]
\end{align}
and for the body force term 
\begin{align}
\label{nm.4}
&F_{\alpha} = \rho \tw_{\alpha} \left[ \left( g_i \xi_{\alpha i} \right)
+  \left( \left( g_i \xi_{\alpha i} \right) \left( u_j \xi_{\alpha j} \right) - u_i g_i \right)
\right.
\nn\\
&\quad+\left. \frac{1}{2 \rho} \left( \Pi_{ij} - \rho \delta_{ij} \right) 
\left( \left( g_k \xi_{\alpha k} \right) \mH^{(2)}_{ij} (\fxi_{\alpha}) - 2 g_i \xi_{\alpha j} \right)
\right]\; .
\end{align}
Here, ${\bf g}$ denotes the acceleration of an external body force field. The numerical 
solution of Eq.\ (\ref{nm.1})
in space ${\bf x}$ and time $t$ can be achieved by integrating along a characteristic for a 
time interval $\Delta t$ using the trapezium rule
\begin{align}
\label{nm.5}
& f_{\alpha} ({\bf x} + {\fxi}_{\alpha} \Delta t , t + \Delta t) - f_{\alpha} ({\bf x} , t )  
\nn\\
& = - \frac{1}{\tau} \int_{0}^{\Delta t}dt'\: \left[ f_{\alpha} ({\bf x} + {\fxi}_{\alpha} t' , t + t') 
\right.
\nn\\
&\quad\quad\quad\quad\left.
- f^{eq}_{\alpha} ({\bf x} + {\fxi}_{\alpha} t' , t + t') \right] \nonumber\\
&= - \frac{1}{\tau}  \frac{\Delta t}{2} \left[ 
\left( f_{\alpha} ({\bf x} , t ) - f^{eq}_{\alpha} ({\bf x} , t )   \right) 
\right.
\nn\\
&\quad+
\left. 
\left( f_{\alpha} ({\bf x} + {\fxi}_{\alpha} \Delta t, t + \Delta t)
- f^{eq}_{\alpha} ({\bf x} + {\fxi}_{\alpha} \Delta t, t + \Delta t)  \right)  
\right] 
\nn\\
&\quad+ 
\mathcal{O} \left( \Delta t^{3} \right)\; ,
\end{align}
resulting in a second-order implicit differencing scheme. By introducing a modified 
distribution function \cite{HeShanDoolen}
\begin{align}
\label{nm.7}
\bar{f}_{\alpha} = f_{\alpha} + \frac{\Delta t}{2 \tau} \left( f_{\alpha} - f^{eq}_{\alpha} \right)
\end{align}
the scheme (\ref{nm.5}) is transformed into a fully explicit scheme of second order
accuracy
\begin{align}
\label{nm.8}
&\bar{f}_{\alpha} ({\bf x} + {\fxi}_{\alpha} \Delta t , t + \Delta t) = \bar{f}_{\alpha} ({\bf x} , t )
\nn\\&\quad\quad- \frac{\Delta t}{\tau + \frac{\Delta t}{2}} \left( \bar{f}_{\alpha} ({\bf x} , t ) -  f^{eq}_{\alpha} ({\bf x} , t )     \right)\; .
\end{align}
The macroscopic density $\rho$ and velocity $\fu$ are given by
\begin{subequations}
\begin{align}
\label{nm.10}
\rho &= \sum_{\alpha} f_{\alpha} = \sum_{\alpha} \bar{f}_{\alpha} \\
\rho u_i &= \sum_{\alpha} f_{\alpha}  \xi_{\alpha i}  = \sum_{\alpha} \bar{f}_{\alpha} \xi_{\alpha i}
+ \frac{\Delta t}{2} \rho g_i
\end{align}
\end{subequations}
where we have used $\sum_{\alpha} F_{\alpha} \xi_{\alpha i} = \rho g_i$.

The relaxation time $\tau$ in the LB--BGK equation \eqref{LBE} is chosen to adjust
the viscosity of the bulk flow. Based on the Navier-Stokes momentum flux tensor \eqref{ce.14} $\tau$
is determined by 
\begin{align}
\label{nm.11}
\tau = \nu = \frac {\tilde{\nu}}{c_0 l_0}
\end{align}
where $\tilde{\nu}$ is the kinematic viscosity in physical units. We write the 
quantities $\nu$, $\lambda$ with tildes if they are expressed in physical units.

Because of the nature of intermolecular collisions there is no well-defined definition of the mean free path. 
The conventional solution to this problem is to consider a model gas with hard sphere molecules,
where the mean free path can be expressed exactly \cite{Chapman,ref1.3,ref1.4}.
For the present study we use Cercignani's definition of the mean free path based on the 
viscosity \cite{ref1.3,ref1.4}
\begin{align}
\label{nm.12}
\tilde{\lambda} = \sqrt{\frac{\pi}{2}} \frac{\tilde{\nu}}{c_0}
\end{align}
which is very close to the analytical result of a hard sphere gas\footnote{
The CE result for the mean free path in a hard sphere gas is
$\tilde{\lambda} = \frac{16}{5 \sqrt{2 \pi}} \frac{\tilde{\nu}}{c_0}$ \cite{Bird2}.}. 
Thus, $\Kn$ is given by
\begin{align}
\label{nm.13}
\Kn = \frac{\tilde{\lambda}}{l_0} = \sqrt{\frac{\pi}{2}} \frac{\tilde{\nu}}{c_0 l_0} 
= \sqrt{\frac{\pi}{2}} \tau
\end{align}
which is---up to a factor---identical to the relaxation time $\tau$.

For the interaction of gas molecules with a solid wall surface we implement the diffuse Maxwell boundary condition (\ref{defPsi}) for a non-moving rigid wall
using the expanded equilibrium function (\ref{nm.3})
\begin{align}
\label{nm.14}
f_{\alpha} ({\bf x}_w, t) = \Psi f^{(0)}_{\alpha} (\rho_w, {\bf u} =0)
\end{align}
with
\begin{align}
\label{nm.15}
\Psi = \frac{\sum_{{\bf n}\cdot{\fxi}_{\alpha} < 0 }  f_{\alpha} \left| {\bf n}\cdot{\fxi}_{\alpha} \right|}
{\sum_{ {\bf n}\cdot{\fxi}_{\beta} > 0 }  f^{(0)}_{\beta} (\rho_w,0) \left| {\bf n}\cdot{\fxi}_{\beta}  \right|} 
\end{align}
Unknown distribution functions near solid walls which cannot be calculated by the standard
propagation step are set to the distribution function of the diffusive Maxwell boundary 
condition (\ref{nm.14}). 
For the test case investigated here, the density is uniform and the flow behavior is steady and unidirectional which implies $\Psi = 1$ \cite{1.10}.
As for the half-way bounce-back scheme, the wall is located 
at a distance of half a lattice spacing from the
first fluid collision center \cite{SofoneaSekerka,Sbragaglia}. 
For a numerical calculation the boundary condition (\ref{nm.14}) needs to be transformed with 
Eqs.\ (\ref{nm.2}) and (\ref{nm.7}) into a corresponding form for the modified distribution 
function $\bar{f}_{\alpha}$.
\section{Poiseuille flow \label{PF}}
\label{sec:Poi}
In this section, we check DVMs for their capability of describing Poiseuille flow 
for various values of $\Kn$. The DVMs considered here are organized in the sets 
${\cal S}_1$, ${\cal S}_3$, and ${\cal S}_4$ shown in Tab.\ \ref{tab:dvmSummary}, 
to be discussed in Sect.\ \ref{sec:PoiS1}, \ref{sec:PoiS3}, and \ref{sec:PoiS4}, 
respectively. This order is chosen with increasing value of the {\wai } 
$\Lambda_{\mathrm{index}}$. It should be possible to find a DVM with all wall 
moments evaluated exactly, i.e.\ {\wao } $\Lambda=Q$, following the procedure 
given in Sect.\ \ref{sec:WallModels} and solving the wall equation \eqref{wallEq} 
as a constraint. However, since the stencil would then become prohibitively large, 
we relax some of these constraints and minimize the remaining wall errors instead.

Two parallel plates are located at $z = \pm l_0/2$ and the flow is driven by a 
constant pressure gradient, $\rho g$, in $x$-direction. We use periodic boundary 
conditions in the $x$- and $y$-directions. The body force is small enough so that 
we can assume low $\Ma$ flow. Due to symmetry, the only non-trivial velocity 
component is $u(z)=u_x(\fx)$. Below, we will discuss the slip velocity at the wall,
\be
\label{defSlip}
u_s=\frac{u\left(
\frac{l_0}{2}
\right)}{u(0)}\; ,
\ee
as well as the normalized mass flow rate
\begin{align}
\label{pf.1}
\dot{m} = \frac{1}{4 u_c \Kn} \frac{1}{B}\int^{B}_{0} dy \; \frac{1}{l_0}\int^{l_0/2}_{-l_0/2} dz\: u(z)\;  ,
\end{align}
where $B$ is the extension of the flow domain in y-direction. The Navier--Stokes equation with 
no-slip boundary condition, $u_s=0$, yields the centerline velocity 
$u_c = u(0)= g l_0^2/(8 \tilde{\nu})$ and the mass flow rate 
\begin{align}
\label{pf.2}
\dot{m} = \frac{1}{6 \Kn} \; .
\end{align}
\begin{figure}  [htb!]
\begin{center}
\epsfig{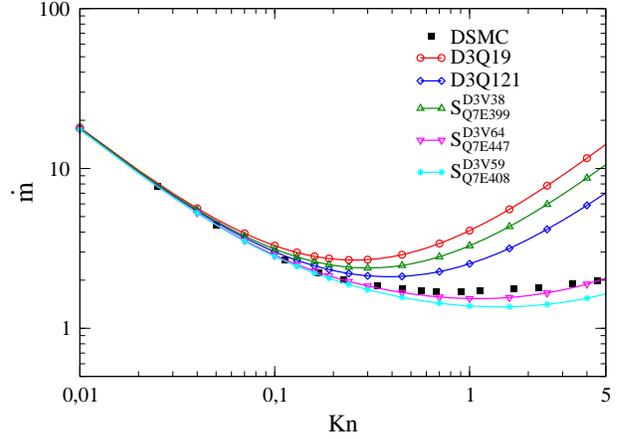}
\caption{\label{fig2} (Color online) Mass flow rate for common {\LBms }  and some exclusive ${\cal S}_1$ 
bulk flow models ($\Lambda_{\mathrm{index}}=0$).}
\end{center}
\end{figure}
\begin{figure}  
\begin{center}
\epsfig{file=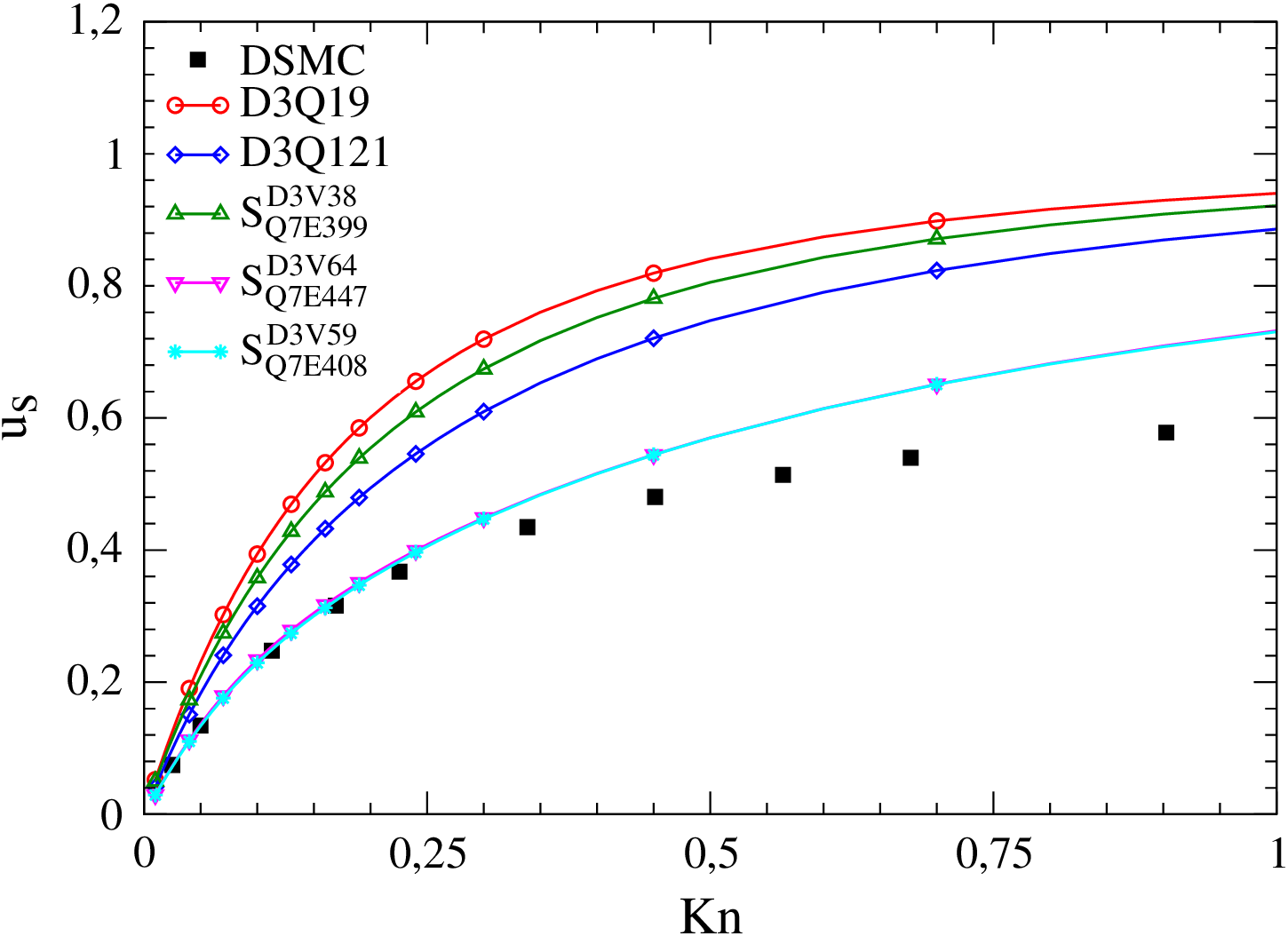,width=8.0cm,clip=,bb=10 26 710 532}
\caption{\label{fig3} (Color online) Slip velocity for common {\LBms } and some exclusive ${\cal S}_1$ 
bulk flow models ($\Lambda_{\mathrm{index}}=0$). }
\end{center}
\end{figure}

\subsection{Using discrete velocity models for bulk flow with $\Lambda_{\mathrm{index}}=0$ \label{sec:PoiS1}}
Figures \ref{fig2} and\ \ref{fig3} show the mass flow rate and the slip velocity at the wall
for two known {\LBms},
$D3Q19$ and $D3Q121$, and some new models of set ${\cal S}_1$ compared with
DSMC data.
For small $\Kn$ the
$D3Q19$ model (accuracy order $Q=5$) and the $D3Q121$ model (accuracy order $Q=9$) 
agree very well with the reference data. However, for higher $\Kn$ both models
exhibit strong deviations from the DSMC results.
Similarly, the new model $S^{D3V38}_{Q7E399}$ with an accuracy order $Q=7$ and a minimal number of 
$38$ velocities fails
for higher $\Kn$. Although the high-order models $D3Q121$ and $S^{D3V38}_{Q7E399}$
recover flow regimes beyond the Navier-Stokes level (for a detailed discussion see Sect.\ \ref{LBH}) 
the results remain unsatisfactory for finite $\Kn$.

The new {\LBms }  of set ${\cal S}_1$ are of Gaussian quadrature order $Q=7$
and thus able to recover the momentum dynamics for small $\Ma$ up to 
the $4$th flow regime ($\Delta  \Pi^{(4)}_{ij}  = \mathcal{O} ( \Ma^{2} )$).
If we analyze these models we observe for finite $\Kn$, 
nevertheless, quite different results and considerable deviations from the DSMC results 
for both the mass flow rate and the slip velocity.
The Gaussian quadrature order $Q$ is very important to recover high-order flow regimes,
but not sufficient to guarantee accurate results of a {\LBm } for finite $\Kn$.
This is also found for high-order {\LBms }  in $D=2$ dimensions \cite{1.10}.

\begin{figure}  
\begin{center}
\epsfig{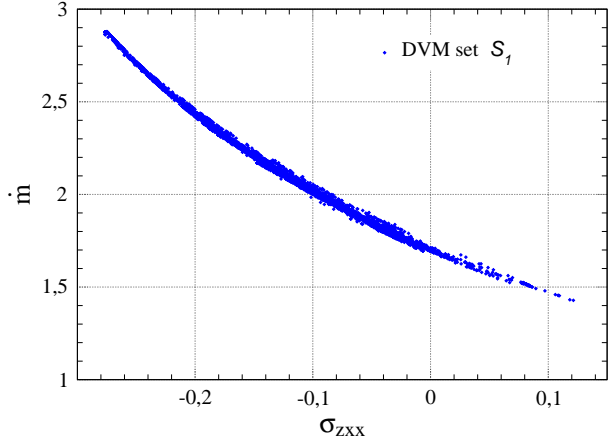}
\caption{\label{fig4} (Color online) Correlation between $\sigma_{zxx}$ and $\dot{m}$ for $\Kn=0.4514$. Every dot represents a DVM within ${\cal S}_1$.}
\end{center}
\end{figure}
\begin{figure}  
\begin{center}
\epsfig{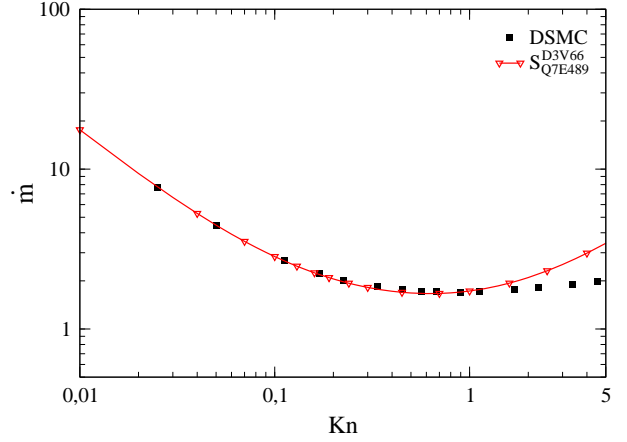}
\caption{\label{fig5} (Color online) Mass flow of the model $S^{D3V66}_{Q7E489}$
with the smallest error $| \sigma_{zxx} |$ within the set ${\cal S}_1$.}
\end{center}
\end{figure}
\begin{figure}  
\begin{center}
\epsfig{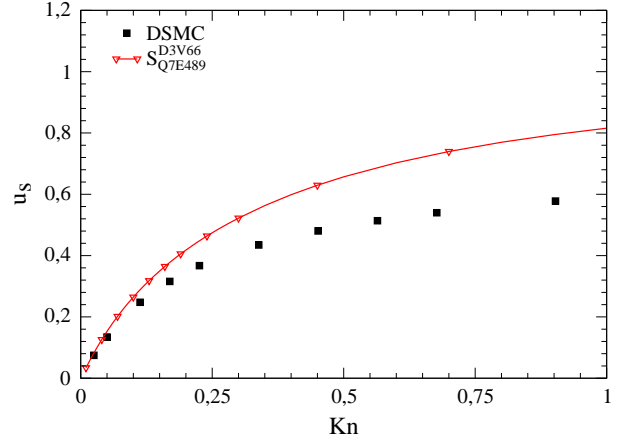}
\caption{\label{fig6} (Color online) Slip velocity of the model $S^{D3V66}_{Q7E489}$
with the smallest error $| \sigma_{zxx} |$ within the set ${\cal S}_1$.}
\end{center}
\end{figure}
\begin{figure}  
\begin{center}
\epsfig{file=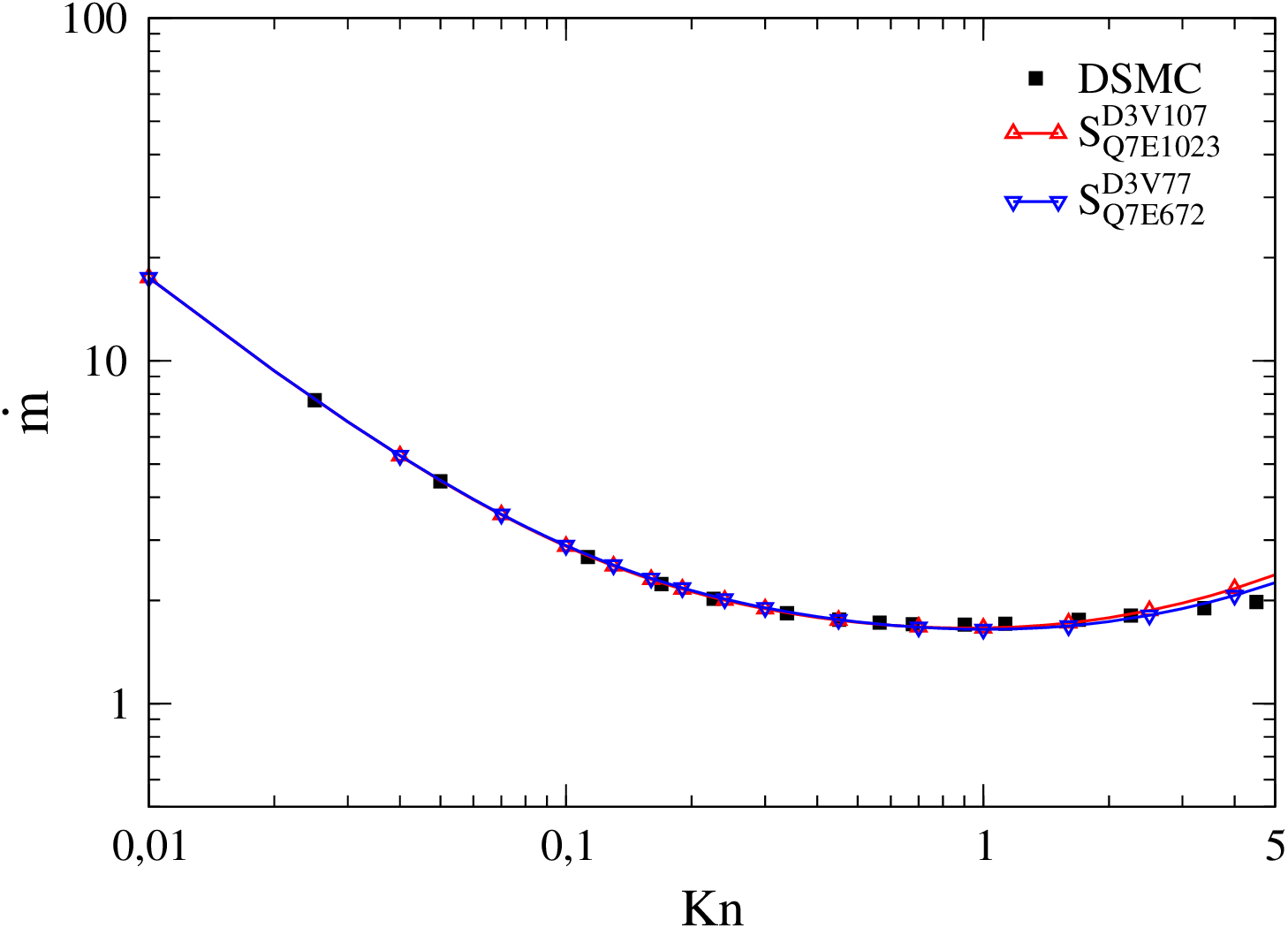,width=8.0cm,clip=,bb=11 26 711 530}
\caption{\label{fig7} (Color online) Mass flow rate of selected ${\cal S}_3$ models ($\Lambda_{\mathrm{index}}=16$). }
\end{center}
\end{figure}
\begin{figure}  
\begin{center}
\epsfig{file=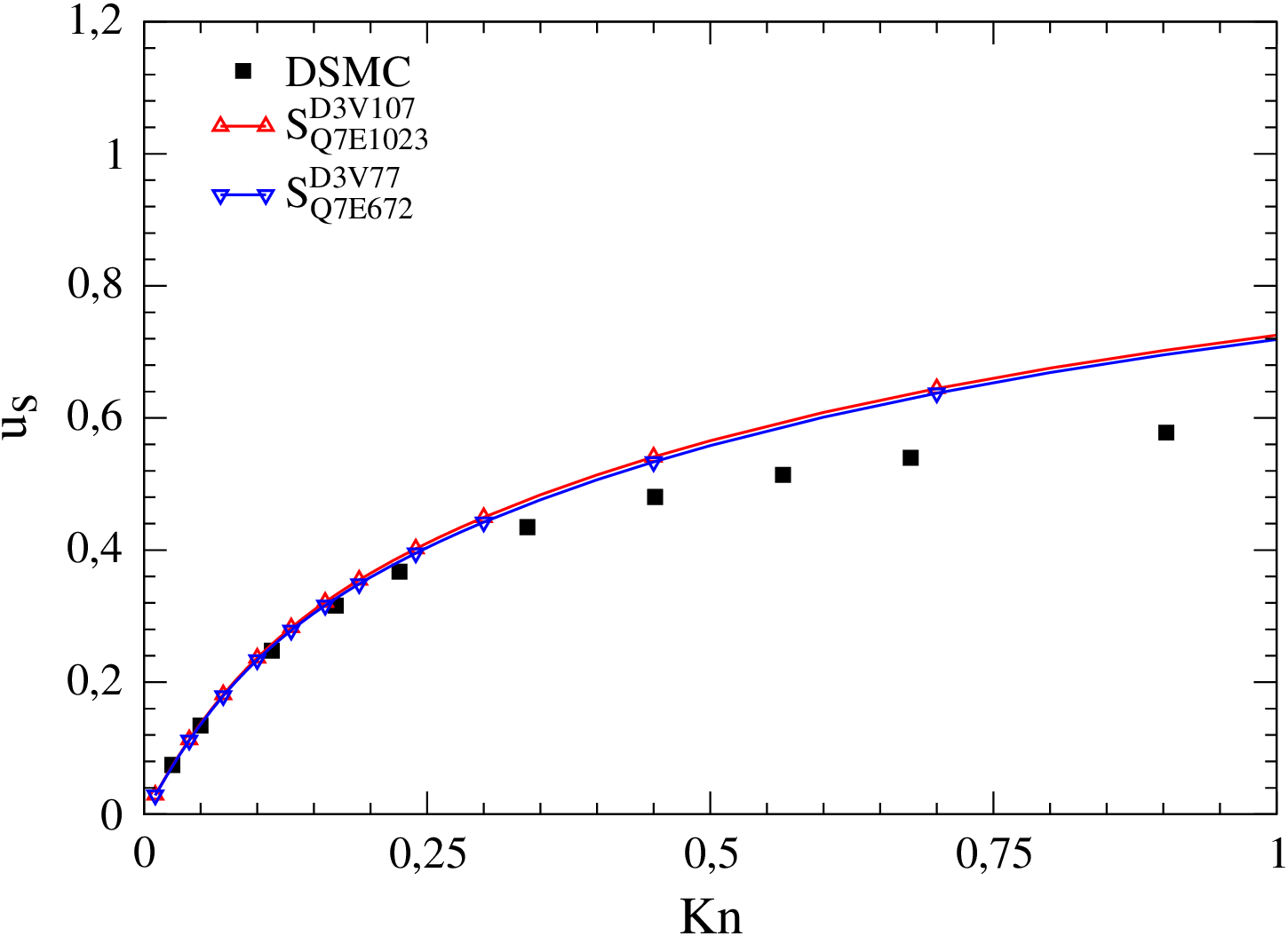,width=8.0cm,clip=,bb= 10 26 710 532}
\caption{\label{fig8} (Color online) Slip velocity of selected ${\cal S}_3$ models ($\Lambda_{\mathrm{index}}=16$). }
\end{center}
\end{figure}
\begin{figure}  
\begin{center}
\epsfig{file=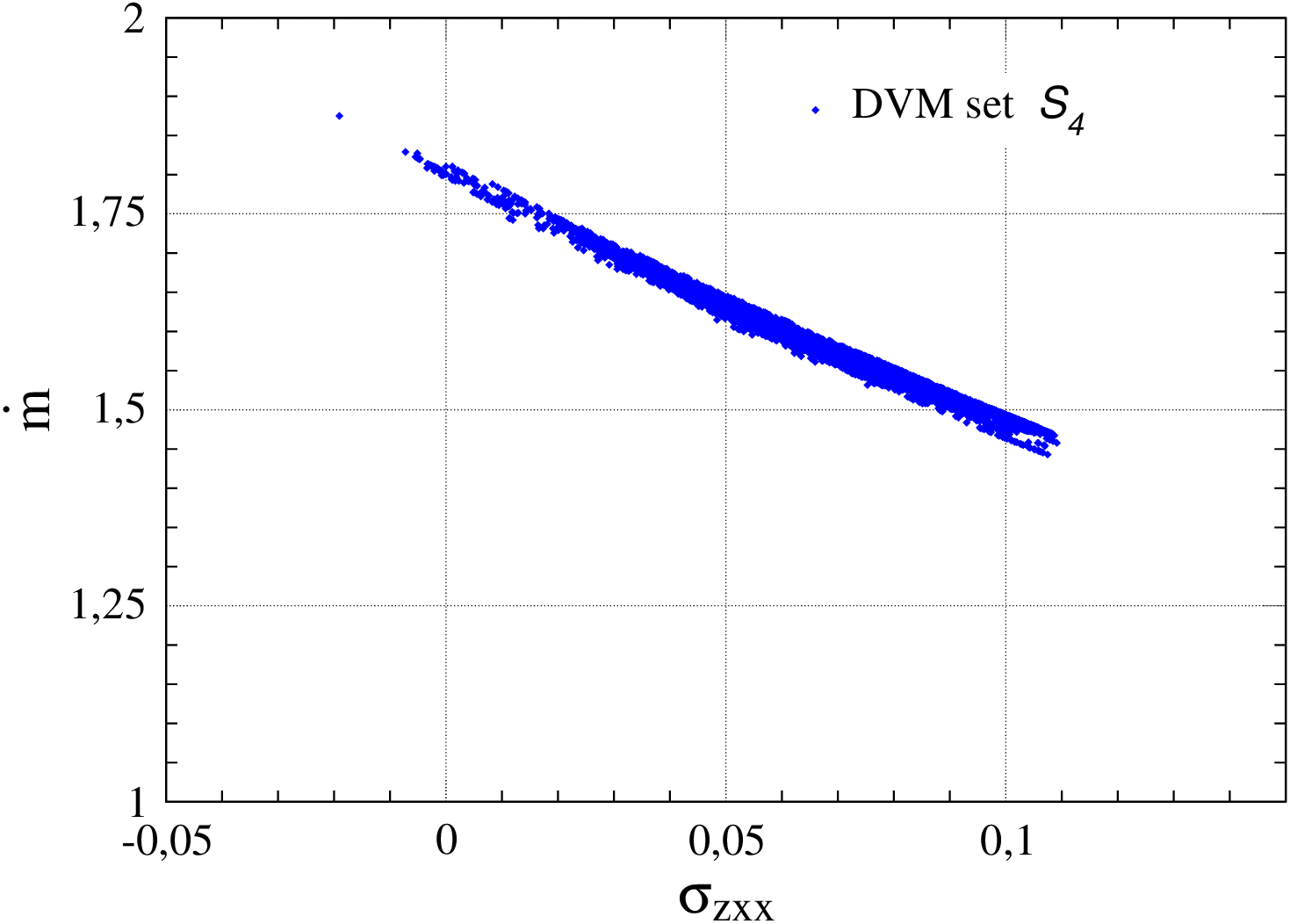,width=8.0cm,clip=,bb=1 25 706 530}
\caption{\label{fig9} (Color online) Correlation between $\sigma_{zxx}$ and $\dot{m}$ for $\Kn=0.4514$. 
Every dot represents a DVM within ${\cal S}_4$.}
\end{center}
\end{figure}
\begin{figure}  
\begin{center}
\epsfig{file=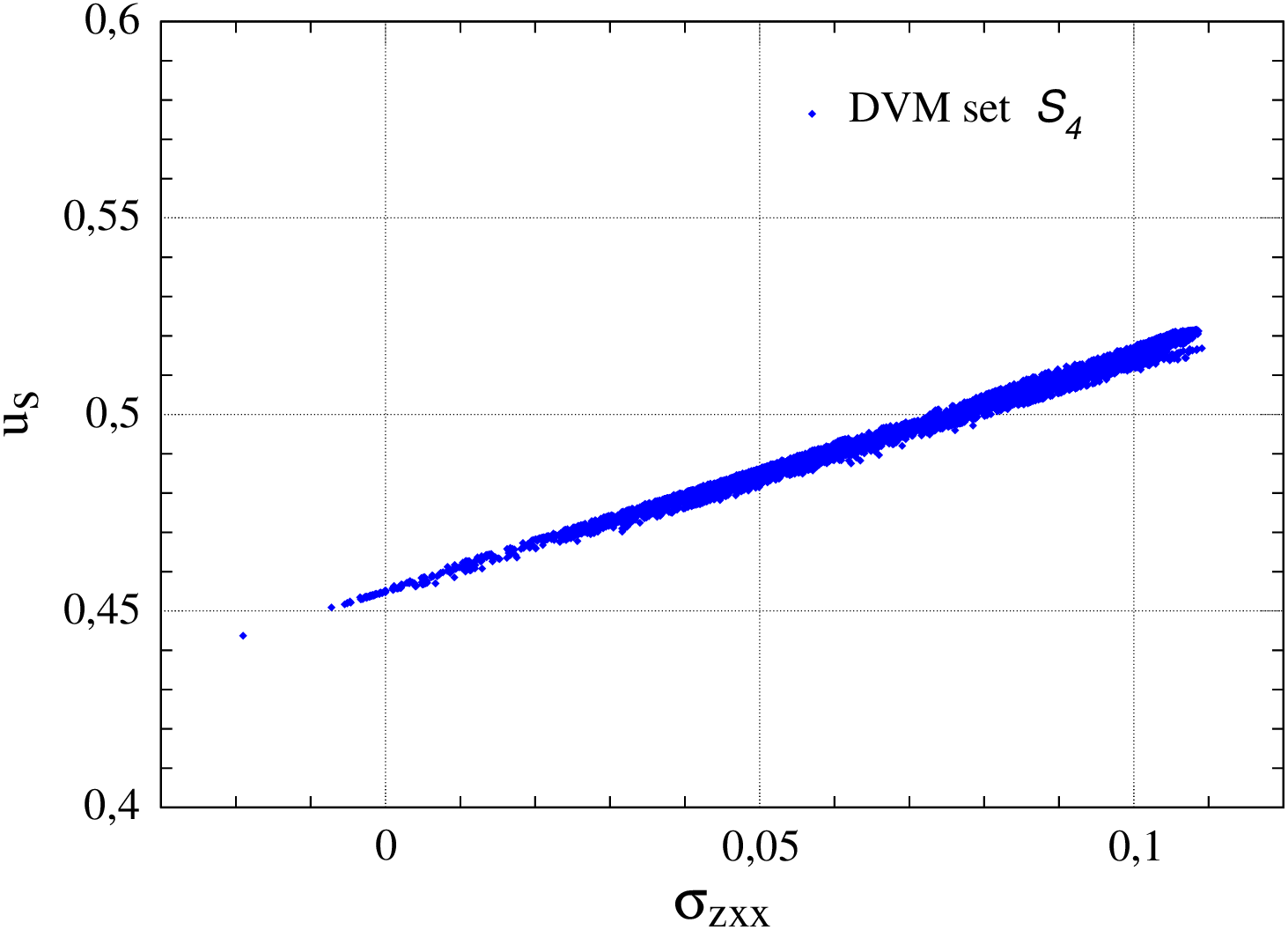,width=8.0cm,clip=,bb=1 25 706 532}
\caption{\label{fig10} (Color online) Correlation between $\sigma_{zxx}$ and $u_s$ for $\Kn=0.4514$. 
Every dot represents a DVM within ${\cal S}_4$.}
\end{center}
\end{figure}
\begin{figure}  
\begin{center}
\epsfig{file=figure_13.eps,width=8.0cm,clip=,bb=11 26 711 530}
\caption{\label{fig11} (Color online) Mass flow rate of selected ${\cal S}_4$ 
models ($\Lambda_{\mathrm{index}}=6245$). }
\end{center}
\end{figure}
\begin{figure}  
\begin{center}
\epsfig{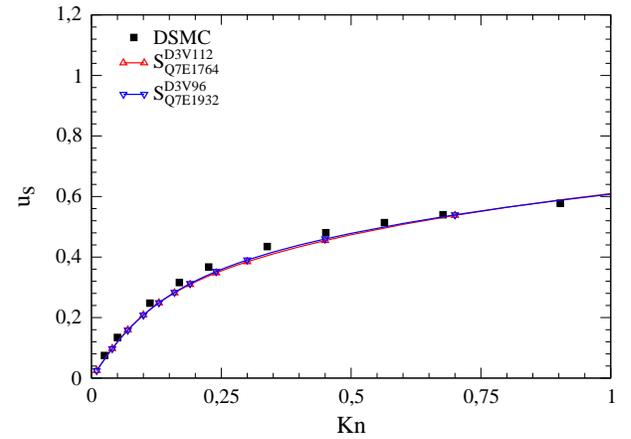}
\caption{\label{fig12} (Color online) Slip velocity of selected ${\cal S}_4$ 
models ($\Lambda_{\mathrm{index}}=6245$). }
\end{center}
\end{figure}

The reason for the failure of many high-order {\LBms }  for finite $\Kn$ 
is their weak ability to recover the {\DMBC }  exactly.
We assess this ability by considering the errors of the wall moments in Table \ref{tab:bulk}. The exact definition of the models listed here can be found in Appendix \ref{sec:3D}.
We here identified DVMs of the set ${\cal S}_1$ 
with wall moment errors as small as possible which thus ensure a good realization 
of the \DMBC.
For the two different models
$S^{D3V64}_{Q7E447}$ and $S^{D3V59}_{Q7E408}$ some
wall moment errors are given in Table \ref{tab:bulk} and the flow results are shown in 
Figs.\ \ref{fig2} and \ref{fig3}. 
Due to smaller wall moment errors we observe for both models a significantly higher accuracy for 
finite $\Kn$.
The errors in the slip velocity are nearly the same whereas the predicted mass flow rate of the model $S^{D3V64}_{Q7E447}$ is more accurate.
By using the CE analysis (see Sect.\ \ref{LBH}) one can show that the relevant low $\Ma$
contributions to the momentum dynamics up to the Super-Burnett flow regime are affected by the 
third and fifth equilibrium moments.
Therefore the higher accuracy of model $S^{D3V64}_{Q7E447}$ in the mass flow prediction is caused 
by a more accurate representation of the third wall moment $W_{zxx}$ (see Table \ref{tab:bulk}).

\begin{table} [htb!]
\renewcommand{\arraystretch}{1.2}
\centering
\begin{tabular}{|c|c|c|c|c|c|c|}
\hline
DVM & Q & $\sigma$ & $\sigma_z$ & $\sigma_{xx}$ & $\sigma_{zxx}$ & $\sigma_{\Sigma}$ \\
\hline
 $D3Q19$ & 5 & $-0.6667$ & $-0.2764$ & $-0.6667$ & $-0.2764$              & $0.5506$ \\ 
 $D3Q121$ & 9 & $-0.4767$ & $-0.1292$ & $-0.4767$ & $-0.1292$             & $0.3685$ \\ 
 $S^{D3V38}_{Q7E399}$ & 7 & $-0.4902$ & $-0.1505$ & $-0.5926$ & $-0.2109$ & $0.4042$ \\ 
\hline
 $S^{D3V64}_{Q7E447}$ & 7 & $-0.0443$ & $-0.0750$ & $-0.1212$ & $-0.0258$ & $0.0621$ \\ 
 $S^{D3V59}_{Q7E408}$ & 7 & $-0.0367$ & $0.0932$ & $-0.0794$ & $0.0658$   & $0.0576$ \\ 
\hline
 $S^{D3V66}_{Q7E489} $ & 7 & $-0.3757$ & $-0.0805$ & $-0.2681$ & $-0.0001$ & $0.2720$ \\ 
\hline
\end{tabular}
\caption{\label{tab:bulk} Wall moment errors of standard {\LBm }  $D3Q19$, the common DVM $\st{9}{121}{3}{594}$ also known as $D3Q121$, as well as some exclusive ${\cal S}_1$ bulk flow models. $Q$ is the quadrature order.}
\end{table}

If we consider the influence of the wall moment error $\sigma_{zxx}$ on the mass flow rate within all ${\cal S}_1$
models we observe a strong correlation between $\sigma_{zxx}$ and $\dot{m}$. This is shown in Fig.\ \ref{fig4}
for a Knudsen number of $\Kn=0.4514$. Pearson's linear correlation coefficient yields the value $-0.98$, exceeding in magnitude all other correlations of $\dot{m}$.
It is interesting to notice that all models with a nearly vanishing wall moment error $\sigma_{zxx}$ predict 
a mass flow rate which is very close to the DSMC results of $\dot{m}_{DSMC} = 1.76$ for $\Kn=0.4514$.
The {\LBm } with the smallest error $|\sigma_{zxx}|$ within the set ${\cal S}_1$ is the $S^{D3V66}_{Q7E489}$
model. The wall moment errors are specified in Table \ref{tab:bulk}. As a consequence of the minimal
$|\sigma_{zxx}|$ error the mass flow rate is in excellent agreement with the DSMC results
up to a Knudsen number of $\Kn \approx 1$ (see Fig.\ \ref{fig5}).
On the other hand the slip velocity, shown in Fig.\ \ref{fig6}, is less accurate compared to the other 
selected models ($S^{D3V64}_{Q7E447}$, $S^{D3V59}_{Q7E408}$) due to remaining inaccuracies of other wall moments.

These results confirm that a reliable {\LBm } for finite $\Kn$ flows requires both
a high quadrature order $Q$ and small wall moment errors which guarantees a precise 
realization of the \DMBC.

\subsection{Using discrete velocity models for wall-bounded flow with $\Lambda_{\mathrm{index}}=16$ \label{sec:PoiS3}}
In this section we discuss the set ${\cal S}_3$ of DVMs.
These models are of Gaussian order $Q=7$ and include a wall moment constraint which ensures 
that the $\sigma_{zxx}$ error vanishes. Consequently, the {\wai } has the value $\Lambda_{\mathrm{index}}=16$, 
see Eq.\ \eqref{lambdaS3}. 
From this set of models we select the $S^{D3V107}_{Q7E1023}$ and $S^{D3V77}_{Q7E672}$ models which 
additionally have a small overall wall moment error $\sigma_\Sigma$ (see Table \ref{tab:Wzxx}).
The results for the mass flow rate, shown in Fig.\ \ref{fig7}, are in excellent agreement with the DSMC results due to the wall constraint $\sigma_{zxx} = 0$.
Additionally, the Knudsen minimum at $\Kn \approx 1$ is very well reproduced.
For the slip velocity (Fig.\ \ref{fig8}) we observe slight differences to the reference data 
which may be caused by remaining inaccuracies of other wall moments.
\begin{widetext}
\begin{center}
\begin{table} [htb!]
\renewcommand{\arraystretch}{1.2}
\centering
\begin{tabular}{|c|c|c|c|c|c|c|c|c|}
\hline
\LBm & Q & $\sigma$ & $\sigma_z$ & $\sigma_{xx}$ & $\sigma_{zxx}$ & $\sigma_{zzzxx}$ & $\sigma_{xxzzzzz}$ & $\sigma_{\Sigma}$  \\
\hline
 $S^{D3V107}_{Q7E1023}$ & 7 & $-0.0998$ & $0.0247$ & $-0.1482$ & $0.0$ & $0.0027$ & $-0.0083$ & $0.0849$ \\ 
 $S^{D3V77}_{Q7E672}$  & 7 & $-0.0816$ & $0.0343$ & $-0.1339$ & $0.0$ & $0.0132$ & $-0.0710$ & $0.0765$ \\ 
\hline
\end{tabular}
\caption{\label{tab:Wzxx} Wall moment errors of some exclusive ${\cal S}_3$ models, where $Q$ is the quadrature order.}
\end{table}
\end{center}
\end{widetext}
\subsection{Using discrete velocity models for wall-bounded flow with $\Lambda_{\mathrm{index}}=6245$ \label{sec:PoiS4}}
Another set of wall moment \LBms, ${\cal S}_4$, is characterized by a Gaussian 
quadrature order $Q=7$ and scattering stencils which guarantee that all even 
wall moments up to the quadrature order $Q=7$ are represented exactly,
\begin{align}
\label{pf.3}
\sigma = \sigma_{i_1 i_2} = \sigma_{i_1 \dots i_4} = \sigma_{i_1 \dots i_6} = 0 .
\end{align}
Consequently, the {\wai } has the value $\Lambda_{\mathrm{index}}=6245$, see Eq.\ \eqref{lambdaS4}. Similar to the model sets ${\cal S}_1$ and ${\cal S}_3$ discussed previously, we observe
for ${\cal S}_4$ models a strong correlation between the wall
moment $W_{zxx}$ and the flow results as well. Figure \ref{fig9} shows the correlation between
$\sigma_{zxx}$ and the mass flow and Fig.\ \ref{fig10} shows the correlation between
$\sigma_{zxx}$ and the slip velocity.
Because of the strengths of these correlations we select the models 
$S^{D3V112}_{Q7E1764}$ and $S^{D3V96}_{Q7E1932}$ 
with small errors $|\sigma_{zxx}|$ (see Table \ref{tab:W0}).
The Poiseuille flow results for these models are shown in Figs.\ \ref{fig11} and\ \ref{fig12}.
Both the mass flow rate and the slip velocity at the wall are in excellent agreement
with the DSMC results up to $\Kn=1$.
The crucial point for the success of these models are very low errors for all wall moments
up to the quadrature order ($Q=7$).
However these models cannot predict the Knudsen minimum which was reproduced by models of
groups ${\cal S}_1$ and ${\cal S}_3$.
\begin{table}[htb!]
\renewcommand{\arraystretch}{1.2}
\centering
\begin{tabular}{|c|c|c|c|c|c|c|}
\hline
\LBm & Q & $\sigma$ & $\sigma_z$ & $\sigma_{xx}$ & $\sigma_{zxx}$ & $\sigma_{\Sigma}$ \\
\hline
 $S^{D3V112}_{Q7E1764}$ & 7 & $0.0$ & $0.0373$ & $0.0$ & $0.000005$ & $0.0090$ \\ 
 $S^{D3V96}_{Q7E1932}$  & 7 & $0.0$ & $0.0386$ & $0.0$ & $0.0114$   & $0.0101$ \\ 
\hline
\end{tabular}
\caption{\label{tab:W0} Wall moment errors of some exclusive ${\cal S}_4$ models, where $Q$ is the quadrature order.}
\end{table}

\subsection{Velocity profiles \label{PF.4}}
\begin{figure}   
\begin{center}
\epsfig{file=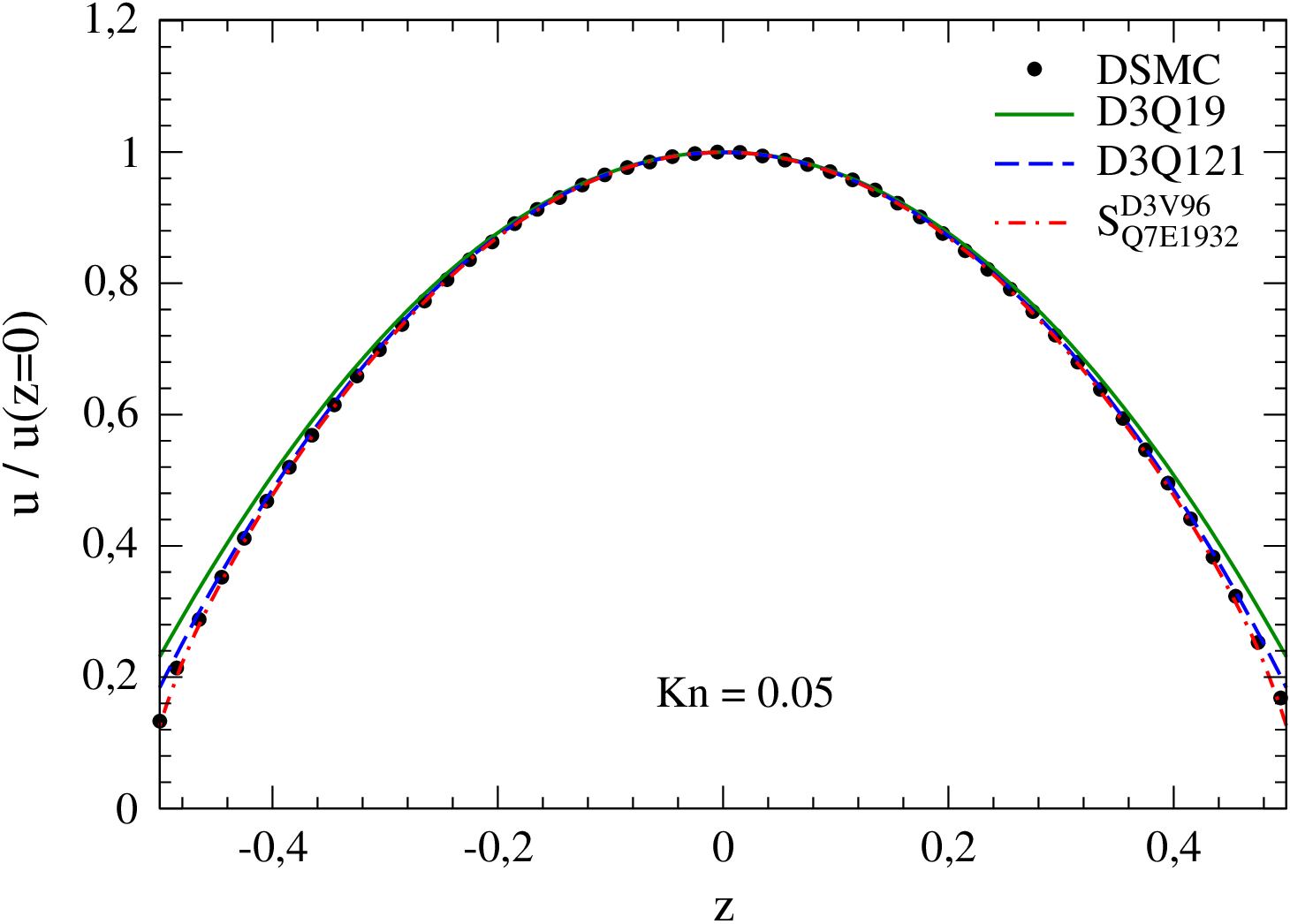,width=8.0cm,clip=,bb=4 29 706 522}
\caption{\label{fig13} (Color online) Streamwise velocity profile for  $\Kn=0.05$. }
\end{center}
\end{figure}

Figs.\ \ref{fig13} -\ \ref{fig18} show the streamwise velocities of the considered 
{\LBms }  for several Knudsen numbers.
At $\Kn = 0.05$ all models agree well with the DSMC results, however 
$D3Q19$ and $S^{D3V38}_{Q7E399}$ slightly over-predict the slip velocity at the walls.
For higher $\Kn$ the models $D3Q19$, $D3Q121$ and $S^{D3V38}_{Q7E399}$
no longer perform well due to errors of the wall moments.
On the other hand, the models $S^{D3V64}_{Q7E447}$, $S^{D3V77}_{Q7E672}$ and $S^{D3V96}_{Q7E1932}$
show a significantly better prediction of the velocity  field because of more
accurate wall moments yielding a more precise realization of
the \DMBC.

In particular, the $S^{D3V96}_{Q7E1932}$ model guarantees a high accuracy for all wall
moments up to the Gaussian quadrature order $Q=7$ and therefore remains quantitatively
accurate at least up to $\Kn \approx 1$.
Only for Knudsen numbers higher than $\Kn = 2 $ we observe a slight overestimation
of the slip velocity.
\begin{figure}  
\begin{center}
\epsfig{file=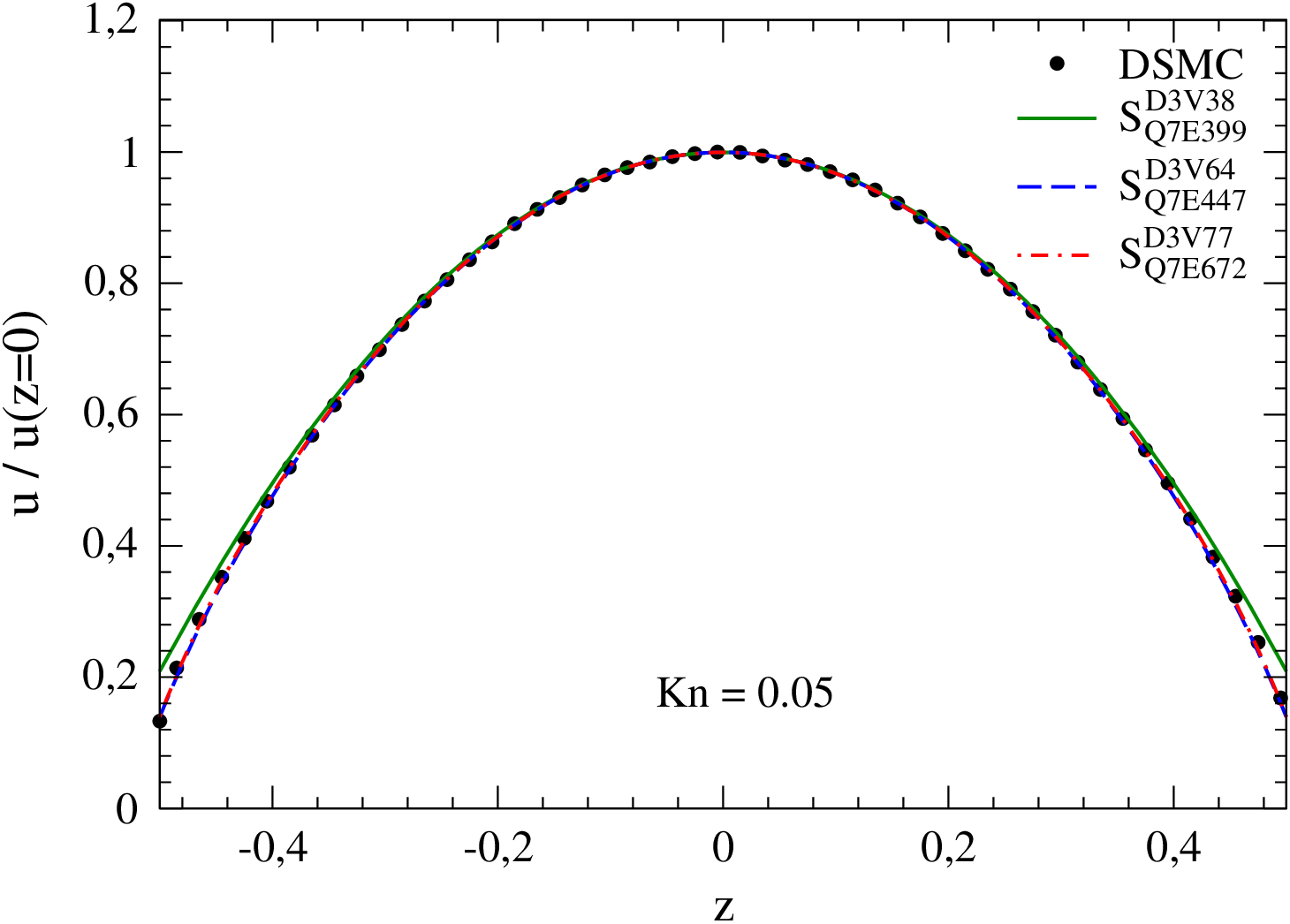,width=8.0cm,clip=,bb=4 29 706 522}
\caption{\label{fig13_2} (Color online) Streamwise velocity profile for  $\Kn=0.05$. }
\end{center}
\end{figure}
\begin{figure}  
\begin{center}
\epsfig{file=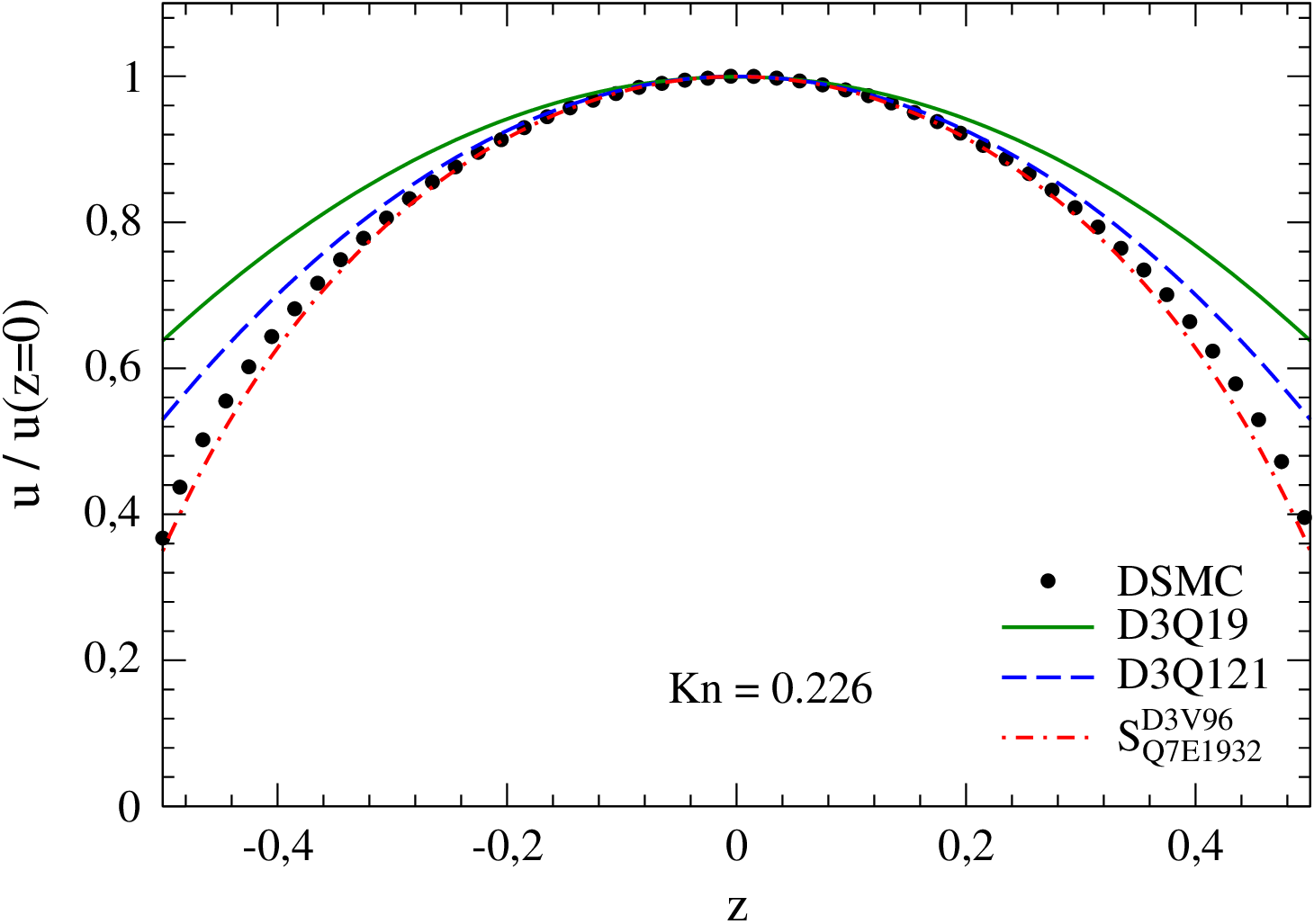,width=8.0cm,clip=,bb=4 29 706 522}
\caption{\label{fig14} (Color online) Streamwise velocity profile for  $\Kn=0.226$. }
\end{center}
\end{figure}
\begin{figure}  
\begin{center}
\epsfig{file=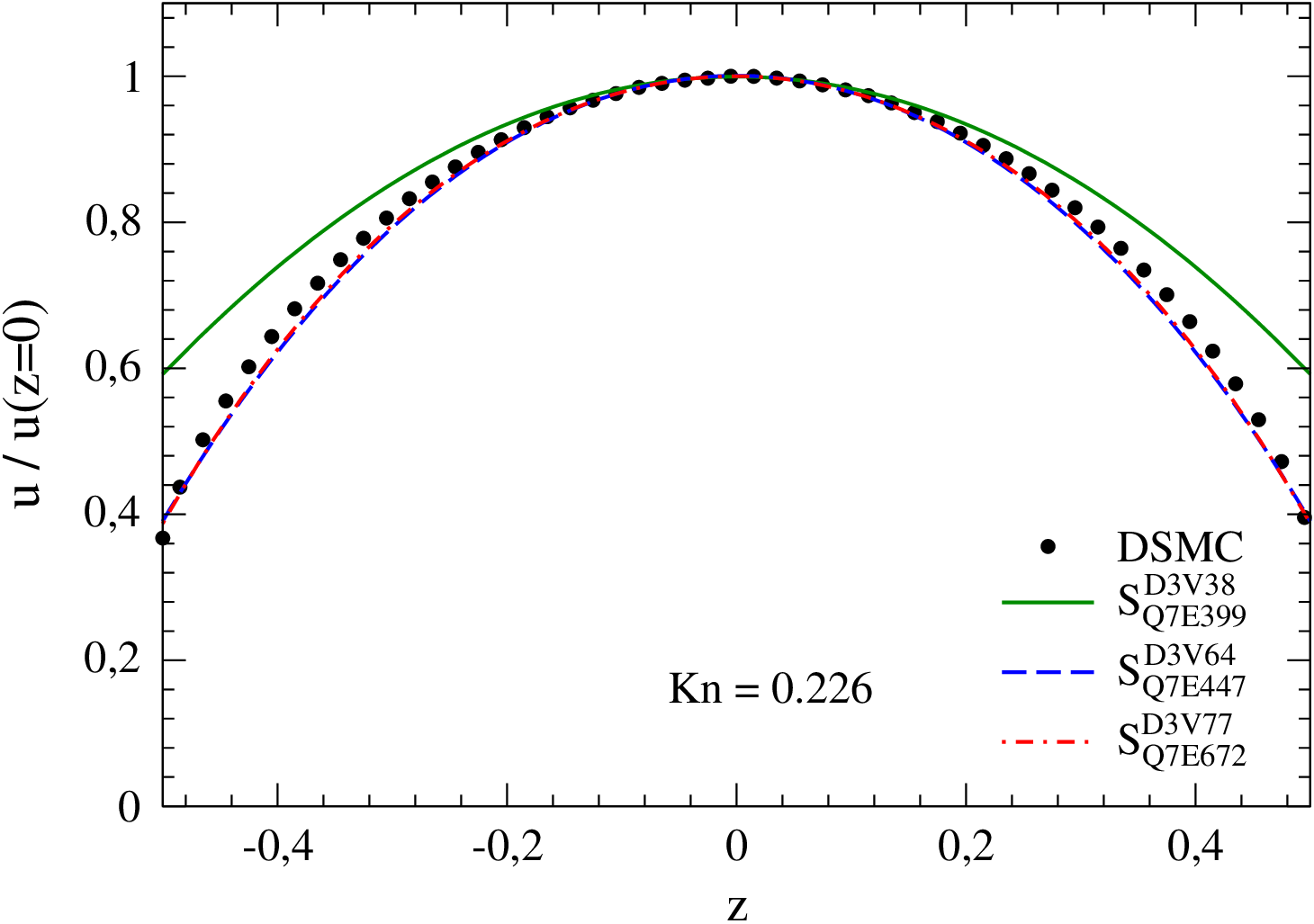,width=8.0cm,clip=,bb=4 29 706 522}
\caption{\label{fig14_2} (Color online) Streamwise velocity profile for  $\Kn=0.226$. }
\end{center}
\end{figure}
\begin{figure}  
\begin{center}
\epsfig{file=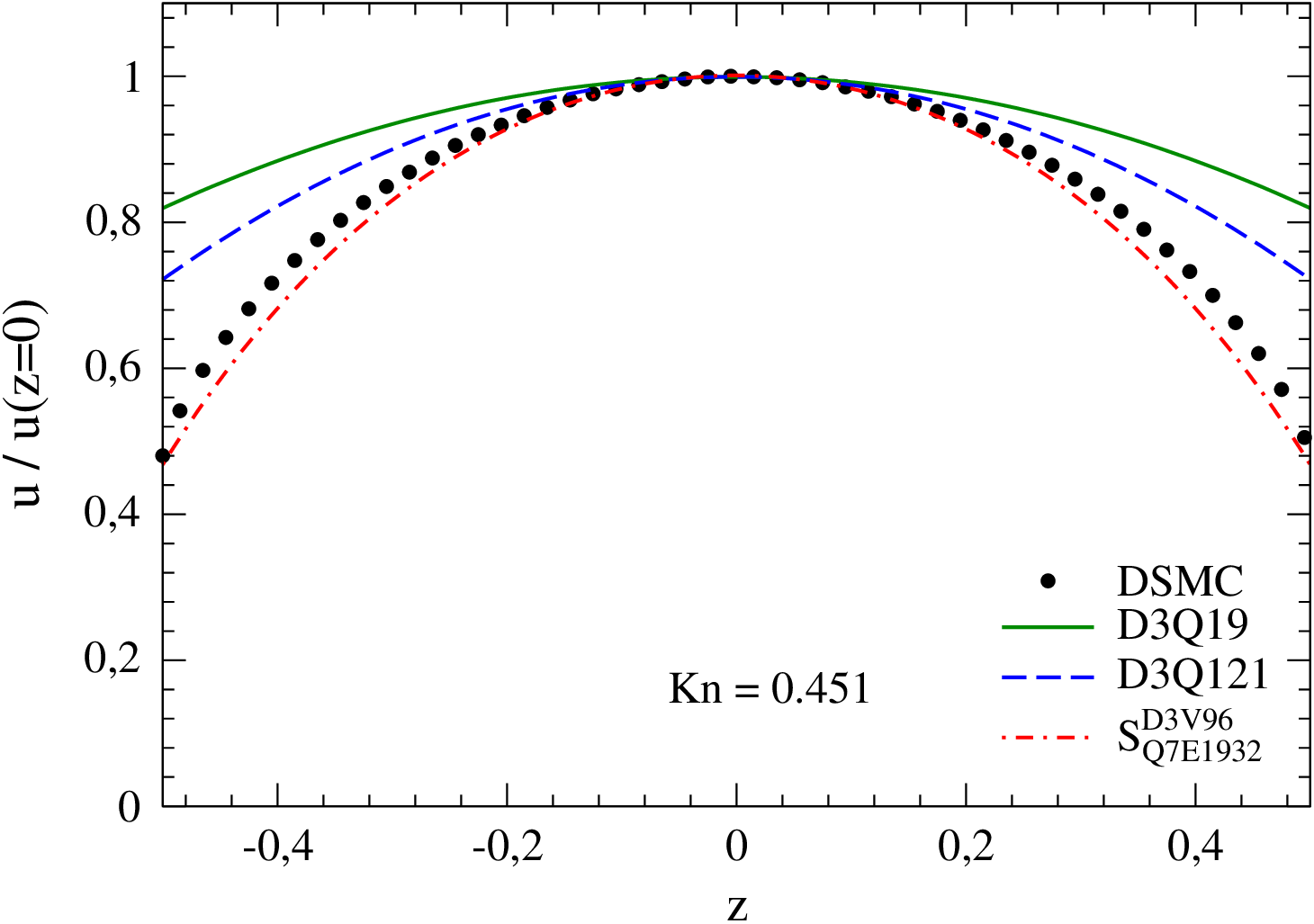,width=8.0cm,clip=,bb=4 29 706 522}
\caption{\label{fig15} (Color online) Streamwise velocity profile for  $\Kn=0.451$. }
\end{center}
\end{figure}
\begin{figure}  
\begin{center}
\epsfig{file=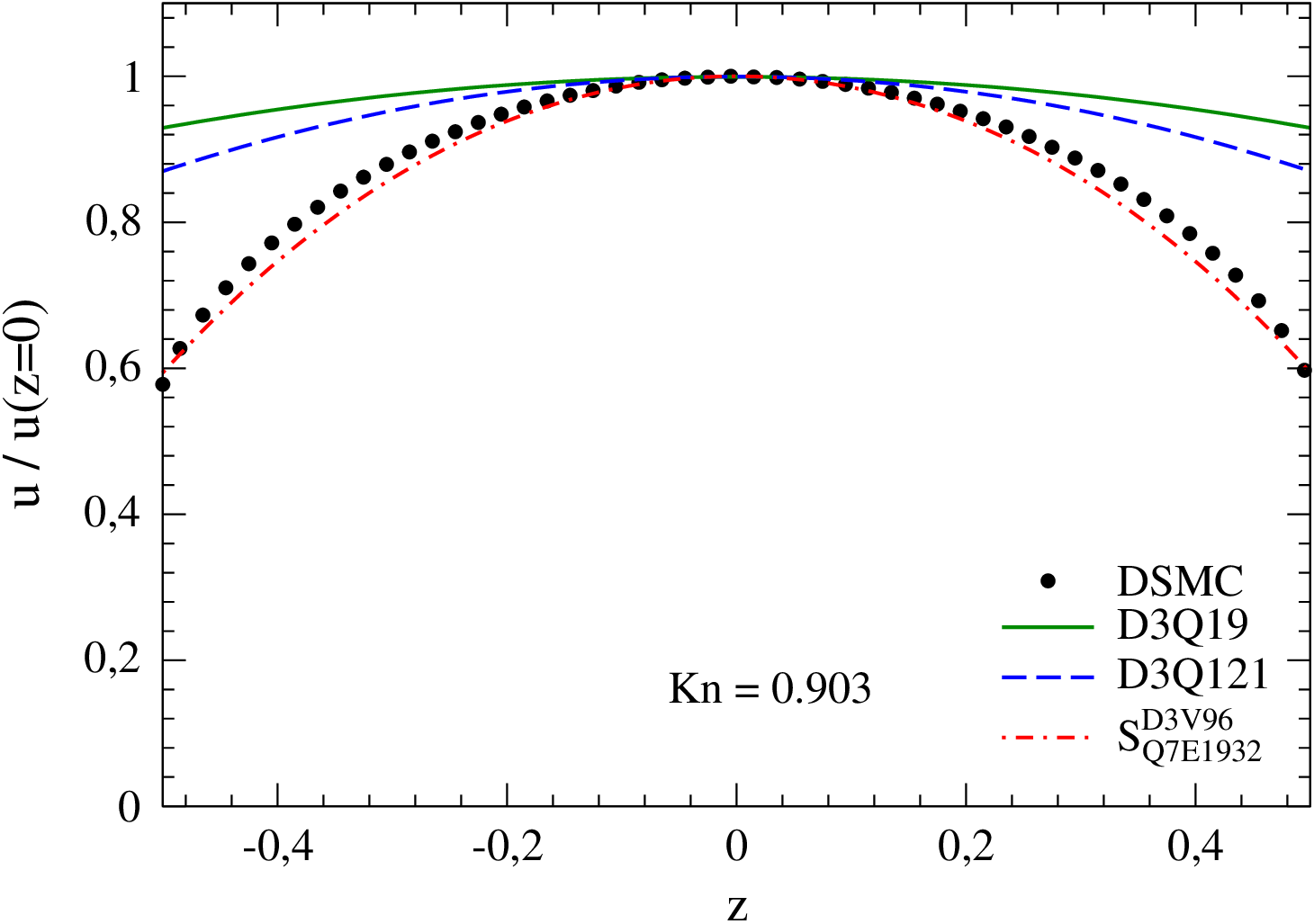,width=8.0cm,clip=,bb=4 29 706 522}
\caption{\label{fig16} (Color online) Streamwise velocity profile for  $\Kn=0.903$. }
\end{center}
\end{figure}
\begin{figure}  
\begin{center}
\epsfig{file=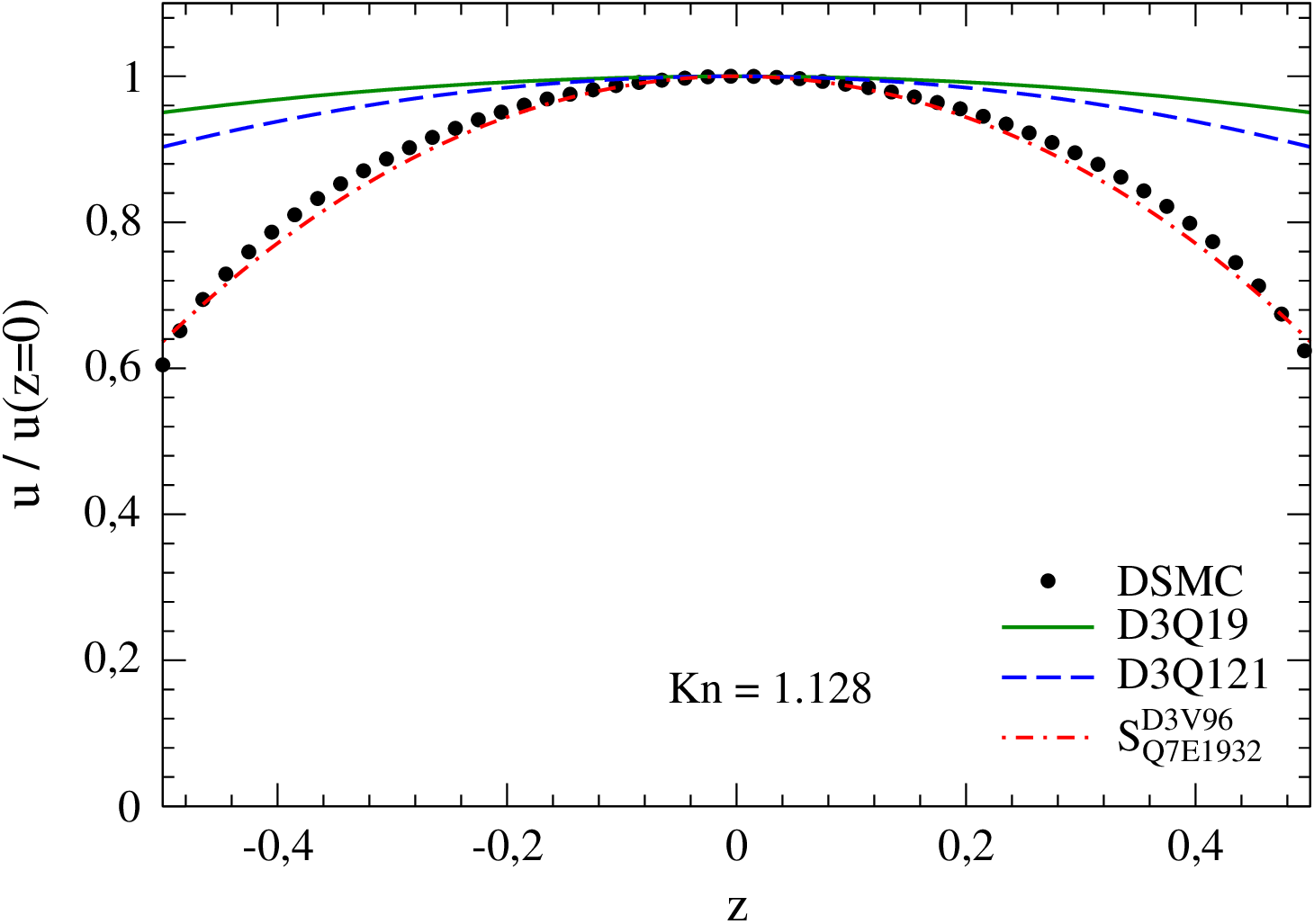,width=8.0cm,clip=,bb=4 29 706 522}
\caption{\label{fig17} (Color online) Streamwise velocity profile for  $\Kn=1.128$. }
\end{center}
\end{figure}
\begin{figure}  
\begin{center}
\epsfig{file=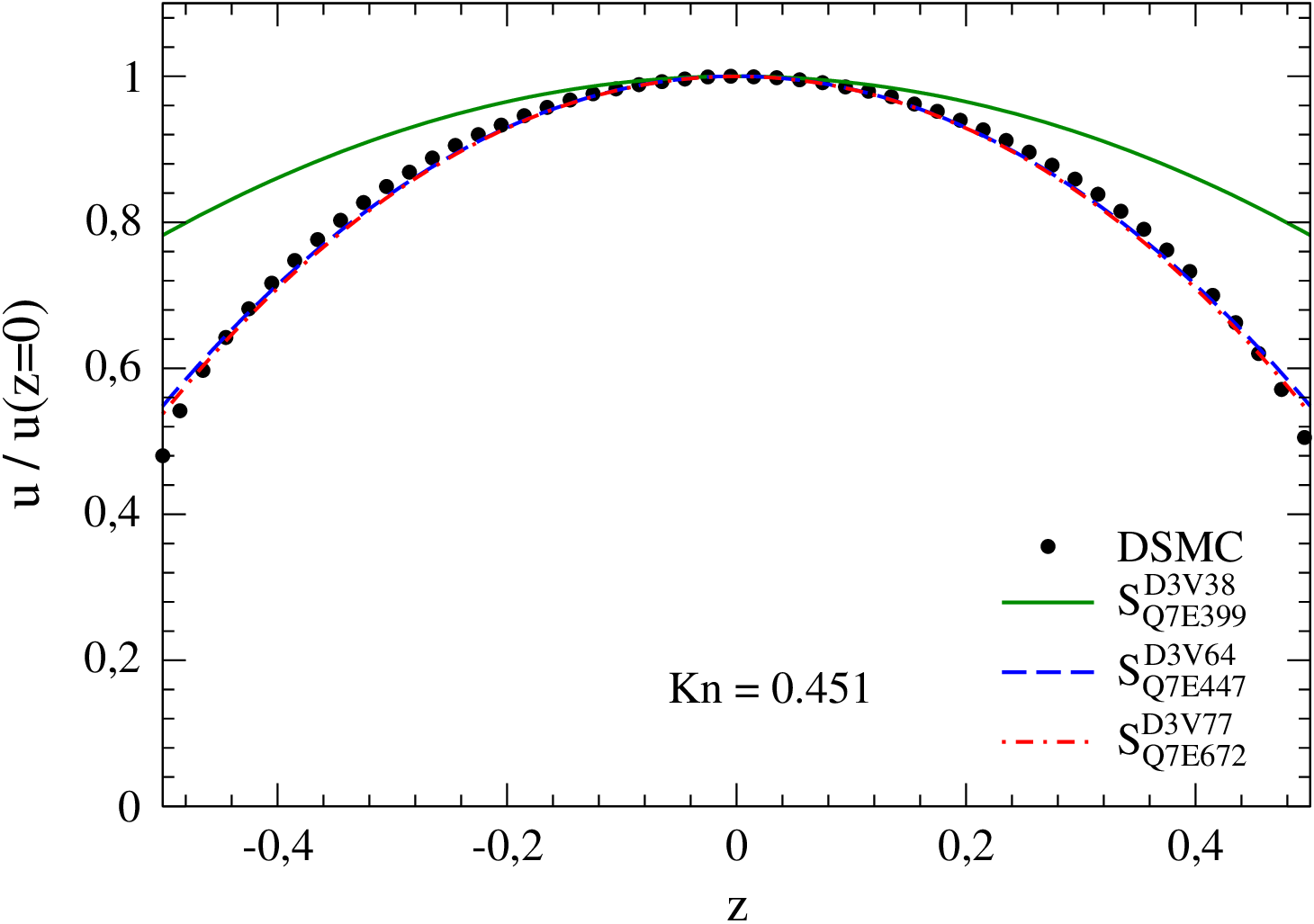,width=8.0cm,clip=,bb=4 29 706 522}
\caption{\label{fig15_2} (Color online) Streamwise velocity profile for  $\Kn=0.451$. }
\end{center}
\end{figure}
\begin{figure}  
\begin{center}
\epsfig{file=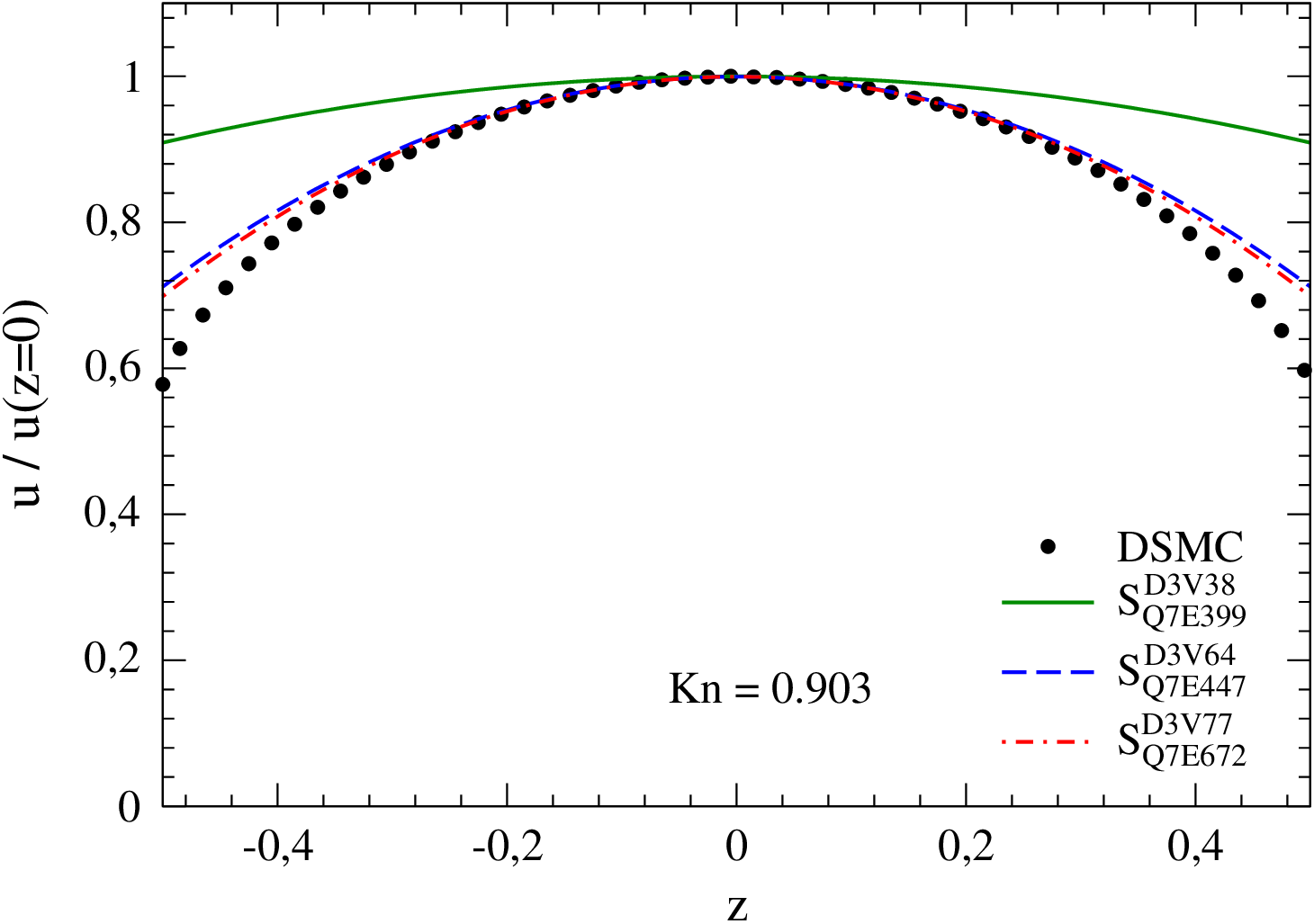,width=8.0cm,clip=,bb=4 29 706 522}
\caption{\label{fig16_2} (Color online) Streamwise velocity profile for  $\Kn=0.903$. }
\end{center}
\end{figure}
\begin{figure}  
\begin{center}
\epsfig{file=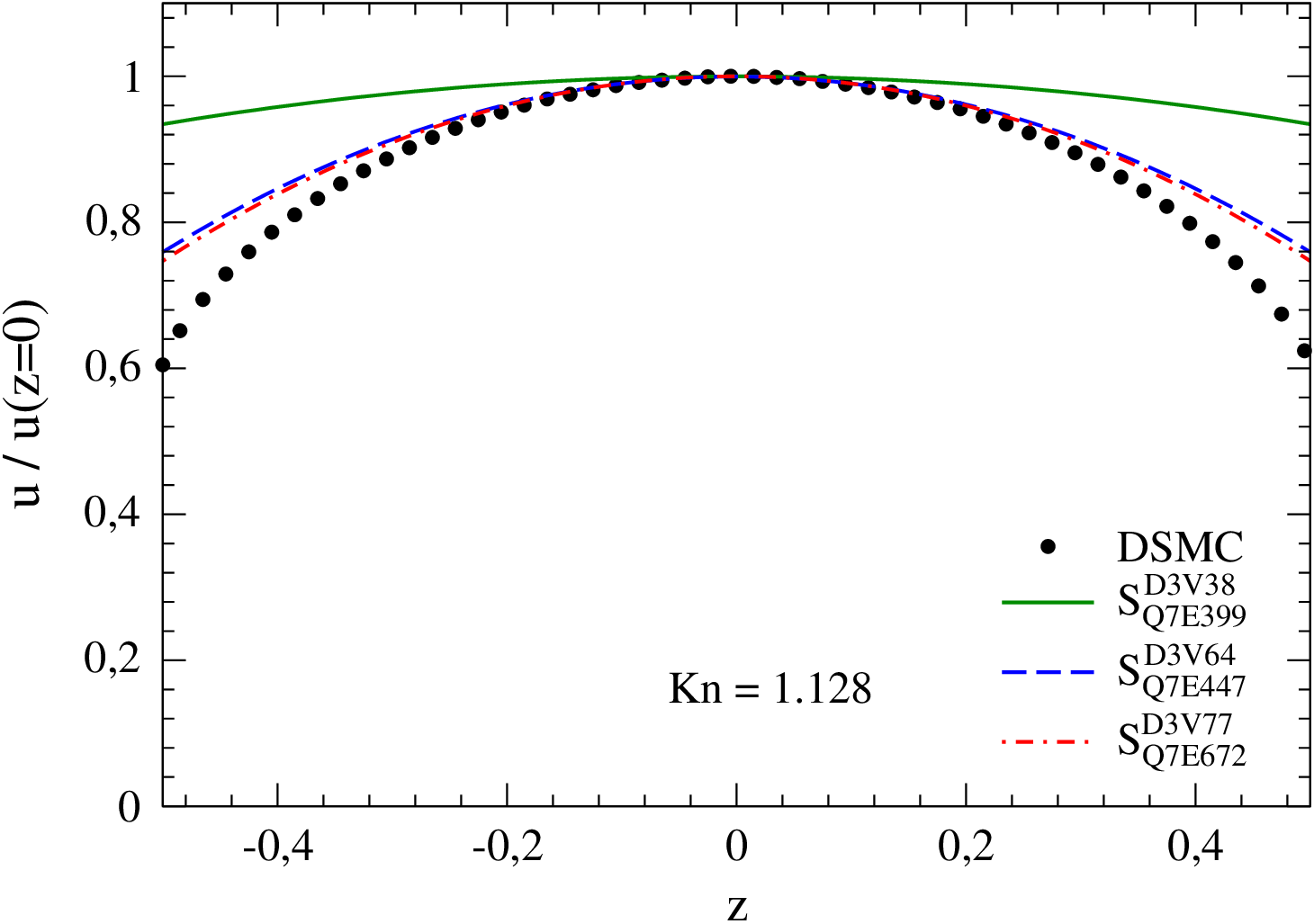,width=8.0cm,clip=,bb=4 29 706 522}
\caption{\label{fig17_2} (Color online) Streamwise velocity profile for  $\Kn=1.128$. }
\end{center}
\end{figure}
\begin{figure}  
\begin{center}
\epsfig{file=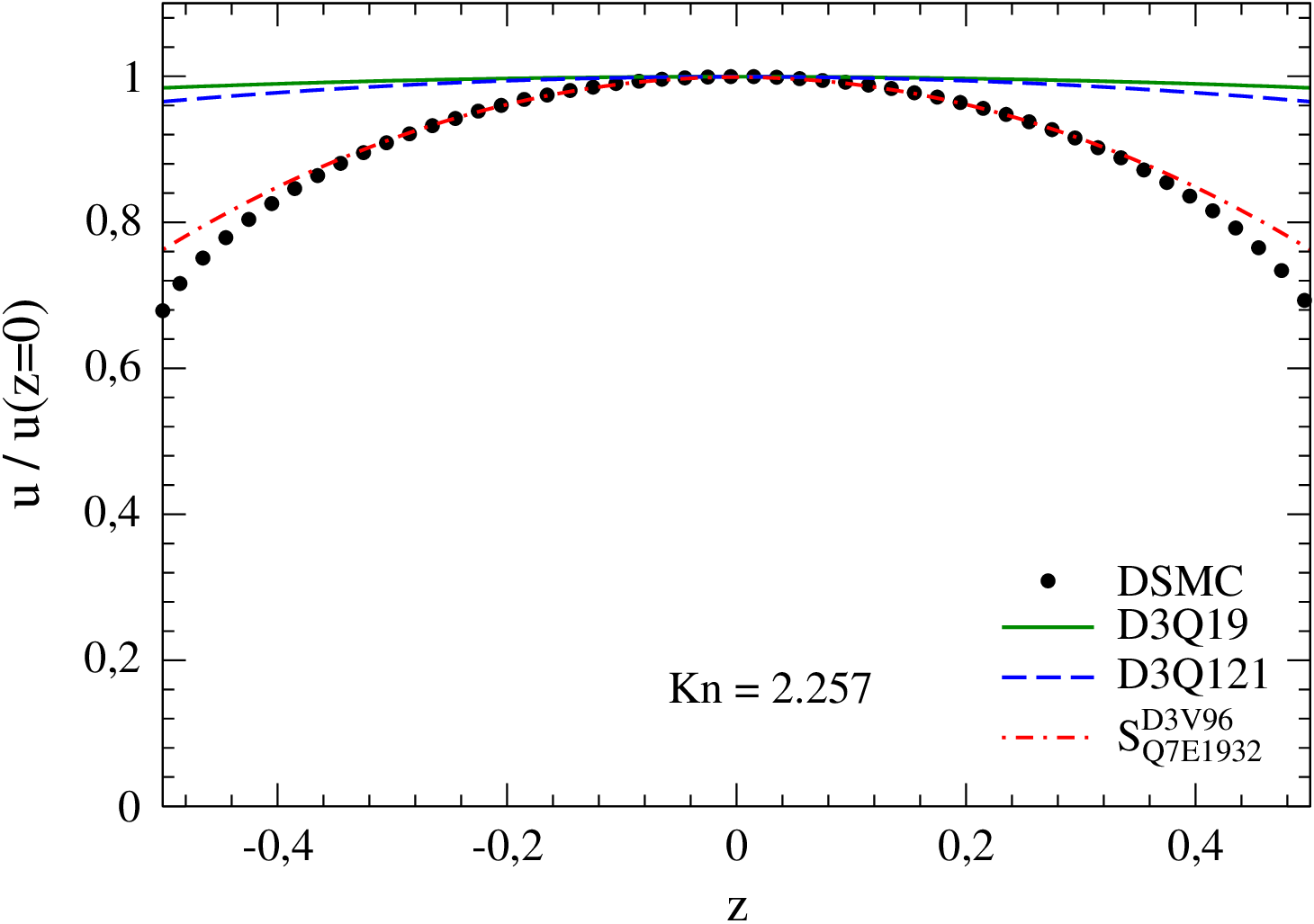,width=8.0cm,clip=,bb=4 29 706 522}
\caption{\label{fig18} (Color online) Streamwise velocity profile for  $\Kn=2.257$. }
\end{center}
\end{figure}
\begin{figure}  
\begin{center}
\epsfig{file=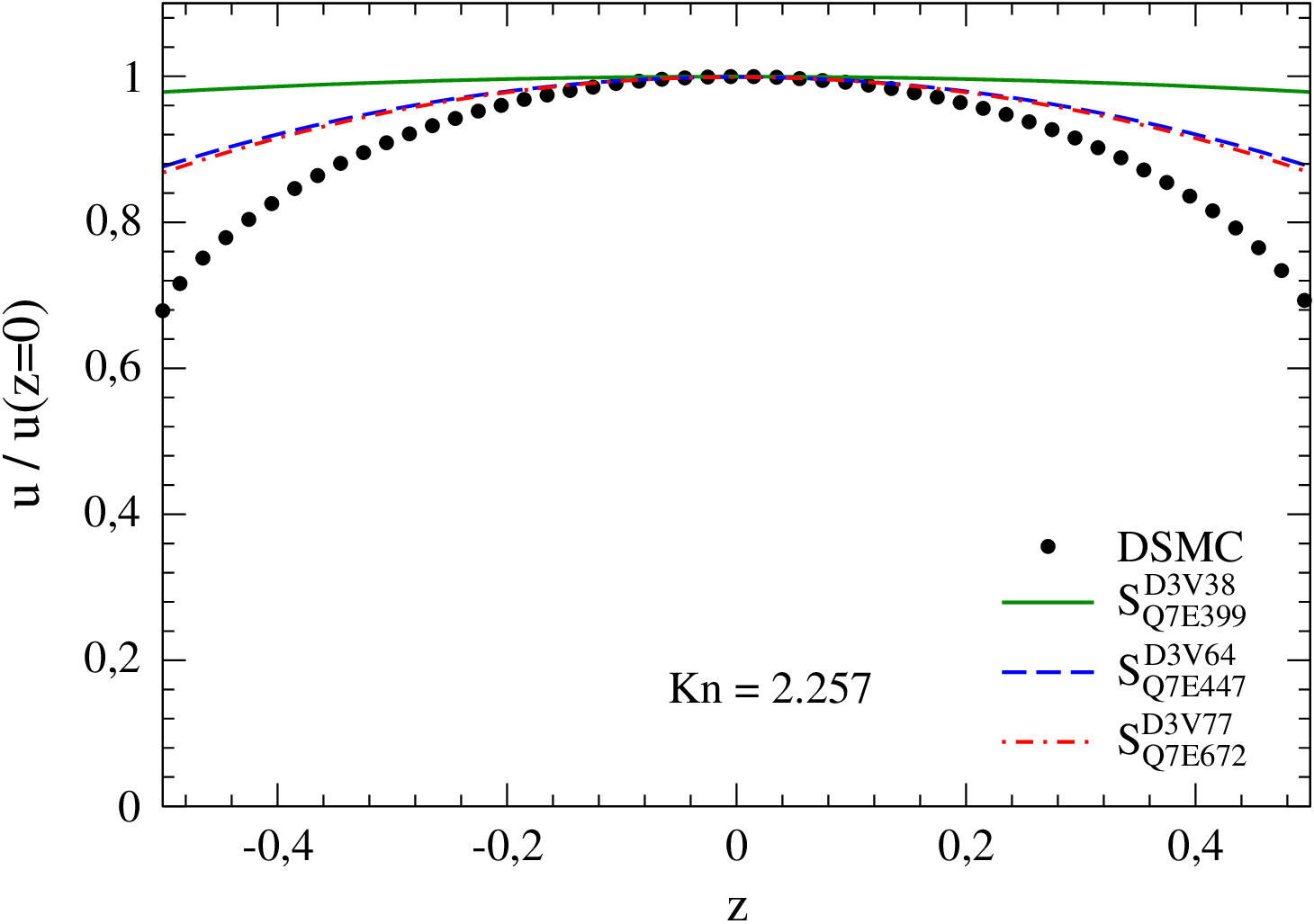,width=8.0cm,clip=,bb=4 29 706 522}
\caption{\label{fig18_2} (Color online) Streamwise velocity profile for  $\Kn=2.257$. }
\end{center}
\end{figure}

\subsection{Knudsen layer}
\label{sec:Knudsen}
An even stronger requirement for the DVMs than correct evaluation of mass flow and slip velocity is to yield the correct velocity profile beyond the Navier--Stokes flow regime. For this purpose, we define for each solution $u(z)$ a quadratic velocity profile by

\begin{figure}  [htb!]
\begin{center}
\epsfig{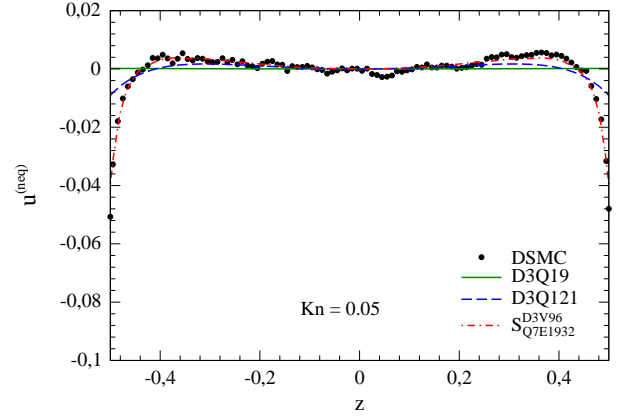}
\caption{\label{fig:KnudsenLayer005} (Color online) Non-equilibrium velocity profile for  $\Kn=0.05$. }
\end{center}
\end{figure}
\begin{figure}  [htb!]
\begin{center}
\epsfig{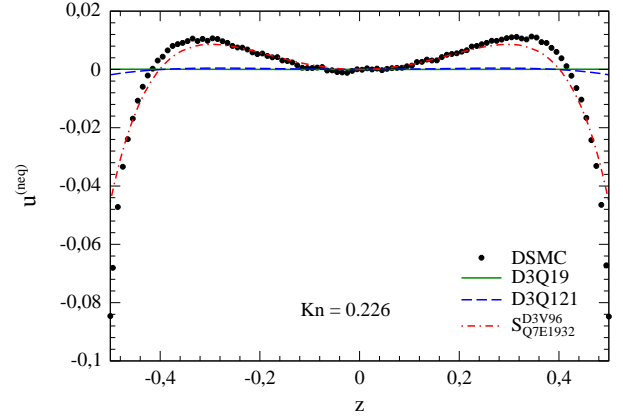}
\caption{\label{fig:KnudsenLayer0226} (Color online) Non-equilibrium velocity profile for  $\Kn=0.226$. }
\end{center}
\end{figure}
\begin{figure} [htb!]
\begin{center}
\epsfig{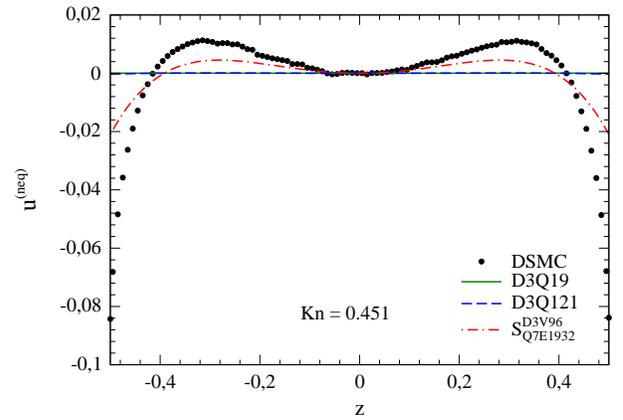}
\caption{\label{fig:KnudsenLayer0451} (Color online) Non-equilibrium velocity profile for  $\Kn=0.451$. }
\end{center}
\end{figure}

\begin{align}
\label{uNS}
&u^{(NS)}(z)=u(0)\nn\\
&\qquad\times\left(
1-3
\left(
1-\frac{1}{l_0}\int_{-l_0/2}^{l_0/2}dz'\: \frac{u(z')}{u(0)}
\right)
\left(\frac{z}{l_0/2}\right)^2\right)
\end{align}
which gives the same mass flow as $u(z)$. The non-equilibrium content of $u(z)$ is measured by its deviation from the quadratic profile \eqref{uNS},
\be
\label{uNEQ}
u^{(neq)}(z)=u(z)-u^{(NS)}(z)\; .
\ee

Using the DSMC data for $u(z)$, we find a reference function $u^{(neq)}(z)$ for 
each $\Kn$ number the DVMs compete with. As shown for $\Kn=0.05$ in Fig.\ \ref{fig:KnudsenLayer005}, 
the DSMC data show a small velocity defect at the vicinity of the 
wall---the Knudsen layer---such that $u^{(neq)}<0$. In turn, there is an exceed 
of non-equilibrium velocity in the bulk, $u^{(neq)}>0$, by definition of 
Eq.\ \eqref{uNS}. The standard model $D3Q19$ is not able to show a Knudsen layer, 
due to its low quadrature order, $Q=5$. It yields a strictly quadratic profile 
with $u^{(neq)}(z)=0$. The model $D3Q121$  with higher quadrature order, $Q=9$, 
shows some velocity defect, however, it does not quite match the DSMC result for 
its wall moments are inaccurate. On the other hand, the DVM $S_{Q7E1932}^{D3V96}$, 
which has both a high quadrature order and (almost) correct wall moments, shows 
excellent agreement with DSMC data in the Knudsen layer for $\Kn=0.05$.

This hierarchy of DVMs remains for higher $\Kn$ numbers. For $\Kn=0.226$, the 
Knudsen layer is more pronounced, see Fig.\ \ref{fig:KnudsenLayer0226}. While 
the models $D3Q19$ and $D3Q121$ are unsatisfactory, the model 
$S_{Q7E1932}^{D3V96}$ is able to describe the Knudsen layer. Increasing the 
$\Kn$ number further high-order flow regimes gain more importance, causing the 
DVMs presented here to cease to be valid for evaluating the non-equilibrium velocity profile. 
This is exemplified for $\Kn=0.451$ in Fig.\ \ref{fig:KnudsenLayer0451}.

\section{Conclusion \label{C}}
In this paper we presented high-order Lattice--Boltzmann (LB) models for solving the Boltzmann--BGK equation for finite $\Kn$ number flows. 

It was shown how to derive new discrete velocity models (DVMs) for any quadrature order 
using an efficient algorithm. The energy of a stencil was bounded from above in order to be 
able to define complete sets of minimal DVMs. These sets comprise more 
than $50000$ models with $7$th quadrature order in $D=3$ spatial dimensions. In future 
investigations, the collection of minimal discrete velocity models for high-order LB simulations can be systematically extended.

We analytically derived a theorem via the Chapman--Enskog expansion which enables us 
to identify the recovered flow regimes of any LB model, 
i.e.\ Navier--Stokes, Burnett, Super--Burnett, and so on. For isothermal flows, this theorem rigorously relates the velocity space discretization error of a velocity moment for low $\Ma$ values with the quadrature order $Q$ of a LB model. Thus, one can tell from the quadrature order $Q$ which flow regime is exactly recovered by the LB model 
for $\Ma\to 0$. In particular, it was shown for isothermal flows that even the standard models with $Q=5$ recover the Burnett momentum dynamics for $\Ma\to 0$. The $7$th order LB models we focused on here recover the momentum flux tensor of non-equilibrium gas flows up to the $4$th flow regime for $\Ma\to 0$.

Aside from non-equilibrium effects in the bulk, wall-bounded high $\Kn$ number flows require a LB model to correctly describe the gas-surface interaction. We observed that several high-order LB models show significant deviations from reference results because of their poor ability to recover the diffuse Maxwell boundary condition accurately. In order to characterize this capability we defined an analytical criterion for a DVM for the exact implementation of the \DMBC. It was shown how to generate in a systematic way high-order DVMs with the inherent ability to fulfill
the \DMBC. Alternatively, it is possible to arrive at discrete velocity models which obey the criterion for even velocity moments at the wall by simply using scattering stencils. The {\wai } $\Lambda_{\mathrm{index}}$ was put forward to label the velocity moments at the wall evaluated correctly by a discrete velocity model. The $7$th order model $\st{7}{96}{3}{1932}$, e.g., was found with a wall accuracy index $\Lambda_{\mathrm{index}}=6245$ and an almost exact representation of the wall moments up to the $7$th order.
The circumstance that models with even numbers of discrete velocities perform better than for odd numbers
can be due to the low net weight of erroneous wall-parallel (and zero) velocities in half-space quadratures. For an exact integration of particle velocities emitted by the wall, we derived a sufficient condition on the wall accuracy order $\Lambda$.

At finite $\Kn$, we compared our  Poiseuille flow LB results with those of DSMC. We found the agreement is strongly correlated with the wall criterion on the one hand and requires the use of high-order DVMs ($Q\geq 7$) on the other. Therefore, the correct evaluation of high $\Kn$ number Poiseuille flow requires both, a sufficiently high quadrature order and an exact representation of the \DMBC. Consequently, the model $\st{7}{96}{3}{1932}$ yields an excellent agreement with DSMC for both the mass flow and the slip velocity, using $\Kn$ numbers up to $\Kn\simeq 1$. In the velocity profiles one can clearly recognize that the discrete velocity model $\st{7}{96}{3}{1932}$ recovers the Knudsen layer in the vicinity of the wall for $\Kn\lesssim 0.3$, while standard discrete velocity models utterly fail. Thus we recommend using $\st{7}{96}{3}{1932}$ for finite $\Kn$ number flows.

For $\Kn\gtrsim0.3$, the results for the non-equilibrium velocity profile departs from the DSMC results, because the flow regimes are only recovered up to the $4$th Chapman--Enskog level and there are remaining errors in high-order wall moments. However, we expect that for LB models with a sufficiently high quadrature order and wall accuracy order, a more precise representation of the Knudsen layer phenomenon can be obtained. This will be part of our future investigations. Moreover, it remains to be seen whether the Knudsen layer turns out to be a nonperturbative phenomenon in the sense of the Chapman--Enskog expansion.
\begin{acknowledgments}
We are grateful to Tim Reis and Kurt Langfeld for giving helpful comments on the 
manuscript. We would also like to thank Alexander Stief for providing DSMC data.
\end{acknowledgments}
\appendix
\section{Three-dimensional DVMs}
\label{sec:appDVM}
Here we present DVMs generated by the algorithm shown in Sect.\ \ref{sec:dvm} and used in the introductory example and
the numerical test case in Sect.\ \ref{PF}. The stencil's symbol is explained in Eq.\ \eqref{Ssymbol}, $c$ is the lattice speed, $g$ 
counts the stencil groups $S_g$ generated by the symmetries of the lattice from a single 
velocity $\fxi^{(g)}$ yielding $V_g$ velocities. Each velocity in $S_g$ is weighted 
by $\overline{\tw}_g$, see Eq.\ \eqref{wdach}, when using the quadrature prescription \eqref{genQua}.
\begin{table}[htb!]
\centering
\begin{tabular}{|c|c|c|c|}
\hline
\multicolumn{2}{|c|}{$\boldsymbol{\st{5}{15}{3}{24}}$}&
\multicolumn{2}{c|}{$c = 1.2247448713915890$}\\
\hline
g & $\fxi^{(g)}/c$ & $V_g$ & $\overline{\tw}_g$ \\
\hline
 1 & \text{(0,0,0)} & 1 & \text{3.8888888888888889e-1} \\
 2 & \text{(2,0,0)} & 6 & \text{2.7777777777777778e-2} \\
 3 & \text{(1,1,-1)} & 8 & \text{5.5555555555555556e-2} \\
\hline
\end{tabular}
\caption{\label{tab:ExampleIntro} DVM used for the introductory example.}
\end{table}

\label{sec:3D}
\begin{table}[htb!]
\centering
\begin{tabular}{|c|c|c|c|}
\hline
\multicolumn{2}{|c|}{$\boldsymbol{\st{7}{38}{3}{399}}$}&
\multicolumn{2}{c|}{$c = 0.75000000000000000$}\\
\hline
g & $\fxi^{(g)}/c$ & $V_g$ & $\overline{\tw}_g$ \\
\hline
1 & \text{(1,0,0)} & 6 & \text{7.2239858906525573e-2} \\
 2 & \text{(4,0,0)} & 6 & \text{6.3648834019204390e-3} \\
 3 & \text{(2,2,0)} & 12 & \text{4.3895747599451303e-2} \\
 4 & \text{(6,0,0)} & 6 & \text{4.1805473904239336e-5} \\
 5 & \text{(4,4,4)} & 8 & \text{1.7146776406035665e-4} \\
\hline
\end{tabular}
\caption{\label{tab:dvmSet1Amin} DVM with minimal velocity count within the set ${\cal S}_1$.}
\end{table}
\begin{table}[htb!]
\centering
\begin{tabular}{|c|c|c|c|}
\hline
\multicolumn{2}{|c|}{$\boldsymbol{\st{7}{38}{3}{219}}$}&
\multicolumn{2}{c|}{$c = 0.86602540378443865$}\\
\hline
g & $\fxi^{(g)}/c$ & $V_g$ & $\overline{\tw}_g$ \\
\hline
 1 & \text{(1,0,0)} & 6 & \text{6.7724867724867725e-2} \\
 2 & \text{(2,0,0)} & 6 & \text{5.5555555555555556e-2} \\
 3 & \text{(2,2,2)} & 8 & \text{4.6296296296296296e-3} \\
 4 & \text{(2,2,0)} & 12 & \text{1.8518518518518519e-2} \\
 5 & \text{(6,0,0)} & 6 & \text{1.7636684303350970e-4} \\
\hline
\end{tabular}
\caption{\label{tab:dvmSet1Aminn} DVM with minimal velocity count within the set ${\cal S}_1$.}
\end{table}
\begin{table}[htb!]
\centering
\begin{tabular}{|c|c|c|c|}
\hline
\multicolumn{2}{|c|}{$\boldsymbol{\st{7}{59}{3}{408}}$}&
\multicolumn{2}{c|}{$c = 0.74685634388439233$}\\
\hline
g & $\fxi^{(g)}/c$ & $V_g$ & $\overline{\tw}_g$ \\
\hline
 1 & \text{(0,0,0)} & 1 & \text{2.0080700829205231e-2} \\
 2 & \text{(1,1,1)} & 8 & \text{8.9344539381631413e-2} \\
 3 & \text{(4,0,0)} & 6 & \text{1.8287317703258027e-3} \\
 4 & \text{(3,3,3)} & 8 & \text{5.2273139486813077e-4} \\
 5 & \text{(3,3,0)} & 12 & \text{2.3272910797607214e-3} \\
 6 & \text{(3,1,1)} & 24 & \text{9.2533853908214560e-3} \\
\hline
\end{tabular}
\caption{\label{tab:dvmSet1Kn} Selection of DVMs within the set ${\cal S}_1$. See Sect.\ \ref{sec:Poi} for a discussion.}
\end{table}
\begin{table}[htb!]
\centering
\begin{tabular}{|c|c|c|c|}
\hline
\multicolumn{2}{|c|}{$\boldsymbol{\st{7}{64}{3}{447}}$}&
\multicolumn{2}{c|}{$c = 0.69965342816864754$}\\
\hline
g & $\fxi^{(g)}/c$ & $V_g$ & $\overline{\tw}_g$ \\
\hline
 1 & \text{(2,0,0)} & 6 & \text{5.9646397884737016e-3} \\
 2 & \text{(1,1,1)} & 8 & \text{8.0827437008387392e-2} \\
 3 & \text{(5,0,0)} & 6 & \text{1.1345266793939999e-3} \\
 4 & \text{(3,3,3)} & 8 & \text{9.5680047874015889e-4} \\
 5 & \text{(3,3,0)} & 12 & \text{3.9787631334632013e-3} \\
 6 & \text{(3,1,1)} & 24 & \text{1.0641080987258957e-2} \\
\hline
\end{tabular}
\caption{\label{tab:dvmSet1Kn2} Selection of DVMs within the set ${\cal S}_1$. See Sect.\ \ref{sec:Poi} for a discussion. }
\end{table}
\begin{table}[htb!]
\centering
\begin{tabular}{|c|c|c|c|}
\hline
\multicolumn{2}{|c|}{$\boldsymbol{\st{7}{66}{3}{489}}$}&
\multicolumn{2}{c|}{$c = 0.91181414856781201$}\\
\hline
g & $\fxi^{(g)}/c$ & $V_g$ & $\overline{\tw}_g$ \\
\hline
 1 & \text{(1,0,0)} & 6 & \text{8.5874589554625067e-2} \\
 2 & \text{(3,0,0)} & 6 & \text{7.7875559017476937e-3} \\
 3 & \text{(2,2,0)} & 12 & \text{1.1890500729870710e-4} \\
 4 & \text{(2,1,1)} & 24 & \text{1.8125533818334885e-2} \\
 5 & \text{(7,0,0)} & 6 & \text{2.7809545430205753e-6} \\
 6 & \text{(4,4,0)} & 12 & \text{1.3089748390696516e-4} \\
\hline
\end{tabular}
\caption{\label{tab:dvmSet1Kn4} Selection of DVMs within the set ${\cal S}_1$. See Sect.\ \ref{sec:Poi} for a discussion.}
\end{table}
\begin{table}[htb!]
\centering
\begin{tabular}{|c|c|c|c|}
\hline
\multicolumn{2}{|c|}{$\boldsymbol{\st{9}{79}{3}{471}}$}&
\multicolumn{2}{c|}{$c = 1.0000000000000000$}\\
\hline
g & $\fxi^{(g)}/c$ & $V_g$ & $\overline{\tw}_g$ \\
\hline
 1 & \text{(0,0,0)} & 1 & \text{1.0570987654320988e-1} \\
 2 & \text{(1,0,0)} & 6 & \text{3.8095238095238095e-2} \\
 3 & \text{(2,0,0)} & 6 & \text{2.8645833333333333e-2} \\
 4 & \text{(1,1,1)} & 8 & \text{4.9479166666666667e-2} \\
 5 & \text{(2,2,2)} & 8 & \text{5.2083333333333333e-4} \\
 6 & \text{(2,2,0)} & 12 & \text{5.2083333333333333e-3} \\
 7 & \text{(6,0,0)} & 6 & \text{2.7557319223985891e-6} \\
 8 & \text{(3,3,3)} & 8 & \text{9.6450617283950617e-6} \\
 9 & \text{(3,1,1)} & 24 & \text{1.3020833333333333e-3} \\
\hline
\end{tabular}
\caption{\label{tab:dvmSet2Amin} DVM with minimal velocity count within the set ${\cal S}_2$. For a description of 
the symbols, see Tab.\ \ref{tab:dvmSet1Amin}.}
\end{table}
\begin{table}[htb!]
\centering
\begin{tabular}{|c|c|c|c|}
\hline
\multicolumn{2}{|c|}{$\boldsymbol{\st{9}{121}{3}{594}}$}&
\multicolumn{2}{c|}{$c = 1.1969797703930744$}\\
\hline
g & $\fxi^{(g)}/c$ & $V_g$ & $\overline{\tw}_g$ \\
\hline
 1 & \text{(0,0,0)} & 1 & \text{3.0591622029486006e-2} \\
 2 & \text{(0,0,-1)} & 6 & \text{9.8515951037263392e-2} \\
 3 & \text{(1,1,-1)} & 8 & \text{2.7525005325638124e-2} \\
 4 & \text{(0,0,-3)} & 6 & \text{3.2474752708807381e-4} \\
 5 & \text{(2,2,-2)} & 8 & \text{1.8102175157637424e-4} \\
 6 & \text{(2,0,-2)} & 12 & \text{4.2818359368108407e-4} \\
 7 & \text{(1,0,-2)} & 24 & \text{6.1110233668334243e-3} \\
 8 & \text{(3,3,-3)} & 8 & \text{6.9287508963860285e-7} \\
 9 & \text{(1,1,-3)} & 24 & \text{1.0683400245939109e-4} \\
 10 & \text{(2,0,-3)} & 24 & \text{1.4318624115480294e-5} \\
\hline
\end{tabular}
\caption{\label{tab:dvmSet2standard} Common DVM ($D3Q121$ \cite{Meng}) within the set ${\cal S}_2$. For a description of 
the symbols, see Tab.\ \ref{tab:dvmSet1Amin}.}
\end{table}
\begin{table}[htb!]
\centering
\begin{tabular}{|c|c|c|c|}
\hline
\multicolumn{2}{|c|}{$\boldsymbol{\st{7}{77}{3}{672}}$}&
\multicolumn{2}{c|}{$c = 0.62590566441325041$}\\
\hline
g & $\fxi^{(g)}/c$ & $V_g$ & $\overline{\tw}_g$ \\
\hline
 1 & \text{(2,0,0)} & 6 & \text{6.5178619315224175e-3} \\
 2 & \text{(1,1,1)} & 8 & \text{5.8638132347907334e-2} \\
 3 & \text{(6,0,0)} & 6 & \text{1.0066312290789269e-3} \\
 4 & \text{(3,3,3)} & 8 & \text{2.1962967289288607e-3} \\
 5 & \text{(3,3,0)} & 12 & \text{4.8042518606023833e-3} \\
 6 & \text{(3,1,1)} & 24 & \text{1.5528601406828448e-2} \\
 7 & \text{(0,0,0)} & 1 & \text{2.9560614762917757e-2} \\
 8 & \text{(4,4,0)} & 12 & \text{6.8996146397277359e-4} \\
\hline
\end{tabular}
\caption{\label{tab:dvmD3Q7Set3} Selection of DVMs within the set ${\cal S}_3$. See Sect.\ \ref{sec:WallModels} for a discussion. }
\end{table}
\begin{table}[htb!]
\centering
\begin{tabular}{|c|c|c|c|}
\hline
\multicolumn{2}{|c|}{$\boldsymbol{\st{7}{107}{3}{1023}}$}&
\multicolumn{2}{c|}{$c = 0.61887631323925978$}\\
\hline
g & $\fxi^{(g)}/c$ & $V_g$ & $\overline{\tw}_g$ \\
\hline
 1 & \text{(1,1,1)} & 8 & \text{5.4242093013777495e-2} \\
 2 & \text{(2,1,0)} & 24 & \text{2.4637212114133877e-3} \\
 3 & \text{(5,0,0)} & 6 & \text{1.5469959015615954e-3} \\
 4 & \text{(3,3,3)} & 8 & \text{1.9425678925800591e-3} \\
 5 & \text{(3,3,0)} & 12 & \text{7.2337497640759286e-3} \\
 6 & \text{(3,1,1)} & 24 & \text{1.4384326790070621e-2} \\
 7 & \text{(0,0,0)} & 1 & \text{4.4998866720663948e-2} \\
 8 & \text{(6,2,2)} & 24 & \text{2.1182172560744541e-4} \\
\hline
\end{tabular}
\caption{\label{tab:dvmD3Q7Set3b} Selection of DVMs within the set ${\cal S}_3$. See Sect.\ \ref{sec:WallModels} for a discussion. }
\end{table}
\begin{table}[htb!]
\renewcommand{\arraystretch}{1.0}
\centering
\begin{tabular}{|c|c|c|c|}
\hline
\multicolumn{2}{|c|}{$\boldsymbol{\st{7}{96}{3}{1932}}$}&
\multicolumn{2}{c|}{$c = 0.37787639086813054$}\\
\hline
g & $\fxi^{(g)}/c$ & $V_g$ & $\overline{\tw}_g$ \\
\hline
 1 & \text{(1,1,1)} & 8 & \text{1.2655649299880090e-3} \\
 2 & \text{(3,3,3)} & 8 & \text{2.0050978770655310e-2} \\
 3 & \text{(3,1,1)} & 24 & \text{2.7543347614356814e-2} \\
 4 & \text{(4,4,4)} & 8 & \text{4.9712543563172566e-3} \\
 5 & \text{(7,1,1)} & 24 & \text{3.6439016726158895e-3} \\
 6 & \text{(6,6,1)} & 24 & \text{1.7168180273737716e-3} \\
\hline
\end{tabular}
\caption{\label{tab:dvmD3Q7W0} Selection of DVMs within the set ${\cal S}_4$. See Sect.\ \ref{sec:WallModels} for a discussion. }
\end{table}
\begin{table}[htb!]
\renewcommand{\arraystretch}{1.0}
\centering
\begin{tabular}{|c|c|c|c|}
\hline
\multicolumn{2}{|c|}{$\boldsymbol{\st{7}{112}{3}{1764}}$}&
\multicolumn{2}{c|}{$c = 0.40531852273291520$}\\
\hline
g & $\fxi^{(g)}/c$ & $V_g$ & $\overline{\tw}_g$ \\
\hline
 1 & \text{(1,1,1)} & 8 & \text{3.3503407500643648e-3} \\
 2 & \text{(3,1,1)} & 24 & \text{2.8894128958152456e-2} \\
 3 & \text{(4,4,4)} & 8 & \text{4.5930345162087793e-3} \\
 4 & \text{(3,2,2)} & 24 & \text{4.4163148398082762e-3} \\
 5 & \text{(7,1,1)} & 24 & \text{2.3237070220062610e-3} \\
 6 & \text{(5,5,1)} & 24 & \text{3.3847240912752922e-3} \\
\hline
\end{tabular}
\caption{\label{tab:dvmD3Q7W0b} Selection of DVMs within the set ${\cal S}_4$. See Sect.\ \ref{sec:WallModels} for a discussion. }
\end{table}

\section{Two-dimensional DVMs}

In this Appendix, we present {\minDVMs } for $D=2$. As shown in Tab.\ \ref{tab:dvmSummary} above, there are three sets of DVMs concerned with $D=2$. 
The $y$ component is absent.
For all but the {\wai } $\Lambda_{\mathrm{index}}$, the definitions are the same as for $D=3$. Since there are no $y$ components of wall moments, we have to set $m_y=0$ in Eq.\ \eqref{nontriv}. The {\wai } is then defined as before, see Eq.\ \eqref{defbin}, but there are only $L=14$ components in the vector
\be
\label{defT2D}
{\bf T}^\alpha&=&(
1,\:
\xi_{\alpha z},\:
\xi_{\alpha x}^2,\:
\xi_{\alpha z}^3,\:
\xi_{\alpha z}\xi_{\alpha x}^2,\:
\xi_{\alpha x}^4,\:
\nn\\&&\:
\xi_{\alpha z}^5,\:
\xi_{\alpha z}^3\xi_{\alpha x}^2,\:
\xi_{\alpha z}\xi_{\alpha x}^4,\:
\xi_{\alpha x}^6,\:
\nn\\&&\:
\xi_{\alpha z}^7,\:
\xi_{\alpha z}^5\xi_{\alpha x}^2,\:
\xi_{\alpha z}^3\xi_{\alpha x}^4,
\xi_{\alpha z}\xi_{\alpha x}^6
)\; ,
\ee
whereas $b_k$ are the components of the vector
\be
\label{defb2D}
{\bf b}&=&
(
W,\:
W_{z},\: 
W_{xx},\: 
W_{zzz},\: 
W_{zxx},\: 
W_{xxxx},\: 
W_{zzzzz},\: 
\nn\\&&\:
W_{zzzxx},\: 
W_{zxxxx},\: 
W_{xxxxxx},\: 
W_{zzzzzzz},\: 
\nn\\&&\:
W_{zzzzzxx},\: 
W_{zzzxxxx},\: 
W_{zxxxxxx}
)
\; ,
\ee
cf.\ Eqs.\ \eqref{defT} and \eqref{defb}.

The set ${\cal S}_5$ comprises all $1188$  {\minDVMs } with quadrature order $Q=7$ in the energy sphere given by $E\leq 250$. These are discrete velocity models for bulk flow, i.e.\ the {\wai } $\Lambda_{\mathrm{index}}$ is zero. The DVM with lowest velocity count $V_{min}=16$ is $\st{7}{16}{2}{58}$ and it is shown in Tab.\ \ref{tab:dvmSet5Amin}. Increasing the quadrature order to $Q=9$, we find $592$ DVMs in the set ${\cal S}_6$ where $E_{max}=300$. The DVM with the lowest velocity count in ${\cal S}_6$ is $\st{7}{32}{2}{944}$, see Tab.\ \ref{tab:dvmSet6Amin}, and it yields $V=33$. Finally, we identified $21952$ DVMs with quadrature order $Q=7$ which have scattering stencils, i.e.\ their {\wai } is $\Lambda_{\mathrm{index}} = 549$. The stencil energy is bounded for these models by $E_{max}=1000$ to define the set ${\cal S}_7$. The DVM with the lowest velocity count gives $V_{min}=20$. Optimizing for both $\sigma_{zx}$ and $\sigma_\Sigma$ yields the DVM $\st{7}{32}{2}{944}$ shown in Tab.\ \ref{tab:dvmSet7opt}. The latter is expected to give back excellent results for planar Poiseuille flow.

\label{sec:2D}
\begin{table}[htb!]
\centering
\begin{tabular}{|c|c|c|c|}
\hline
\multicolumn{2}{|c|}{$\boldsymbol{\st{7}{16}{2}{58}}$}&
\multicolumn{2}{c|}{$c = 0.86602540378443865$}\\
\hline
g & $\fxi^{(g)}/c$ & $V_g$ & $\overline{\tw}_g$ \\
\hline
 1 & \text{(1,0)} & 4 & \text{1.5802469135802469e-1} \\
 2 & \text{(2,0)} & 4 & \text{6.1728395061728395e-2} \\
 3 & \text{(2,2)} & 4 & \text{2.7777777777777778e-2} \\
 4 & \text{(4,0)} & 4 & \text{2.4691358024691358e-3} \\
\hline
\end{tabular}
\caption{\label{tab:dvmSet5Amin} DVM with the lowest velocity count $V_{min}=16$ in the set ${\cal S}_5$ where $D=2$, $Q=7$, and $E\leq 250$ (see Tab.\ \ref{tab:dvmSummary}). }
\end{table}
\begin{table}[htb!]
\centering
\begin{tabular}{|c|c|c|c|}
\hline
\multicolumn{2}{|c|}{$\boldsymbol{\st{9}{33}{2}{132}}$}&
\multicolumn{2}{c|}{$c = 1.1587791906520175$}\\
\hline
g & $\fxi^{(g)}/c$ & $V_g$ & $\overline{\tw}_g$ \\
\hline
 1 & \text{(0,0)} & 1 & \text{1.6198651186147246e-1} \\
 2 & \text{(1,0)} & 4 & \text{1.4320396528198750e-1} \\
 3 & \text{(1,1)} & 4 & \text{3.3883996404301766e-2} \\
 4 & \text{(2,0)} & 4 & \text{5.5611157082744134e-3} \\
 5 & \text{(2,2)} & 4 & \text{8.4479885070276616e-5} \\
 6 & \text{(3,0)} & 4 & \text{1.1325437650467775e-3} \\
 7 & \text{(2,1)} & 8 & \text{1.2816907733721003e-2} \\
 8 & \text{(4,4)} & 4 & \text{3.4555225091487045e-6} \\
\hline
\end{tabular}
\caption{\label{tab:dvmSet6Amin} DVM with the lowest velocity count $V_{min}=33$ in the set ${\cal S}_6$ where $D=2$, $Q=9$, and $E\leq 300$ (see Tab.\ \ref{tab:dvmSummary}). }
\end{table}
\begin{table}[htb!]
\centering
\begin{tabular}{|c|c|c|c|}
\hline
\multicolumn{2}{|c|}{$\boldsymbol{\st{7}{32}{2}{944}}$}&
\multicolumn{2}{c|}{$c = 0.34040702226615838$}\\
\hline
g & $\fxi^{(g)}/c$ & $V_g$ & $\overline{\tw}_g$ \\
\hline
 1 & \text{(1,1)} & 4 & \text{8.4201053650845727e-2} \\
 2 & \text{(2,2)} & 4 & \text{5.0708714918479963e-2} \\
 3 & \text{(5,1)} & 8 & \text{4.4541157350549509e-2} \\
 4 & \text{(6,4)} & 8 & \text{1.2646044225450934e-2} \\
 5 & \text{(12,3)} & 8 & \text{3.5791413933671213e-4} \\
\hline
\end{tabular}
\caption{\label{tab:dvmSet7opt} DVM with optimal values for the wall errors in the set ${\cal S}_7$ where $D=2$, $Q=7$, $E\leq 1000$, and $\sigma_{i_1\dots i_{2n}}=0$. (see Tab.\ \ref{tab:dvmSummary}). }
\end{table}

\section{Error estimate of equilibrium moments \label{ap1}}
For a detailed analysis of the quadrature error of an equilibrium moment $\Delta M^{(0)}_{i_1 \dots i_k}$ 
we distinguish four different cases.
\subsection*{{\bf Case 1: $k \le N$ and $k+N \le Q$}}
The Hermite order $N$ is high enough to capture all contributions of the Hermite 
polynomials $\mH^{(k)}_{i_1\dots i_k}, \mH^{(k-2)}_{i_1\dots i_{k-2}}, \dots$ and 
the quadrature order guarantees an exact evaluation of all terms in Eq.\ (\ref{ce.29}). 
Thus the moment $M^{(0)}_{i_1 \dots i_k} $ is exactly recovered and 
\be
\Delta M^{(0)}_{i_1 \dots i_k} = 0\; .
\ee
\subsection*{{\bf  Case 2: $k \le N$ and $k+N > Q$}}
Although the Hermite order is high enough, the quadrature 
order $Q=n_0+k$, with $n_0<N$, is not high enough to evaluate 
$M^{(0)}_{i_1 \dots i_k}$ exactly. For low $\Ma$ values, the 
leading term of the quadrature error in Eq.\ \eqref{ce.29} is the 
smallest value of $n$ exceeding $n_0$, due to the relation \eqref{ce.27}. 
This term is the one with $n=n_0+1$ in Eq.\ \eqref{ce.29} and we can write
\vspace{1mm}
\begin{widetext}
\begin{align}
\label{ce.30}
\Delta M^{(0)}_{i_1 \dots i_k} &= \frac{1}{(Q-k+1)!} a^{(0)}_{i_1 \dots i_{Q-k+1}} 
\sum_{\alpha} \tw_{\alpha} \mH^{(Q-k+1)}_{i_1 \dots i_{Q-k+1}} (\fxi_{\alpha})
\mH^{(k)}_{i_1 \dots i_k} (\fxi_{\alpha}) \quad
+ \textit{subleading terms} \nonumber\\
&=\mathcal{O} \left( \Ma^{Q-k+1} \right) \; ,
\end{align}
\end{widetext}
where the subleading terms contain errors with a higher $\Ma$ power compared 
to the first term.
Obviously, the quadrature error $\Delta M^{(0)}_{i_1 \dots i_k} $ is finite for low $\Ma$
values, see Eq.\ (\ref{ce.29}). Consequently we obtain 
$\Delta M^{(0)}_{i_1 \dots i_k} = \mathcal{O} \left( \Ma^0 \right)$ for $Q-k+1<0$.

\subsection*{{\bf Case 3: $k > N$ and $k+N \le Q$}}
The Hermite order $N$ is not high enough to capture the moment $M^{(0)}_{i_1 \dots i_k}$
completely whereas the quadrature order guarantees the exact evaluation of all terms in 
Eq.\ (\ref{ce.29}). Therefore errors are produced by the absence of terms in the equilibrium 
function which are beyond the Hermite order $N$. The leading term in the quadrature 
error is thus $\sim a^{(0)}_{i_1 \dots i_{N+1}}$ and with Eq.\ \eqref{ce.27} we get
\begin{align}
\label{ce.32}
\Delta M^{(0)}_{i_1 \dots i_k} = \mathcal{O} \left( \Ma^{N+1} \right) .
\end{align}

\subsection*{{\bf Case 4: $k > N$ and $k+N > Q$}}
Neither the Hermite order $N$ is high enough to recover $M^{(0)}_{i_1 \dots i_k}$ nor
the quadrature order $Q$ guarantees an exact evaluation. The relevant quadrature error
in the sense of the $\Ma$ expansion of the equilibrium function $f^{(0)}$ is determined analogously to case 2 by
\begin{align}
\label{ce.33}
\Delta M^{(0)}_{i_1 \dots i_k} = \mathcal{O} \left( \Ma^{Q-k+1} \right) 
\end{align}
for $Q-k+1\ge0$ and $\Delta M^{(0)}_{i_1 \dots i_k} = \mathcal{O} \left( \Ma^0 \right)$ otherwise.

These cases can be summarized into a more compact relation
\begin{widetext}
\begin{align}
\label{ce.34b}
   \Delta M^{(0)}_{i_1 \dots i_k} =
   \begin{cases}
     0                                 & \text{for   }  k+N \le Q \text{   and   } k \le N \\
     \mathcal{O} \left( \Ma^{N+1} \right) & \text{for   }  k+N \le Q \text{   and   } k > N \\
     \mathcal{O} \left( \Ma^{Q-k+1} \right) & \text{for   } k-1 \le Q < k+N \\
     \mathcal{O} \left( \Ma^0 \right) & \text{for   } Q < k-1 \; .
   \end{cases} 
\end{align}
\end{widetext}

\section{Proof of the quadrature error theorem \label{ap2}}
The theorem in Sect.\ \ref{LBH.3} is proved by induction. Based on Eq.\ (\ref{ce.21b}) the 
truncation error of $M^{(1)}_{i_1 \dots i_k}$ is determined by
\begin{align}
\label{ce.52}
\Delta M^{(1)}_{i_1 \dots i_k} &= - \tau \left[  \Delta \partial^{(0)}_t  M^{(0)}_{i_1 \dots i_k}
+ \partial_{j_1} \Delta M^{(0)}_{i_1 \dots i_k j_1} \right]  .
\end{align}
Because of Eq.\ (\ref{ce.35})
the relevant error with respect to the $\Ma$ power is produced by the highest 
equilibrium moment which is contained in the second term 
on the right-hand side. The first term does not contribute to the dominant truncation error, 
because it consists of a lower equilibrium moment and the multiple-scale derivative $\partial^{(0)}_t$ 
does not change the $\Ma$ power of this error term. Consequently we find
\begin{align}
\label{ce.53}
\Delta M^{(1)}_{i_1 \dots i_k} &= - \tau  \partial_{j_1} \Delta M^{(0)}_{i_1 \dots i_k j_1}
+ \textit{subleading terms}
\end{align}
and with Eq.\ (\ref{ce.35}) the estimate
\begin{widetext}
\begin{align}
\label{ce.54}
   \Delta M^{(1)}_{i_1 \dots i_k} =
   \begin{cases}
     0                                 & \text{for   }  2(k+1) + 1 \le Q  \\
     \mathcal{O} \left( \Ma^{Q-k} \right) & \text{for   }   k \le Q < 2(k+1) + 1 \\
     \mathcal{O} \left( \Ma^0 \right)     & \text{for   }   Q < k 
   \end{cases} 
\text{  .}
\end{align}
For the next CE moment we obtain from Eq.\ (\ref{ce.21b})
\begin{align}
\label{ce.55}
M^{(2)}_{i_1 \dots i_k} &= - \tau \left[  \partial^0_t  M^{(1)}_{i_1 \dots i_k} 
- \frac{1}{\rho} \sum_{q} \frac{\partial M^{(0)}_{i_1 \dots i_k}}{\partial (\partial_{j_1 \dots j_q}^{(q)} u_s)} \partial_{j_1 \dots j_q}^{(q)}
\left( \partial_r \Pi^{(1)}_{rs} \right) 
+ \partial_j M^{(1)}_{i_1 \dots i_k j} \right] .
\end{align}
\end{widetext}
The error of the first term on the right-hand side is given by Eq.\ (\ref{ce.54})
where the time derivative $\partial_t^{(0)}$ does not affect the $\Ma$ power
\begin{align}
\label{ce.56}
&\Delta \partial_t^{(0)} M^{(1)}_{i_1 \dots i_k} =
   \partial_t^{(0)} \Delta M^{(1)}_{i_1 \dots i_k} \nn\\&\quad=
   \begin{cases}
     0                                 & \text{for   }  2(k+1) + 1 \le Q  \\
     \mathcal{O} \left( \Ma^{Q-k} \right) & \text{for   }   k \le Q < 2(k+1) + 1 \\
     \mathcal{O} \left( \Ma^0 \right)     & \text{for   }   Q < k 
   \end{cases} 
\text{  .}
\end{align}
With respect to Eq.\ (\ref{ce.35}) we get for the error of the next term
\begin{align}
\label{ce.57}
   \Delta \frac{\partial M^{(0)}_{i_1 \dots i_k}}{\partial (\partial_{j_1 \dots j_q}^{(q)} u_s)} =
   \begin{cases}
      0                                 & \text{for   }  2k + 1 \le Q \\
      \mathcal{O} \left( \Ma^{Q-k} \right) & \text{for   }  k \le Q < 2k+1 \\
      \mathcal{O} \left( \Ma^0     \right) & \text{for   }  Q< k
   \end{cases} 
\end{align}
where we have to take into account that the $\Ma$ power of the error 
term is decreased by one by the derivative $\partial / \partial ( \partial_{j_1 \dots j_q}^{(q)} u_s)$.
Based on Eq.\ (\ref{ce.54}) we obtain
\begin{align}
\label{ce.58}
   \Delta \Pi^{(1)}_{r s} =
   \begin{cases}
     0                                 & \text{for   }  7 \le Q  \\
     \mathcal{O} \left( \Ma^{Q-2} \right) & \text{for   } 2 \le Q < 7 \\
     \mathcal{O} \left( \Ma^0     \right) & \text{for   } Q < 2 
   \end{cases} 
\end{align}
and for the last term in Eq.\ (\ref{ce.55})
\begin{widetext}
\begin{align}
\label{ce.59}
   \Delta M^{(1)}_{i_1 \dots i_k j} =
   \begin{cases}
     0                                   & \text{for   }  2(k+2) + 1 \le Q  \\
     \mathcal{O} \left( \Ma^{Q-k-1} \right) & \text{for   }   k+1 \le Q < 2(k+2) + 1 \\
     \mathcal{O} \left( \Ma^0 \right) & \text{for   }   Q < k+1 
   \end{cases} 
\text{  .}
\end{align}
\end{widetext}
Due to $k \ge 2$ the error estimate of all terms in Eq.\ (\ref{ce.55}) implies 
that the relevant quadrature error of $M^{(2)}_{i_1 \dots i_k}$ is contained in the
last term of Eq.\ (\ref{ce.55}) and thus we get
\begin{align}
&\Delta M^{(2)}_{i_1 \dots i_k} = - \tau  \partial_{j_2} \Delta M^{(1)}_{i_1 \dots i_k j_2}
+ \textit{subleading terms} \nonumber \\
\label{ce.60}
&\quad= (- \tau)^2  \partial_{j_1} \partial_{j_2} \Delta M^{(0)}_{i_1 \dots i_k j_1 j_2}
+ \textit{subleading terms} .
\end{align}
In the last step of the proof we have to show the relation (\ref{ce.50}) for an integer $n$
assuming that the relation is valid for $n-1, n-2, \dots , 1$.
We analyze the error of all terms on the right-hand side of Eq.\ (\ref{ce.21b}).
\begin{widetext}
\begin{align}
\label{ce.21bdelta}
&\Delta M^{(n)}_{i_1 \dots i_k} = - \tau \left[ \Delta  \partial^{(0)}_t  M^{(n-1)}_{i_1 \dots i_k} 
- \Delta \left\{ \frac{1}{\rho} \sum_{m=1}^{n-1} \sum_{q} \frac{\partial M^{(n-m-1)}_{i_1 \dots i_k}}{\partial (\partial_{j_1 \dots j_q}^{(q)} u_s)}
\partial_{j_1 \dots j_q}^{(q)} \left( \partial_r \Pi^{(m)}_{rs} \right) \right\} + \partial_j \Delta  M^{(n-1)}_{i_1 \dots i_k j} \right] .
\end{align}
Using Eq.\ (\ref{ce.50}) for $n-1$ we find 
\begin{align}
\Delta \partial_t^{(0)} M^{(n-1)}_{i_1 \dots i_k} 
&= \partial_t^{(0)} \Delta M^{(n-1)}_{i_1 \dots i_k} 
= (- \tau)^{n-1}  \partial_t^{(0)}
\partial_{j_1} \dots \partial_{j_{n-1}} \Delta M^{(0)}_{i_1 \dots i_k j_1 \dots j_{n-1}}
+ \textit{subleading terms} \nonumber \\
\label{ce.61}
   &=
   \begin{cases}
     0                                 & \text{for   }  2(k+n-1) + 1 \le Q \\
     \mathcal{O} \left( \Ma^{Q-k-n+2} \right) & \text{for   } k+n-2 \le Q < 2(k+n-1) + 1 \\
     \mathcal{O} \left( \Ma^0 \right) & \text{for   } Q < k+n-2
   \end{cases} 
\end{align}
where $\partial_t^{(0)}$ does not affect the $\Ma$ power of the error term.
Furthermore, we can estimate the error of 
$\partial M^{(n-m-1)}_{i_1 \dots i_k} / \partial (\partial_{j_1 \dots j_q}^{(q)} u_s)$
for $m=1, \dots , n-1$ by using Eq.\ (\ref{ce.50}) for $n-m-1$ in combination with 
Eq.\ (\ref{ce.35})
\begin{align}
\label{ce.62}
   \Delta \frac{\partial M^{(n-m-1)}_{i_1 \dots i_k}}{\partial (\partial_{j_1 \dots j_q}^{(q)} u_s)} =
   \begin{cases}
     0                                 & \text{for   }  2(k+n-m-1) + 1 \le Q \\
     \mathcal{O} \left( \Ma^{Q-k-n+m+1} \right) & \text{for   } k+n-m-1 \le Q < 2(k+n-m-1) + 1 \\
     \mathcal{O} \left( \Ma^0           \right) & \text{for   } Q < k+n-m-1 
   \end{cases} 
\end{align}
where the derivatives $\partial / \partial (\partial_{j_1 \dots j_q}^{(q)} u_s)$ reduce the $\Ma$ power of 
the error term by one.
The error of the quantity $\Pi^{(m)}_{rs}$ for $m=1, \dots, n-1$ can be determined 
by applying the theorem (\ref{ce.50}) for $m$
\begin{align}
\Delta \Pi^{(m)}_{rs} &= (- \tau)^{m+2}  
\partial_{j_1} \dots \partial_{j_m} \Delta M^{(0)}_{r s j_1 \dots j_m}
+ \textit{subleading terms} \nonumber \\
\label{ce.63}
   &=
   \begin{cases}
     0                                 & \text{for   }  2(m+2) + 1 \le Q  \\
     \mathcal{O} \left( \Ma^{Q-m-1} \right) & \text{for   } m+1 \le Q < 2(m+2) + 1 \\
     \mathcal{O} \left( \Ma^0 \right) & \text{for   }      Q < m+1 
   \end{cases} 
\end{align}
where $m=1, \dots , n-1$. The error of the last term of Eq.\ (\ref{ce.21b}) is 
given by 
\begin{align}
&\Delta  M^{(n-1)}_{i_1 \dots i_k j} = (- \tau)^{n-1}  
\partial_{j_1} \dots \partial_{j_{n-1}} \Delta M^{(0)}_{i_1 \dots i_k j j_1 \dots j_{n-1}}
+ \textit{subleading terms} \nonumber \\
\label{ce.64}
   &=
   \begin{cases}
     0                                 & \text{for   }     2(k+n) + 1 \le Q \\
     \mathcal{O} \left( \Ma^{Q-k-n+1} \right) & \text{for   } k+n-1 \le Q < 2(k+n) + 1 \\
     \mathcal{O} \left( \Ma^0         \right) & \text{for   } Q < k+n-1 
   \end{cases} 
\end{align}
\end{widetext}
where we have used Eq.\ (\ref{ce.35}) and applied Eq.\ (\ref{ce.50}) for $n-1$.
Because of $k \ge 2$ and $m=1, \dots, n-1$ the estimate of all errors 
occurring in Eq.\ (\ref{ce.21bdelta}) implies that the last term on the right-hand 
side of Eq.\ (\ref{ce.21bdelta}) contains the relevant error with respect to the 
$\Ma$ power. This yields
\begin{align}
&\Delta M^{(n)}_{i_1 \dots i_k} = - \tau  \partial_{j_n} \Delta M^{(n-1)}_{i_1 \dots i_k j_n}
+ \textit{subleading terms} \nonumber \\
\label{ce.65}
&= (- \tau)^n  \partial_{j_1} \dots \partial_{j_n} \Delta M^{(0)}_{i_1 \dots i_k j_1 \dots j_n}
+ \textit{subleading terms} 
\end{align}
and therefore the theorem is proved for any CE level $n$.

\end{document}